\documentclass[12pt,bibliography=totoc,DIV=21]{scrartcl}
\pdfoutput=1
\usepackage[utf8]{inputenc}

\usepackage{amsmath,amssymb,color,graphicx,array}
\usepackage[symbol]{footmisc}

\usepackage{amsmath,amsfonts,mathtools}
\let\refeq\undefined
\usepackage[mathscr]{euscript}
\usepackage{enumitem}
\usepackage{longtable}
\usepackage{booktabs}
\usepackage{multirow}
\usepackage{cite}
\usepackage{graphicx, ulem, hhline}
\usepackage[table]{xcolor}
\usepackage{xspace}
\usepackage{needspace}
\usepackage{siunitx}
\usepackage{soul}
\usepackage{wasysym}              
\usepackage{hyphenat}             

\usepackage{afterpage}
\usepackage{etoolbox,xstring,xspace,calc,xifthen}
\usepackage{centernot}

\usepackage{tikz}
\makeatletter
\newif\iftag@here
\newcommand*{\taghere}[1][0pt]
{\ifmeasuring@\else
  \global\tag@heretrue
  \tikz[remember picture,overlay]{\coordinate (taghere) at (0pt,#1);}%
\fi}

\def\place@tag{%
    \iftagsleft@
      \kern-\tagshift@
      \iftag@here
        \global\tag@herefalse
        \tikz[remember picture,overlay]%
          {\path (taghere) -| node[anchor=base]{\rlap{\boxz@}} (0pt,0pt);}%
      \else
        \if1\shift@tag\row@\relax
            \rlap{\vbox{%
                \normalbaselines
                \boxz@
                \vbox to\lineht@{}%
                \raise@tag
            }}%
        \else
            \rlap{\boxz@}%
        \fi
        \kern\displaywidth@
      \fi
    \else
      \kern-\tagshift@
      \iftag@here
        \global\tag@herefalse
        \tikz[remember picture,overlay]%
          {\path  (taghere) -|  node[anchor=base]{\llap{\boxz@}} (0pt,0pt);}%
      \else
        \if1\shift@tag\row@\relax
            \llap{\vtop{%
                \raise@tag
                \normalbaselines
                \setbox\@ne\null
                \dp\@ne\lineht@
                \box\@ne
                \boxz@
            }}%
        \else \llap{\boxz@}%
        \fi
      \fi
    \fi
}
\makeatother

\long\def\symbolfootnote[#1]#2{\begingroup%
\def\thefootnote{\fnsymbol{footnote}}\footnote[#1]{#2}\endgroup}

\makeatletter
\def\@fnsymbol#1{\ensuremath{\ifcase#1\or%
\ast\or \dagger\or \ddagger\or \mathsection\or \parallel\or \nparallel\or%
\mathparagraph\or \cap\or \cup\or \subset\or \supset\or%
\wedge\or \vee\or <\or >\or \diamond\or \circ\or%
\vartriangle\or \triangledown\or \triangleleft\or \triangleright\or%

\else\@ctrerr\fi}}
\makeatother

\makeatletter
\newlength{\fnhskip}
\renewcommand\@makefntext[1]{
  \settowidth{\fnhskip}{\@makefnmark}
  \leftskip=\fnhskip
  \hskip-\fnhskip
  \@makefnmark#1
}
\makeatother

\renewenvironment{subequations}[1][]{
  \refstepcounter{equation}%
  \setcounter{parentequation}{\value{equation}}
  \setcounter{equation}{0}
  \def\theequation{\theparentequation\alph{equation}}%
  \let\parentlabel\label
  \ifx\\#1\\\relax\else\label{#1}\fi
  \ignorespaces
}{%
  \setcounter{equation}{\value{parentequation}}
  \ignorespacesafterend
}

\newcommand*{\nextParentEquation}[1][]{
  \refstepcounter{parentequation}
  \setcounter{equation}{0}
  \ifx\\#1\\\relax\else\parentlabel{#1}\fi
}

\usepackage{setspace}

\usepackage[numbers,sort&compress]{natbib}
\makeatletter
\def\NAT@spacechar{\,}
\makeatother
\usepackage[
colorlinks=true
,urlcolor=blue
,anchorcolor=blue
,citecolor=blue
,filecolor=blue
,linkcolor=blue
,menucolor=blue
,linktoc=all
,unicode
,backref=page
]{hyperref}
\usepackage{hypernat}

\newrobustcmd*{\tocref}[1]{\hyperref[TOC]{\color{black}{#1}}}
\newcommand{\tocsection}[2][]{\section[\boldmath #2]{\boldmath\tocref{#2#1}}}
\newcommand{\tocsubsection}[2][]{\subsection[#2]{\boldmath\tocref{#2#1}}}

\usepackage[all]{hypcap}
\usepackage{footnotebackref}

\usepackage[format=plain, margin=0.5cm, font=footnotesize, labelfont=bf]{caption}

\renewcommand*{\backref}[1]{}
\renewcommand*{\backrefalt}[4]{%
  \ifcase #1%
  \or [p\,#2]%
  \else [pp\,#2]%
  \fi%
}

\newif\ifbackrefshowonlyfirst
\backrefshowonlyfirsttrue
\makeatletter
\let\BR@direct@old@hyper@natlinkstart\hyper@natlinkstart
\renewcommand*{\hyper@natlinkstart}{\phantomsection\BR@direct@old@hyper@natlinkstart}
\let\BR@direct@oldBR@citex\BR@citex
\renewcommand*{\BR@citex}{\phantomsection\BR@direct@oldBR@citex}%

\long\def\hyper@page@BR@direct@ref#1#2#3{\hyperlink{#3}{#1}}

\ifx\backrefxxx\hyper@page@backref
    \let\backrefxxx\hyper@page@BR@direct@ref
    \ifbackrefshowonlyfirst
    \fi
\else
    \ifbackrefshowonlyfirst
    \fi
\fi

\patchcmd{\Hy@backout}{Doc-Start}{\@currentHref}{}{\errmessage{I can't seem to patch backref}}
\makeatother

\let\theparentequation\theequation
\patchcmd{\theparentequation}{equation}{parentequation}{}{}

\apptocmd{\thebibliography}{\scriptsize}{}{}
\let\OLDthebibliography\thebibliography
\renewcommand\thebibliography[1]{
  \OLDthebibliography{#1}
  \setlength{\parskip}{1pt}
  \setlength{\itemsep}{1pt plus 0.3ex}
}

\addtokomafont{disposition}{\rmfamily}
\BeforeTOCHead[toc]{{\pdfbookmark[1]{\contentsname}{toc}}}

\newcommand{\unicodescriptO}{^^f0^^9d^^92^^aa}
\newcommand{\unicodesupfour}{^^e2^^81^^b4}
\newcommand{\unicodesubt}{^^e2^^82^^9c}
\newcommand{\unicodeitalicY}{^^f0^^9d^^91^^8c}

\makeatletter
\patchcmd{\upbracefill}{\m@th}{\scriptstyle\m@th}{}{}
\patchcmd{\upbracefill}{$\braceld$}{$\scriptstyle\braceld$}{}{}
\patchcmd{\upbracefill}{\bracelu}{\bracelu\mkern-1mu}{}{}
\patchcmd{\upbracefill}{\hfill\braceru}{\hfill\mkern-1mu\braceru}{}{}
\DeclareOldFontCommand{\rm}{\normalfont\rmfamily}{\mathrm}
\DeclareOldFontCommand{\sf}{\normalfont\sffamily}{\mathsf}
\DeclareOldFontCommand{\tt}{\normalfont\ttfamily}{\mathtt}
\DeclareOldFontCommand{\bf}{\normalfont\bfseries}{\mathbf}
\DeclareOldFontCommand{\it}{\normalfont\itshape}{\mathit}
\makeatother

\newlength{\floatwidth}
\def\beq{\begin{equation}}
\def\eeq{\end{equation}}

\newcommand{\Anull}[1]{A_{0}{\left(#1\right)}}
\newcommand{\Bnull}[1]{B_{0}{\left(#1\right)}}

\newcommand{\g}[1]{g_{\scriptscriptstyle #1}}
\newcommand{\AtoB}[2]{\mbox{$#1\to #2$}}
\newcommand{\Amp}[4][\mathcal{A}]{#1^{\mbox{\tiny #2}}_{\mbox{\tiny #3}}\ifthenelse{\isempty{#4}}{}{{\left[#4\right]}}}
\newcommand{\fsl}[1]{{\centernot{#1}}}

\def\twomat[#1,#2][#3,#4]{\left( \begin{array}{cc} #1 & #2 \\ #3 & #4 \end{array} \right)}
\def\threemat[#1,#2,#3][#4,#5,#6][#7,#8,#9]{\left( \begin{array}{ccc} #1 & #2 & #3\\ #4 & #5 & #6 \\ #7 & #8 & #9 \end{array} \right)}
\def\twovec[#1,#2]{\left( \begin{array}{c} #1  \\ #2 \end{array} \right)}
\def\thv[#1,#2,#3]{\left( \begin{array}{c} #1 \\ #2 \\ #3 \end{array} \right)}
\def\twv[#1,#2]{\left( \begin{array}{c} #1 \\ #2 \end{array} \right)}

\newcommand{\IE}{\textit{i.\,e.}\xspace}
\newcommand{\EG}{\textit{e.\,g.}\xspace}
\newcommand{\refeq}[1]{Eq.\,\eqref{#1}}
\newcommand{\refeqs}[1]{Eqs.\,\eqref{#1}}

\newcommand{\fig}[1]{Fig.\,\ref{#1}}

\newcommand{\citere}[1]{Ref.\,\cite{#1}}
\newcommand{\citeres}[1]{Refs.\,\cite{#1}}
\newcommand{\simord}{\mathord{\sim}\,}

\newcommand{\yint}[3]{Y_{#1}^{[#2]}\ifthenelse{\isempty{#3}}{}{{\left(#3\right)}}}
\newcommand{\ysum}[3]{Y_{#1}^{#2}\ifthenelse{\isempty{#3}}{}{{\left(#3\right)}}}
\newcommand{\yonesum}[3]{{^0}Y_{#1}^{#2}\ifthenelse{\isempty{#3}}{}{{\left(#3\right)}}}
\newcommand{\tint}[2]{\ifthenelse{\isempty{#2}}{T_{#1}}{T_{#1}^{\left|#2\right.}}}

\newcommand{\toneint}[2]{\ifthenelse{\isempty{#2}}{{^0}T_{#1}}{^0T^{\left|#2\right.}_{#1}}}

\newcommand{\yoneint}[3]{{^0}Y_{#1}^{[#2]}\ifthenelse{\isempty{#3}}{}{{\left(#3\right)}}}
\newcommand{\bint}[1]{B_{0}\ifthenelse{\isempty{#1}}{}{{\left(#1\right)}}}
\newcommand{\aint}[1]{A_{0}\ifthenelse{\isempty{#1}}{}{{\left(#1\right)}}}
\newcommand{\binteps}[2]{\ifthenelse{\isempty{#2}}{B_{0}^{\left|#1\right.}}{B_{0}^{\left|#1\right.}{\!\left(#2\right)}}}

\newcommand{\CP}{\ensuremath{\mathcal{CP}}\xspace}

\newcommand{\Real}[1]{\Re\hspace{-1pt}\mathfrak{e}{\left[#1\right]}}

\newcounter{notecount}
\hypersetup{
  pdfauthor={Florian Domingo and Sebastian Paßehr}
  ,pdftitle={Towards Higgs masses and decay widths\\
    satisfying the symmetries in the (N)MSSM
  }
  ,pdfsubject={}
  ,pdfkeywords={}
}

\begin{document}

\newcommand*{\mytitle}[1]{%
  \parbox{\linewidth}{\setstretch{1.5}\centering\Large\textsc{\textbf{\boldmath #1}}}
}

\thispagestyle{empty}

\def\thefootnote{\fnsymbol{footnote}}

\begin{flushright}
  BONN-TH-2020-05\\
  TTK-20-23
\end{flushright}

\vfill

\begin{center}

\mytitle{
  Towards Higgs masses and decay widths\\
  satisfying the symmetries in the (N)MSSM
}

\vspace{1cm}

Florian Domingo$^{1}$\footnote{email: florian.domingo@csic.es}
and
Sebastian Pa{\ss}ehr$^{2}$\footnote{email: passehr@physik.rwth-aachen.de}

\vspace*{1cm}

\textsl{
$^1$Bethe Center for Theoretical Physics \&
Physikalisches Institut der Universit\"at Bonn,\\
Nu\ss allee 12, D--53115 Bonn, Germany
}

\medskip
$^2$\textsl{Institute for Theoretical Particle Physics and Cosmology,}\\
\textsl{RWTH Aachen University, Sommerfeldstra{\ss}e 16, 52074 Aachen, Germany.}

\end{center}

\vfill

\begin{abstract}
In models with an extended Higgs sector, such as the (N)MSSM, scalar
states mix with one another. Yet, the concept of Higgs mixing is
problematic at the radiative level, since it introduces both a scheme
and a gauge dependence. In particular, the definition of Higgs masses
and decay amplitudes can be impaired by the presence of
gauge-violating pieces of higher order. We discuss in depth the origin
and magnitude of such effects and consider two strategies that
minimize the dependence on the gauge-fixing parameter and
field-renormalization of one-loop order in the definition of the mass
and decay observables, both in degenerate and non-degenerate
scenarios. In addition, the intuitive concept of mixing and the
simplicity of its definition in terms of two-point diagrams can make
it tempting to include higher-order corrections on this side of the
calculation, irrespectively of the order achieved in vertex
diagrams. Using the global $SU(2)_{\mathrm{L}}$-symmetry in the
decoupling limit, we show that no improvement can be expected from
such an approach at the level of the Higgs decays, but that, on the
contrary, the higher-order terms may lead to numerically large
spurious effects.
\end{abstract}

\vfill

\def\thefootnote{\arabic{footnote}}
\setcounter{page}{0}
\setcounter{footnote}{0}

\hypersetup{linkcolor=black}
\tableofcontents\label{TOC}
\hypersetup{linkcolor=blue}

\tocsection{Introduction}

Many models of new physics suggest the existence of an extended Higgs
sector. In such a context, the Higgs boson discovered at the
LHC\,\cite{Aad:2012tfa,Chatrchyan:2012xdj,Aad:2015zhl} and presenting
characteristics that are approximately consistent with a Standard
Model~(SM)
interpretation\,\cite{Khachatryan:2016vau,Sirunyan:2018koj,Aad:2019mbh},
would be regarded as only one of many scalar states. The absence of
conclusive evidence for the additional Higgs bosons admittedly
constrains the available parameter space but continues to spare
multiple scenarios where, in general, the decoupling of light
new-physics states from the SM particles requires a careful handling
of the Higgs mixing (see
\EG\ \citere{Bahl:2018zmf}). On the other hand, the production of
heavy states at colliders is kinematically suppressed, hence leaving
little constraints on new Higgs bosons beyond the
TeV-range\,\cite{Aad:2020zxo}. The prototype of such extensions of the
SM is the Two-Higgs-Doublet Model~(THDM)\,\cite{Branco:2011iw} but
further enlargement through supersymmetric~(SUSY)
sectors\,\cite{Nilles:1983ge,Haber:1984rc} or singlet fields are easy
to motivate.

The presence of extended Higgs sectors opens up the possibility for
mixing between the Higgs fields. This concept appears as relatively
intuitive at the tree level, but is in fact ill-defined when
considering radiative corrections. Indeed, corresponding definitions
depend on the renormalization scheme and on the chosen gauge. The
effective-potential
approximation\,\cite{Okada:1990vk,Haber:1990aw,Ellis:1990nz,Ellis:1991zd,Brignole:1991pq}
offers a popular definition of the loop-corrected mixing, preserving
the unitarity of the mixing matrix, but the missing momentum-dependent
corrections make it inappropriate---or insufficient---for a consistent
description of external legs in Feynman amplitudes, in particular when
considering the decays of such mixed states. On the other hand, one
may directly use the LSZ reduction formula in order to define the
Higgs mixing in terms of the loop corrections applying on an external
Higgs leg in a physical amplitude: this has been described in
\EG\ \citeres{Williams:2011bu,Domingo:2017rhb,Baglio:2019nlc}. Among
the advantages of this approach, loop diagrams applying on the
external leg should be automatically contained within the mixing
matrix, making it formally suitable for a `pseudo-on-shell' treatment
of the external legs. On the other hand, unitarity of the mixing
matrix is lost in the inclusion of momentum-dependent pieces.

In this paper, we outline several shortcomings in the use of a mixing
matrix as a substitute to the loop expansion derived from the LSZ
reduction formula. These problems rest less in the principle of the
procedure than in its technical implementation. Indeed, the formalism
described in \EG\ \citere{Williams:2011bu} is designed in such a way
that it should coincide with the LSZ expansion, at least at the order
of the calculation. However, the separation of the contributions of
one-loop order between mixing effects on one side, and vertex
corrections (\EG\ in the case of a two-body Higgs decay) on the other,
lends itself to a misleading step, namely working with different
orders (or numerical parameters) on each side. While the resulting
mismatch is formally a higher-order effect, it can be numerically
significant due to imperfect cancellations. Reasons for distrusting
these partial higher-order contributions appear clearly when they
violate a symmetry that is expected to hold, either exactly or
approximately in a given regime. The electroweak gauge symmetry or the
global $SU(2)_{\mathrm{L}}$-symmetry in the limit of Higgs masses far
above the electroweak scale are examples of such handles on the
validity of the calculation, providing a guideline for the resolution
of the unphysical effects or at least an estimate of the associated
uncertainties.

In practice, we work in the context of the Next-to-Minimal
Supersymmetric Standard Model\,\cite{Ellwanger:2009dp,Maniatis:2009re}
and aim at improving our previous work on the Higgs decays at one-loop
electroweak order in this
model\,\cite{Domingo:2017rhb,Domingo:2018uim,Domingo:2019vit}---we
refer the reader
to \EG~\cite{Goodsell:2017pdq,Belanger:2017rgu,Krause:2018wmo,Krause:2019oar,Kanemura:2019kjg,Baglio:2019nlc,Krause:2019qwe}
for similar projects. However, our discussion in this paper should be
valid for a large class of extensions of the~SM based on a
THDM~framework (in particular the~MSSM). In fact, as the symmetry
arguments control the properties of doublet, but not of singlet
states, we focus below on scenarios with doublet-dominated, MSSM-like
Higgs bosons. In addition, the SUSY~context induces some additional
complications related to the connection between the gauge and the
quartic Higgs couplings. Indeed there are then too few degrees of
freedom available to simultaneously renormalize all the Higgs masses
on-shell. One advantage is the gain in predictivity, since the mass of
the SM-like Higgs boson cannot be set to an arbitrary value, while the
mass-splitting between heavy doublet states is determined by
electroweak effects. On the other hand, this causes a difficulty in
evaluating Feynman amplitudes involving a loop-corrected mass as
kinematical input and simultaneously preserving gauge
invariance. \textit{A priori}, such an issue also exists in a~THDM,
but there it can be easily evaded when working in an on-shell scheme
(see \EG~\citere{Pilaftsis:1997dr}). Consequently, we study in some
detail the gauge dependence in the observables and suggest possible
means of restoring a manifest gauge invariance. All our one-loop
calculations are performed with the assistance
of \texttt{FeynArts}\,\cite{Kublbeck:1990xc}, \texttt{FormCalc}\,\cite{Hahn:2000kx,Hahn:1998yk}
and \texttt{LoopTools}\,\cite{Hahn:1998yk}. Results at the two-loop
order are derived with the help
of \texttt{TwoCalc}\,\cite{Weiglein:1993hd}, \texttt{TSIL}\,\cite{Martin:2005qm}
and \texttt{TLDR}\,\cite{Goodsell:2019zfs}.

In the following sections, we first analyze the electroweak gauge
dependence in Higgs masses and decays. This provides us with a first
formal argument to disfavor the naive inclusion and/or resummation of
higher-order corrections in the Higgs-propagator matrix. Then, we
consider the global $SU(2)_{\mathrm{L}}$-symmetry in the limit of
massive doublet-like Higgs states and compare the consequences of this
symmetry at the analytical and the numerical level for heavy-Higgs
mass-splittings and several decay channels. Finally, we conclude as to
a cautious and consistent use of mixing formalisms and the combination
of mixing contributions with the vertex corrections in decay
amplitudes.

\tocsection{Aspects of gauge invariance on the determination of Higgs
masses and decays}

The loop-induced mixing in the Higgs sector is defined at the level of
the Higgs self-energies. However, the latter are not gauge-invariant
objects, in general, highlighting the artificiality of the
loop-corrected mixing matrix. Below, we detail how gauge invariance is
ensured at the one-loop order in observable quantities such as the
Higgs masses and decays, or how one could attempt to remedy its
violation by terms of higher order.

\tocsubsection{Self-energies and gauge dependence}

We first consider the gauge-dependent contributions to the
self-energies of one-loop order between two external Higgs legs~$h_i$
and~$h_j$ (\textit{a priori} mass eigenstates at the tree level) that
convey an external momentum~$p$. The calculation applies to extensions
of the~SM of type~THDM---details concerning the THDM Higgs sector are
provided in appendix\,\ref{ap:renTHDM}---though additional fields
(\EG\ scalar singlets) can be mixed to this sector as long as they do
not generate additional breaking of the electroweak symmetry. For
simplicity, we focus on the terms associated with the electroweak
charged current (the generalization to the neutral current is
straightforward). The terms that depend on the gauge-fixing
parameter~$\xi$ read
\begin{align}
  \label{eq:selfxi}
  16\,\pi^2\,\Sigma_{h_ih_j}(p^2) &\supset
  \left\{2\,M_W^2\,\Real{\g{h_iG^+W^-}\,\g{h_jG^+W^-}^*} - \g{h_iW^+W^-}\,\g{h_jW^+W^-}\!\right\}\notag\\
  &\quad \begin{aligned}[t] \,\times\,\Bigg\{
    & \!\!\left[\xi - \frac{p^2}{M_W^2}\right]^2
      \Bnull{p^2,\,\xi\, M_W^2,\,\xi\, M_W^2}\\
    &{-} \left[\xi - \left(1+\frac{p^2}{M_W^2}\right)\right]^2
      \Bnull{p^2,\,M_W^2,\,\xi\, M_W^2}
      -\frac{\xi}{M_W^2}\, \Anull{M_W^2}
    \!\Bigg\}
  \end{aligned}\notag\\
  &\quad-\left\{2\,\Real{\g{h_iH^+G^-}\,\g{h_jH^+G^-}^*}
    -\frac{2\left(m_{H^{\pm}}^2 - p^2\right)^2}{M_W^2}\,
    \Real{\g{h_iH^+W^-}\,\g{h_jH^+W^-}^*}\!\right\}\notag\\
    &\qquad \times \Bnull{p^2,\,m_{H^{\pm}}^2,\,\xi\, M_W^2}\notag\\
  &\quad-\left\{\g{h_iG^+G^-}\,\g{h_jG^+G^-} - \frac{p^4}{4\,M_W^4}\,
    \g{h_iW^+W^-}\,\g{h_jW^+W^-}\right\}
    \Bnull{p^2,\,\xi M_W^2,\,\xi\, M_W^2}\notag\\
  &\quad-\left\{2\,p^2\,\Real{\g{h_iG^+W^-}\,\g{h_jG^+W^-}^* + \g{h_iH^+W^-}\,\g{h_jH^+W^-}^*} + C_A\right\} \Anull{\xi\, M_W^2}\,.
\end{align}
All the~$\g{xyz}$ ($x,y,z\in\{h_{i,j},G^{\pm},H^{\pm},W^{\pm}\}$) represent Higgs couplings, while~$M_W$,
$m_{H^{\pm}}$, and~$m_{h_i}$ symbolize the~$W$, charged Higgs, and
neutral Higgs masses, respectively. We checked that \refeq{eq:selfxi}
agrees with expressions available in the literature, \EG\ Eq.\,(3.1)
of \citere{Martin:2003it} (see also
\citeres{Martin:2004kr,Pierce:1996zz}). Here, we have not explicitly
written the momentum-independent terms in~$\Anull{\xi\, M_W^2}$ and
just collected them within~$C_A$. In particular, we expect additional
gauge-dependent terms of this form from the counterterms, so that the
expression for~$C_A$ before renormalization is of limited interest.

In order to exploit \refeq{eq:selfxi}, it is useful to consider the
various relevant couplings, which are fully determined by the gauge
symmetry:
\begin{subequations}\allowdisplaybreaks
\label{eq:gaugecoup}
\begin{align}
  \g{h_iW^+W^-} &= g_2\,M_W \left[c_{\beta}\,X^R_{id} + s_{\beta}\,X^R_{iu}\right],\\
  \g{h_iG^+W^-} &= -\frac{g_2}{2} \left[c_{\beta}\left(X^R_{id} - \imath\, X^I_{id}\right) + s_{\beta} \left(X^R_{iu} + \imath\, X^R_{iu}\right)\right],\\
  \g{h_iH^+W^-} &= -\frac{g_2}{2} \left[s_{\beta} \left(X^R_{id} - \imath\, X^I_{id}\right) - c_{\beta} \left(X^R_{iu} + \imath\, X^R_{iu}\right)\right],\\
  \g{h_iG^+G^-} &= -\frac{g_2}{2}\,\frac{m_{h_i}^2}{M_W} \left[c_{\beta}\,X^R_{id} + s_{\beta}\,X^R_{iu}\right],\\
  \g{h_iH^+G^-} &= \left(\g{h_iG^+H^-}\right)^* = -\frac{g_2}{2}\,\frac{m_{H^{\pm}}^2-m_{h_i}^2}{M_W} \left[s_{\beta}\left(X^R_{id} - \imath\, X^I_{id}\right) - c_{\beta} \left(X^R_{iu} + \imath\, X^R_{iu}\right)\right].
\end{align}
\end{subequations}
Here, $m_{h_i}$ or~$m_{H^{\pm}}$ represent the tree-level Higgs masses
(in contrast with loop-corrected masses~$M_{h_i}$ or~$M_{H^{\pm}}$
that we consider later on). The symbol~$g_2$ is the
$SU(2)_{\mathrm{L}}$-gauge coupling and~$X_{iu/d}^{R/I}$ (we also
define~\mbox{$X_{ik}\equiv X_{ik}^R+\imath\,X^I_{ik}$}) correspond to
the components of the external Higgs~$h_i$ projecting onto the
real~($R$) or imaginary~($I$) part of the doublet fields~$H_d$
or~$H_u$ that take on a vacuum expectation
value~(v.e.v.)~$v\,c_{\beta}$ or~$v\,s_{\beta}$, respectively,
with~\mbox{$v=\sqrt{2}\,M_W/g_2$}. Then,
the first three lines of
\refeq{eq:selfxi} obviously vanish as long as~$h_i$ (or~$h_j$) is
orthogonal to the neutral Goldstone boson. The other terms amount to
\begin{align}
  \label{eq:selfxidep}
  16\,\pi^2\,\Sigma_{h_ih_j}(p^2) &\supset
  \frac{\g{h_iW^+W^-}\,\g{h_jW^+W^-}}{4\,M_W^4}
  \left(p^4 - m_{h_i}^2\,m_{h_j}^2\right) \Bnull{p^2,\,\xi\, M_W^2,\,\xi\, M_W^2}\notag\\
  &\quad+ \frac{2}{M_W^2}\,\Real{\g{h_iH^+W^-}\,\g{h_jH^+W^-}^*}
    \left[p^4 - m_{h_i}^2\,m_{h_j}^2 - m_{H^{\pm}}^2 \left(2\,p^2 - m_{h_i}^2 - m_{h_j}^2\right)\right]\notag\\
    &\qquad\times\Bnull{p^2,\,m_{H^{\pm}}^2,\,\xi\, M_W^2}\notag\\
  &\quad- \frac{g_2^2}{4\,M_W^2}\,\Real{X_{id}\,X_{jd}^* + X_{iu}\,X_{ju}^*}
      \left(2\,p^2 - m_{h_i}^2 - m_{h_j}^2+\tilde{C}_A\right) \Anull{\xi\, M_W^2}\,.
\end{align}
The dependence of this expression on the tree-level Higgs masses
originates in the Higgs--Goldstone couplings, except for
the~$A_0$~term. For this latter term, one systematically expects
contributions of the same form from other origins, \EG~tadpoles. In
order to clarify, we may consider the example of a SUSY framework with
on-shell renormalization conditions for the $W$, $Z$ and charged-Higgs
masses, and vanishing tadpoles (see \EG \citere{Frank:2006yh} for
details in the~MSSM): in such a `physical' setup, all the mass
parameters in the doublet sector are determined by observable
quantities; then the tadpole and $W$-, $Z$-, $H^{\pm}$-mass
counterterms contribute terms~${\propto}\,\Anull{\xi\,M_W^2}$ that
cancel out $\tilde{C}_A$ at the level of the renormalized diagonal
self-energies. Alternatively, still in the SUSY case, we could employ
$\overline{\text{DR}}$ conditions for the $W$, $Z$, $H^{\pm}$ masses,
so that a non-trivial $\tilde{C}_A$ would persist at the level of the
renormalized self-energies. We call such a scheme `unphysical' because
the renormalized tree-level mass parameters are not directly
observable quantities then, but implicitly gauge-dependent
parameters. This would also apply in a THDM with
$\overline{\text{MS}}$ conditions.

\tocsubsection{Gauge dependence and masses\label{sec:gaugedep_mass}}

Since we are interested in the gauge dependence emerging from the
electroweak one-loop order, we restrict ourselves to a determination
of the masses (and later decays) at the strict one-loop~(1L)
level, \IE~all two-loop~(2L) contributions (whether included in the
calculation or not) are regarded as being objects of higher
order. Obviously, it is possible to include fully known two-loop
orders---\EG~$\mathcal{O}{\left(\alpha_{t,b}\alpha_s,\,\alpha_{t,b}^2,\,\text{etc.}\right)}$---in
the picture as well, but what we comment about the gauge dependence
only applies as long as the two-loop electroweak order is not
considered.

In order to discuss gauge invariance, it is convenient to work in a
`physical' scheme, as defined above, \IE~a renormalization scheme
where tree-level parameters are directly related to observable
quantities, making them gauge-independent objects. Then, the
dependence on the gauge-fixing parameters is fully explicit in the
radiative contributions and should explicitly vanish (at the
considered order) in any observable quantity. A similar analysis would
be possible in an `unphysical' scheme as well, but only after
extracting the implicit gauge dependence contained in the tree-level
parameters, \IE~after relating them to observable quantities. Thus, we
assume below that we are working in a `physical' scheme, and that
\EG~mass~parameters are fixed by on-shell renormalization conditions.

\needspace{3ex}
From the explicit form of the gauge-dependent terms in
\refeq{eq:selfxidep}, we observe that the $\xi$-dependent contribution
to~$\Sigma_{h_ih_i}$ vanishes at~\mbox{$p^2=m_{h_i}^2$}---up to the
$\tilde{C}_A$ term, which only disappears after renormalization in a
`physical' scheme. Thus, no gauge-dependent term contaminates the
one-loop correction to the mass of~$h_i$:
\begin{align}\label{eq:massbasic}
  M_{h_i}^2 &\approx m_{h_i}^2 - \Real{\hat{\Sigma}_{h_ih_i}{\left(m_{h_i}^2\right)}}\,,
\end{align}
with~$\hat{\Sigma}$ denoting the renormalized self-energy. If, on the
contrary, the Higgs mass is determined by the implicit condition
\begin{subequations}\label{eq:massrecursion}
\begin{align}
  \mathcal{M}_{h_i}^2 &= m_{h_i}^2
  - \hat{\Sigma}_{h_ih_i}{\left(\mathcal{M}_{h_i}^2\right)}\\
  &\approx m_{h_i}^2 - \hat{\Sigma}_{h_ih_i}{\left(m_{h_i}^2\right)}
  + \hat{\Sigma}_{h_ih_i}{\left(m_{h_i}^2\right)}\,
  \frac{d\hat{\Sigma}_{h_ih_i}}{dp^2}{\left(m_{h_i}^2\right)}
  +\cdots\label{eq:massrecursionb}
\end{align}\end{subequations}
with the complex
pole~\mbox{$\mathcal{M}_{h_i}^2=M_{h_i}^2+\imath\,M_{h_i}\,\Gamma_{h_i}$}
consisting of the (real) pole mass~$M_{h_i}$ and the
width~$\Gamma_{h_i}$ (and derived through \EG\ an iterative
procedure), then the mismatch between~$M_{h_i}^2$ and~$m_{h_i}^2$
generates a gauge-dependent piece of two-loop order due to the terms
in \refeq{eq:selfxidep}: this is formally a contribution of higher
order---as is exhibited in the expansion of
\refeq{eq:massrecursionb}---and, though a source of uncertainty, can
be dismissed in principle in virtue of the expansion. In addition, the
off-shell renormalization of the Higgs self-energy requires the
introduction of field~renormalization, introducing an explicit (and
unphysical) dependence on the latter at the level of the Higgs masses
as defined in \refeqs{eq:massrecursion}; this was already discussed
in \citere{Bahl:2018ykj}. We apply $\overline{\text{DR}}$-conditions
on the Higgs fields, so that the dependence on field~renormalization
is going to translate into a dependence on the renormalization~scale.

It is actually common
practice\,\cite{Frank:2006yh,Degrassi:2009yq,Williams:2011bu,Graf:2012hh,Athron:2014yba,Goodsell:2014bna,Goodsell:2016udb,Drechsel:2016jdg,Domingo:2017rhb,Athron:2017fvs,Bahl:2018qog,Dao:2019nxi}
to include the full propagator matrix in the determination of the
Higgs masses. The latter are then obtained from a complex pole search
using the following condition on the characteristic polynomial of the
inverse propagator matrix:
\begin{align}\label{eq:massimplicit}
  \det{\left[p^2 - \mathrm{diag}{\left(m^2_{h_i}\right)} + \mathbf{\hat{\Sigma}}{\left(p^2\right)}\right]} &= 0\,.
\end{align}
The impact of off-diagonal one-loop self-energies on the Higgs-mass
calculation is in general of two-loop order. Yet, such off-diagonal
terms intervene at one-loop order in the case where they mix nearly
degenerate tree-level states, justifying a more intricate
procedure. Let us first consider the case where there is no
approximate degeneracy with the state~$h_i$ (\IE\ \mbox{$\lvert
m_{h_i}^2 - m_{h_j}^2\rvert\gg\lvert\hat{\Sigma}_{h_ih_j}\rvert$} for
\mbox{$j\neq i$}). Then, the condition defining the Higgs mass amounts
to the following expansion:
\begin{align}\label{eq:massoffdiag}
  \mathcal{M}_{h_i}^2 &= m^2_{h_i}
  - \hat{\Sigma}_{h_ih_i}{\left(\mathcal{M}_{h_i}^2\right)}
  + \sum_{j\neq i} \frac{\hat{\Sigma}_{h_ih_j}{\left(\mathcal{M}_{h_i}^2\right)}\,
    \hat{\Sigma}_{h_jh_i}{\left(\mathcal{M}_{h_i}^2\right)}}
  {\mathcal{M}_{h_i}^2 - m_{h_j}^2} + \cdots\,.
\end{align}
From the expressions in \refeq{eq:selfxidep}, we observe that the
off-diagonal terms generate new gauge-dependent contributions of
two-loop order. In fact, even though the self-energies would be
evaluated at the tree-level mass~$m_{h_i}^2$, these gauge-dependent
terms would persist, indicating the need for a full electroweak
calculation of two-loop order to control gauge invariance at this
level.

Now, let us consider the case where~$h_i$ and~$h_j$ are nearly
degenerate. Then,
$\hat{\Sigma}_{h_ih_i}{\left(\mathcal{M}_{h_i}^2\right)}$,
$\hat{\Sigma}_{h_ih_j}{\left(\mathcal{M}_{h_i}^2\right)}$,
$\hat{\Sigma}_{h_jh_i}{\left(\mathcal{M}_{h_i}^2\right)}$
and~$\hat{\Sigma}_{h_jh_j}{\left(\mathcal{M}_{h_i}^2\right)}$ all
intervene with a weight of one-loop order in the determination of the
Higgs mass~$M_{h_i}^2$. However, since \mbox{$\lvert M_{h_i}^2 -
m_{h_i}^2\rvert\approx\lvert M_{h_i}^2 -
m_{h_j}^2\rvert=\mathcal{O}{(\text{1L})}$}, the gauge-violating
pieces---see \refeq{eq:selfxidep}---originating in the off-diagonal
terms are still of subleading order at the level of the Higgs mass. We
postpone further discussion of the degenerate case to
section\,\ref{sec:neardeg}.

\begin{figure}[p!]
\centering
\includegraphics[width=\textwidth]{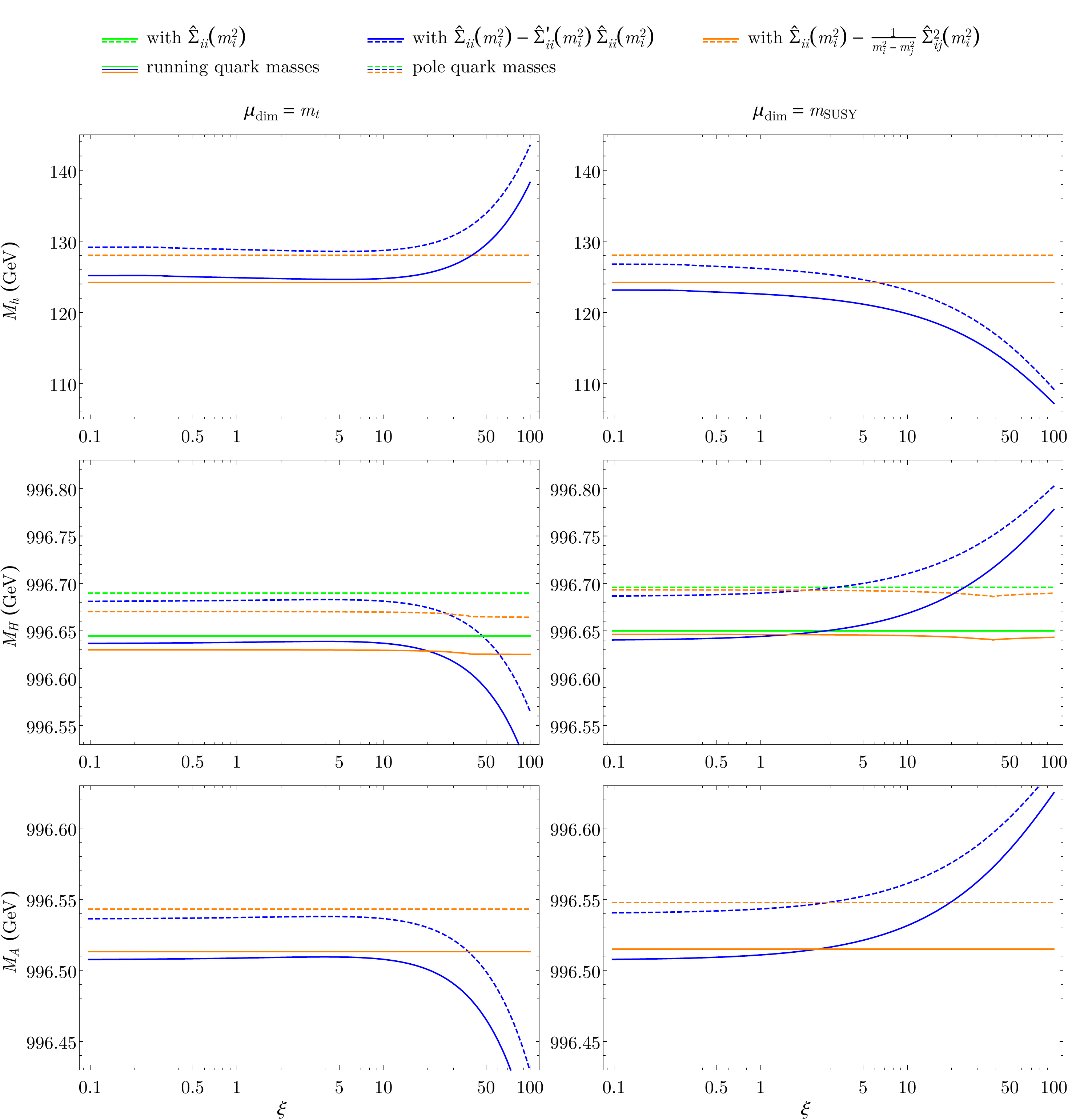}
\caption{The gauge dependence of the MSSM Higgs masses is shown for
  \mbox{$m_{H^{\pm}}=1$}\,TeV, \mbox{$t_{\beta}=10$}.  The horizontal
  green lines correspond to the expansion of one-loop order as
  in \refeq{eq:massbasic}. The blue curves correspond to the
  $\xi$-dependence originating from the diagonal self-energy (the
  off-diagonal terms are set to~$0$) as
  in \refeqs{eq:massrecursion}. In the orange
  curves (on top of the green curves for $M_h$ and
  $M_A$), the momentum in the diagonal self-energy is evaluated at
  the tree-level mass, so that no gauge-dependence is introduced by
  this term, and the $\xi$-dependence originating from the
  off-diagonal terms---see \refeq{eq:massoffdiag}---is studied. The
  plots on the left-hand side are obtained for the renormalization
  scale $\mu_{\text{dim}}=m_t$, while those on the right-hand side use
  $\mu_{\text{dim}}=m_{\text{SUSY}}\equiv 1.5$\,TeV. The input
  parameters ($t_{\beta}$) have been accordingly transformed so that
  the two schemes consider the same point in parameter space. Finally,
  for the Higgs masses at the one-loop order, it is formally
  equivalent to employ Yukawa couplings and quark masses defined
  on-shell (dashed curves) or QCD-corrected $\overline{\text{DR}}$
  running masses, with a scale corresponding to the Higgs mass (solid
  curves): the difference between these mass predictions provides an
  estimate of the magnitude of leading two-loop order
  effects.\label{fig:xidep_MSSM}}
\end{figure}

The presence of gauge-dependent pieces of two-loop order in the
one-loop-corrected Higgs masses obtained from the pole search can be
exploited in order to estimate (a lower bound on) the uncertainty
associated with such a determination. This was already performed
in \EG\ \citere{Dao:2019nxi}, and we only consider one example point
in a non-degenerate scenario for illustration
in \fig{fig:xidep_MSSM}.\footnote{We constrain our analysis to the
case of an~$R_\xi$~gauge with all gauge-fixing parameters being
equal.} There, we consider the MSSM with
\mbox{$m_{H^{\pm}}=1$}\,TeV, \mbox{$t_{\beta}=10$}, squark masses
of~\mbox{$\simord1.5$}\,TeV for the third generation
(\mbox{$\simord2$}\,TeV for the other two) and electroweakino masses
in the range of a few~$100$\,GeV. The $\xi$-dependence from the
diagonal self-energy is found to dominate the $\xi$-variation~(blue
curve), when setting the external momentum to the loop-corrected mass
value. The effect is sizable at large $\xi$ for the SM-like Higgs and
much smaller for the heavy-doublet states (though in fact of
comparable magnitude at the level of the self-energies). This
variation in a broad range of $\xi$ exposes the presence of a
gauge-violating piece in the definition of the Higgs masses
via \refeqs{eq:massrecursion}, hence hinting at the necessity to
separate genuine radiative effects from spurious symmetry-violating
artifacts in the interpretation of the results. However, before this
gauge dependence can be interpreted as an uncertainty applying to the
iterative mass determination with \refeqs{eq:massrecursion}, one needs
to assess the relevant range of $\xi$-variation. Indeed, large values
of~$\xi$ introduce a new scale in the calculation, hence appear less
suited for reliable predictions. Stability in a $\xi$-range of order
unity would thus appear as a sufficient criterion. Restricting
ourselves to~\mbox{$\xi\lesssim5$}, we observe that the typical mass
variation associated with gauge dependence for the SM-like Higgs is of
order~$0.3$--$2$\,GeV (depending on the chosen scale for field
renormalization): in a fixed-order calculation, this gauge uncertainty
can only be reduced after inclusion of the two-loop gauge corrections.

The off-diagonal contribution to the $\xi$-dependence~(orange curve)
is heavily suppressed by the clear hierarchy in the \CP-even
sector. The somewhat more-pronounced $\xi$-dependence from the
off-diagonal terms in $M_{h_2}$ at $\xi\gtrsim30$ is due to the
crossing of thresholds in loop functions (\EG\ originating from
the~$G^+$--$G^-$~loop). In fact, this curve exhibits a more troubling
feature than $\xi$-dependence, which is its shift away from the green
curve, whereas both curves lie on top of each other in the case of
the \CP-odd Higgs. This $SU(2)$-violating effect will be discussed
more in depth in section\,\ref{sec:log} and is due to a non-decoupling
feature of the matrix description in the non-degenerate case,
requiring the inclusion of a 2L charged-Higgs mass counterterm for a
consistent order counting. Admittedly, the impact is numerically quite
small at the level of the mass itself, but competes in magnitude with
2L effects, opening the question of the relevance of applying
higher-order corrections to such states.

In addition, all these definitions of the
mass---from \refeqs{eq:massrecursion} or
\refeq{eq:massoffdiag}---explicitly depend (at two-loop order) on the
renormalization of the Higgs fields, as we commented above. This
introduces an explicit dependence on the renormalization scale, as can
be observed by comparing the left- and right-hand side of
\fig{fig:xidep_MSSM}. In contrast, the tiny scale~dependence in the
masses obtained from \refeq{eq:massbasic} (green curves) is implicit
and originates in higher orders in the conversion of $t_{\beta}$
between the two choices of renormalization scale. In particular, we
stress that the mass shift associated with \refeq{eq:massrecursion}
can be given whatever sign at the level of the SM-like state,
depending on the choice of field renormalization. It is thus devoid of
physical content as long as corresponding 2L effects, neutralizing the
dependence on field renormalization, are not included. From this
analysis, we see that there is no gain in precision, at the one-loop
order in the non-degenerate scenario, in including the shifts of
two-loop order of \refeqs{eq:massrecursion} or \refeq{eq:massoffdiag}
since these introduce a non-physical behavior for the would-be
observable masses: namely an explicit dependence on $\xi$ and the
field~renormalization. The combined impact of these artifacts on the
mass~determination competes with the magnitude of the leading two-loop
corrections of
$\mathcal{O}{\left(\alpha_{t,b}\alpha_s\right)}$---estimated in
\fig{fig:xidep_MSSM} through a variation of the numerical input for
the Yukawa couplings (solid vs.\ dashed lines). Admittedly, the result
obtained for low values of~$\xi$ (\EG\ equal to~$1$)
with \refeqs{eq:massrecursion} looks comparatively close to the
gauge-independent result of \refeq{eq:massbasic}. However, the scale
variation alone evidences a shift of~$\simord2$\,GeV at the level of
the SM-like Higgs mass, hence a sizable uncertainty due to the
explicit dependence on the field renormalization; such contributions
were neglected in the uncertainty estimates
of \citeres{Degrassi:2002fi,Bahl:2019hmm} for `fixed-order'
calculations, obtained without or through a very narrow scale
variation. Yet, we believe that the SUSY~scale should be as legitimate
as the electroweak scale for the renormalization of the Higgs
fields. Thus, the uncertainty originating in the prescription for the
Higgs-pole determination---as already discussed
in \citere{Bahl:2018ykj}---should be added to the previous assessments
at fixed order (after being adapted to the order controlled in the
calculation).\footnote{At \mbox{$\xi\sim1$}, most of the dependence of
$M_{h}$ on the field renormalization in the $\mathcal{O}(\text{2L})$
terms of \refeqs{eq:massrecursionb} and (\ref{eq:massoffdiag}) is
neutralized by the inclusion of the two-loop self-energies of
$\mathcal{O}(\alpha_t^2)$, reducing this source of uncertainty in the
scenario of \fig{fig:xidep_MSSM} in the range of
$\mathcal{O}(100\,\text{MeV})$; this is also the setup
of \citeres{Degrassi:2002fi,Bahl:2019hmm}. Of course, the pole search
generates further field-dependent terms beyond those considered
here. On the other hand, the gauge variation, also of the order of~GeV
for $\xi\lesssim5$, cannot be reduced without inclusion of the 2L
gauge corrections, as we commented above.} We thus believe that the
violation of the gauge symmetry by the inclusion of an incomplete
electroweak two-loop order indicates that the reliability of the
prediction is not improved, but rather that the uncertainty is
inflated by the introduction of symmetry-violating pieces. While we
cannot materially check that the dependence on the gauge-fixing
parameters will continue to vanish by pushing the expansion up to the
two-loop order (we miss the 2L electroweak terms), as suggested
in \citere{Martin:2005eg}, we strongly believe this to be the case (at
least in the non-degenerate scenario). On the other hand, we
explicitly checked the cancellation of the dependence on 1L field
counterterms between 2L and 1L$^2$ pieces (provided the pieces are
perturbatively combined).

\tocsubsection{Gauge dependence and decays in the non-degenerate case}

At the level of the Higgs decays, loop corrections on the external
Higgs leg appear explicitly unless the Higgs fields are renormalized
on-shell---in which case corresponding contributions are entirely
shifted to the vertex counterterms. Contrarily to the case of the
Higgs-mass determination, where only the diagonal self-energy was
formally needed at the one-loop order, both diagonal and off-diagonal
self-energies intervene at this order in the decays. As we discussed
above, the off-diagonal self-energies are gauge dependent at the
one-loop order even when we set~\mbox{$p^2=m_{h_i}^2$}: this indicates
that these objects---or the associated Higgs-mixing matrix at the
one-loop order---are not physical observables (obviously, they are
also scheme dependent), but simply intermediate steps in the
calculation of the decay width.

For definiteness, we can consider the example of the
decay~\AtoB{h_i}{t\bar{t}}. The vertex corrections (including the
on-shell renormalization of the top-quark fields) produce the
following gauge-dependent terms (still restricting ourselves to the
charged currents):
\begin{subequations}\label{eq:vertxi}
\begin{align}\label{eq:vertxiamp}
  \Amp{\text{vert}}{}{\AtoB{h_i}{t\bar{t}\,}}{} &\supset
  \frac{\imath\, g^3\, m_t}{64\,\pi^2\,M_W^3}\,\bar{u}(p_t)\left[
  \left(p^2 - m_{h_i}^2\right) \mathscr{A}_1^{\text{\tiny vert}} +
  \left(p^2 - m_{H^{\pm}}^2\right) \mathscr{A}_2^{\text{\tiny vert}} +
  \mathscr{A}_3^{\text{\tiny vert}}
  \right] v(p_{\bar{t}})\,,\\
  \begin{split}
  \mathscr{A}_1^{\text{\tiny vert}} &= \left(c_{\beta}\,X_{id}^R + s_{\beta}\,X_{iu}^R\right)
    I_1{\left(-p_t,\,p_{\bar{t}},\,m_b^2,\,\xi\, M_W^2,\,\xi\, M_W^2\right)}\\
    &\quad+ \left(-s_{\beta}\,X_{id}^* + c_{\beta}\,X_{iu}\right)
    I_2{\left(-p_t,\,p_{\bar{t}},\,m_b^2,\,m_{H^{\pm}}^2,\,\xi\,M_W^2\right)}\\
    &\quad+ \left(-s_{\beta}\,X_{id} + c_{\beta}\,X_{iu}^*\right)
    I_2{\left(p_{\bar{t}},\,-p_t,\,m_b^2,\,m_{H^{\pm}}^2,\,\xi\, M_W^2\right)}\,,
  \end{split}\\
  \mathscr{A}_2^{\text{\tiny vert}} &=
  \frac{1}{t_{\beta}} \left[\left(-s_{\beta}\,X_{id}^* + c_{\beta}\,X_{iu}\right) P_L
    + \left(-s_{\beta}\,X_{id} + c_{\beta}\,X_{iu}^*\right) P_R\right]
  \Bnull{p^2,\,m_{H^{\pm}}^2,\,\xi\, M_W^2}\,,\\
  \mathscr{A}_3^{\text{\tiny vert}} &=
  \frac{p^2}{2} \left(c_{\beta}\,X_{id}^R + s_{\beta}\,X_{iu}^R\right)
  \Bnull{p^2,\,\xi\, M_W^2,\,\xi\, M_W^2}
  -\frac{1}{2\,s_{\beta}} \left(X_{iu}^R - \imath\,\gamma_5\,X_{iu}^I\right)
  \Anull{\xi\, M_W^2}
\end{align}
\end{subequations}
with the vertex functions
\begin{subequations}\small
\begin{align}
  I_1{\left(-p_t,\,p_{\bar{t}},\,m_b^2,\,\xi\, M_W^2,\,\xi\, M_W^2\right)} &\equiv
  \frac{\imath\,16\,\pi^2}{m_t} \int\frac{d^Dk}{(2\,\pi)^D}
  \frac{\fsl{k}\left(m_t^2\,P_R + m_b^2\,P_L\right) - m_t\,m_b^2}{
    \left[k^2 - m_b^2\right]\left[\left(k - p_t\right)^2 - \xi\, M_W^2\right]
    \left[\left(k + p_{\bar{t}}\right)^2 - \xi\, M_W^2\right]}\,,\\[1ex]
  I_2{\left(-p_t,\,p_{\bar{t}},\,m_b^2,\,\xi\, M_W^2,\,\xi\, M_W^2\right)} &\equiv
  \frac{\imath\,16\,\pi^2}{m_t} \int\frac{d^Dk}{(2\,\pi)^D}
  \frac{\fsl{k}\left(m_t^2\,t_{\beta}^{-1}\,P_R - m_b^2\,t_{\beta}\,P_L\right)
    + m_t\,m_b^2\left(t_{\beta}\,P_R - t_{\beta}^{-1}\,P_L\right)}{
    \left[k^2 - m_b^2\right]\left[\left(k - p_t\right)^2 - m^2_{H^{\pm}}\right]
    \left[\left(k + p_{\bar{t}}\right)^2 - \xi\, M_W^2\right]}\,.\notag\\[-3.5ex]
\end{align}
\end{subequations}
Here, $h_i$ is identified with the field (mass eigenvector) that is
associated with the tree-level mass~$m_{h_i}$. Yet we allow its
external momentum~$p$ to be `free', with~\mbox{$p = p_t +
p_{\bar{t}}$} ($p_{t}$ and~$p_{\bar{t}}$ are the external momenta
associated with the on-shell quark lines). In the expression above,
$m_{h_i}^2$ originates in the Higgs--Goldstone couplings, while~$p^2$
appears in scalar products of external momenta. Obviously, the
$\xi$-dependence in the three-point vertex functions~$I_{1,2}$ only
disappears if the mass appearing in the Higgs--Goldstone coupling
coincides with the kinematical one. Thus, employing a loop-corrected
mass under the claim that it provides a better description of the
kinematical situation would also generate a gauge-violating
contribution of two-loop order from these terms: it is consequently
arguable whether this choice brings any actual improvement at the
numerical level.

The gauge-dependent~$B_0$ and~$A_0$~terms in \refeqs{eq:vertxi} must
be combined with the contributions from loop~corrections on the
external Higgs leg, as prescribed by the LSZ reduction formula. Then,
the terms of \refeq{eq:selfxidep} generate
\begin{subequations}\label{eq:mixxi}\allowdisplaybreaks
\begin{align}\label{eq:mixxiamp}
  \Amp{\text{mix}}{}{\AtoB{h_i}{t\bar{t}\,}}{} &\supset
  -\frac{\imath\, g^3\, m_t}{64\,\pi^2\,M_W^3}\,\bar{u}(p_t) \left[
    \left(p^2-m^2_{h_i}\right) \mathscr{A}_1^{\text{\tiny mix}}
    + \left(p^2 - m_{H^{\pm}}^2\right) \mathscr{A}_2^{\text{\tiny mix}}
    + \mathscr{A}_3^{\text{\tiny mix}} \right] v(p_{\bar{t}})\,,\\
  \label{eq:mixxirem}
  \mathscr{A}_1^{\text{\tiny mix}} &=
  \frac{X_{iu}\,P_L + X_{iu}^*\,P_R}{4\,s_{\beta}} \begin{aligned}[t]\,\Big[
    & 2\left|-s_{\beta}X_{id}^*+c_{\beta}X_{iu}\right|^2
      \left(p^2 + m^2_{h_i} - 2\,m^2_{H^{\pm}}\right)
      \partial_{p^2}\Bnull{p^2,\,m_{H^{\pm}}^2,\,\xi\, M_W^2}\\
    &{+} \left(c_{\beta}\,X_{id}^R + s_{\beta}\,X_{iu}^R\right)^2
      \left(p^2 + m_{h_i}^2\right)
      \partial_{p^2}\Bnull{p^2,\,\xi\, M_W^2,\,\xi\, M_W^2}\Big]
  \end{aligned}\notag\\
  &\quad+\frac{1}{2\,s_{\beta}} \sum_{j\neq i}
  \frac{X_{ju}\,P_L + X_{ju}^*\,P_R}{p^2 - m^2_{h_j}}\notag\\[-.6ex]
  &\qquad\times\begin{aligned}[t]\!\Big[
    & 2\,\Real{\left(-s_{\beta}\,X_{id}^* + c_{\beta}\,X_{iu}\right)
        \!\left(-s_{\beta}\,X_{jd} + c_{\beta}\,X_{ju}^*\right)\!}
      \!\left(m_{h_j}^2 - m_{H^{\pm}}^2\right)
      \!\Bnull{p^2,\,m_{H^{\pm}}^2,\,\xi\, M_W^2}\\
    &{+} \left(c_{\beta}\,X_{id}^R + s_{\beta}\,X_{iu}^R\right)
      \left(c_{\beta}\,X_{jd}^R + s_{\beta}\,X_{ju}^R\right)
      m_{h_j}^2\,\Bnull{p^2,\,\xi\, M_W^2,\,\xi\, M_W^2}\\
    &{-}\,\Real{X_{id}\,X_{jd}^* + X_{iu}\,X_{ju}^*}\, \Anull{\xi\, M_W^2}\Big]\,,
  \taghere\end{aligned}\\
  \mathscr{A}_2^{\text{\tiny mix}} &=
  \frac{1}{t_{\beta}} \left[\left(-s_{\beta}\,X_{id}^* + c_{\beta}\,X_{iu}\right) P_L + \left(-s_{\beta}\,X_{id} + c_{\beta}\,X_{iu}^*\right) P_R\right]
  \Bnull{p^2,\,m_{H^{\pm}}^2,\,\xi\, M_W^2}\\
  \mathscr{A}_3^{\text{\tiny mix}} &=
  \frac{p^2}{2} \left(c_{\beta}\,X_{id}^R + s_{\beta}\,X_{iu}^R\right)
  \Bnull{p^2,\,\xi\, M_W^2,\,\xi\, M_W^2}
  - \frac{1}{2\,s_{\beta}} \left(X_{iu}^R - \imath\,\gamma_5\,X_{iu}^I\right)
  \Anull{\xi\, M_W^2}\,.\hspace{-1em}
\end{align}
\end{subequations}
Adding \refeq{eq:vertxiamp} and \refeq{eq:mixxiamp}, we observe the
cancellation of the $B_0$ and $A_0$~terms up to the remainder
in~$\mathscr{A}_1^{\text{\tiny mix}}$ of \refeq{eq:mixxirem}. The
latter (generically) disappears only if its prefactor is zero, \IE~the
kinematical mass~$p^2$ and the tree-level mass~$m^2_{h_i}$ appearing
in the Higgs--Goldstone couplings coincide.

The $\xi$-dependence from the electroweak neutral current vanishes in
a similar way; additional terms from the mixing of the external Higgs
with internal $Z$ and $G^0$ cancel out separately up to contributions
proportional to~$(p^2 - m^2_{h_i})$. This cancellation has already
been discussed, \EG\ in \citeres{Williams:2011bu,Domingo:2018uim}.

So far, we have checked how the LSZ reduction formula ensures gauge
invariance in the decay amplitude, up to higher-order terms
\mbox{${\propto}\,(p^2 - m^2_{h_i})$}. Now, let us turn to the mixing
formalism of \citeres{Williams:2011bu,Domingo:2017rhb}. The
loop-corrected field is defined as~\mbox{$H_k=Z_{ki}\,h_i$}, where the
loop-corrected mixing matrix~$\mathbf{Z}$ is built out of
eigenvectors~$(Z_k)_{i=1,\ldots}$ of the effective
mass~matrix~$[\mathrm{diag}(m^2_{h_i})-\mathbf{\hat{\Sigma}}(\mathcal{M}^2_{H_k})]$
for the associated eigenvalue~$\mathcal{M}^2_{H_k}$ and satisfying the
normalization
condition~\mbox{$[\delta_{ij}+\hat{\Sigma}^\prime_{ij}{(\mathcal{M}_{H_k}^2)}]\,Z_{ki}\,Z_{kj}=1$}\,\cite{Domingo:2017rhb}. By
construction, in the absence of degeneracies,
$Z_{ki}\,\Amp{\text{tree}}{}{\AtoB{h_i}{t\bar{t}\,}}{}$ coincides with
the expansion
$(\Amp{\text{tree}}{}{}{}+\Amp{\text{mix}}{}{}{})[\AtoB{h_k}{t\bar{t}\,}]$
at the one-loop order. Similarly,
\mbox{$\Amp{\text{vert}}{}{\AtoB{H_k}{t\bar{t}\,}}{}=Z_{ki}\,\Amp{\text{vert}}{}{\AtoB{h_i}{t\bar{t}\,}}{}$}
coincides with~$\Amp{\text{vert}}{}{\AtoB{h_k}{t\bar{t}\,}}{}$ at this
order. If degeneracies are present, the same formal expansion as above
applies (though ill-converging). Gauge invariance is thus satisfied at
the strict one-loop order in all the cases. However, there remain
gauge-dependent pieces of higher order. The
terms~${\propto}\,(p^2-m^2_{h_i})$ of \refeqs{eq:vertxi}
and \refeqs{eq:mixxi}
become~${\propto}\,Z_{ki}\,(\mathcal{M}_{H_k}^2-m_{h_i}^2)$, which is
an object of one-loop order, hence generates gauge-violating
contributions of two-loop order. In addition, there are
gauge-violating effects beyond those contained in the LSZ expansion,
due to the resummation of mixing effects in the mixing matrix and the
inclusion of terms from the product of mixing and vertex
corrections. The restoration of gauge invariance will thus prove more
difficult in this formalism.

\begin{figure}[p!]
  \centering
  \includegraphics[width=\textwidth]{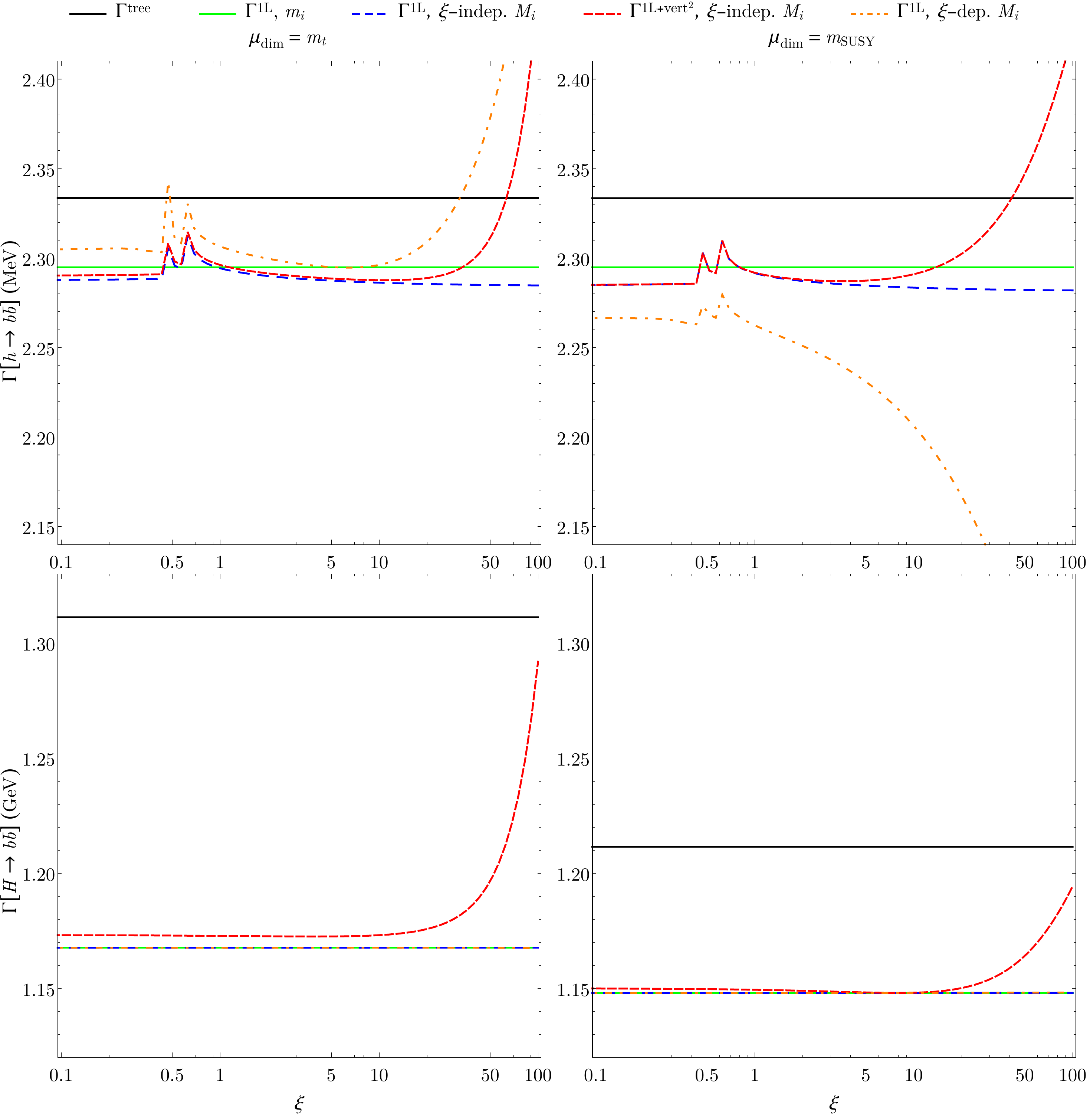}
  \caption{The $\xi$-dependence in the Higgs decay widths into
  $b\bar{b}$ is shown in the scenario of \fig{fig:xidep_MSSM}. The
  solid black curve represents the QCD-corrected tree-level width,
  while the solid green line corresponds to a full one-loop
  calculation where the one-loop functions are evaluated at the
  tree-level value of the Higgs mass. In both cases, the kinematics
  employs the loop-corrected mass of \refeq{eq:massbasic}. For the
  dashed blue curve, this same loop-corrected mass is also used in
  the one-loop functions. In addition, a
  $\lvert\Amp{vert}{}{}\rvert^2$ piece is kept in the squared
  amplitude for the dashed red curve. In the dot-dashed orange curve,
  no explicit contribution to the decay of two-loop order is included,
  but the momentum is set to the $\xi$-dependent Higgs mass
  of \refeqs{eq:massrecursion}. The renormalization scale is set to
  $m_t$ on the left-hand side, and to $m_{\text{SUSY}}$ on the
  right-hand side. The decay widths for the pseudoscalar state have
  been omitted since they are essentially identical to those of the
  heavy \CP-even state (lower row).\label{fig:xidepwidth}}
\end{figure}

In \fig{fig:xidepwidth}, we show the $\xi$-dependence in the decay
widths $\Gamma[h_i\to b\bar{b}]$ in the MSSM for the point considered
in \fig{fig:xidep_MSSM}. The latter is threefold, originating firstly
in the explicit $\xi$-dependence of the decay width from the terms
of \refeqs{eq:vertxi} and \refeqs{eq:mixxi} when the Higgs mass is set
to a loop-corrected value, secondly in a possible processing of mixing
and vertex corrections on different footings, thirdly in the implicit
$\xi$-dependence of the mass when it is derived
from \EG~\refeq{eq:massoffdiag}. In order to illustrate these
features, we plot several definitions of the decay widths:
\begin{itemize}
\item the solid black curves correspond to the (inclusive)
  QCD-corrected tree-level decay widths, where however the kinematical
  factors employ the loop-corrected masses of \refeq{eq:massbasic};
\item a first version of the decay widths of full one-loop order is
  shown in solid~green: there, the amplitudes are evaluated at the
  tree-level Higgs mass, while the kinematics employs the
  mass~determination of \refeq{eq:massbasic}, leading to an explicitly
  gauge-independent result (for the heavy-doublet states, this curve
  is hardly distinguishable from the blue and orange ones); the
  difference with the solid black curves provides the magnitude of the
  (non-QCD) radiative corrections;
\item for the dashed blue curves, the $\xi$-independent Higgs mass
  of \refeq{eq:massbasic} is employed everywhere, leading to explicit
  $\xi$-dependent decay widths due to the terms of \refeqs{eq:vertxi}
  and \refeqs{eq:mixxi}; this gauge-dependence is found to be rather
  mild in the example, showing only at the level of the light Higgs
  through threshold effects;
\item the dashed red curves show the impact of keeping a term $\lvert
  A_{\text{vert}}\rvert^2$ in the decay width: the $\xi$-dependence is
  sizable for all the states, competing with the absolute magnitude of
  the electroweak effects;
\item finally, the dot-dashed orange curve corresponds to a decay width of
  one-loop order employing a $\xi$-dependent loop-corrected Higgs
  mass, \IE~adding to the explicit gauge dependence of the decay width
  the implicit one contained in \refeqs{eq:massrecursion}: the effect
  is mostly relevant for the light Higgs, as could be expected from
  the impact at the level of the mass~determination.
\end{itemize}
This comparison in particular shows that while the explicit
gauge~dependence of the decay width at strict (truncated) one-loop
order remains rather mild, the predictivity of the calculation can be
wasted when vertex and mixing contributions are not consistently
combined so as to neutralize the gauge~dependence. In addition, we
also vary the renormalization scale between $m_t$ (left) and
\mbox{$m_{\text{SUSY}}=1.5$}\,TeV (right): the associated effects appear to be
mostly driven by the dependence on the Higgs mass (dot-dashed orange
curves), hence essentially affect the decays of the light Higgs (where
the mass prediction is most sensitive to the variations
in~$\xi$). There, even at small $\xi$, the fluctuations associated
with the scale dependence compete in magnitude with the absolute size
of the electroweak corrections. For the heavy-doublet Higgs states,
the percent-level shift of the decay widths of strict one-loop order
(solid green curve) should be seen as belonging to the higher-order
uncertainty and remains much smaller than the magnitude of the
one-loop corrections, which are dominated by effects of Sudakov type
as explained in \citere{Domingo:2019vit}. The shift of the tree-level
widths (solid black curves) by~$10\%$ is associated with the
scheme~conversion, \IE~the modified value of~$t_{\beta}$.

\tocsubsection{Restoring gauge invariance in the non-degenerate case}

From the perspective of a strict order-counting, it can appear
superfluous to worry about the gauge-violating pieces described above,
since they correspond to a higher order in the expansion. In fact,
gauge dependence can be exploited as a means to estimate part of the
theoretical uncertainty, as we showed at the level of the Higgs masses
in \fig{fig:xidep_MSSM}. On the other hand, the gauge-violating
effects can numerically dominate the radiative contributions to the
decays, because internal cancellations caused by symmetries are no
longer enforced. Thus, the violation of the Ward~identity
in~\AtoB{h_i}{\gamma\gamma}\,\cite{Domingo:2018uim,Benbrik:2012rm} can
sizably unsettle the corresponding estimate of the decay
width. In~\AtoB{h_i}{WW}, infrared~(IR) divergences do not cancel
between virtual QED~corrections and soft photon
radiation\,\cite{Gonzalez:2012mq,Goodsell:2017pdq,Domingo:2018uim}. We
thus believe that it is meaningful for reliable predictions to employ
decay amplitudes that satisfy the symmetry principles.

A first obvious method neutralizing explicit gauge dependence would
simply consist in expanding the loop functions in terms of the
external Higgs masses appearing as argument, in the vicinity of the
tree-level value, then truncating the expansion at the order achieved
in the calculation. In this way, the $\xi$-dependent terms in
\refeqs{eq:massrecursion}, \refeqs{eq:vertxi} and \refeqs{eq:mixxi}
would explicitly vanish indeed. We stress that only the `amplitudes'
or `form-factors' need to be expanded and truncated in this fashion:
the kinematical factors continue to be written in terms of kinematical
(\IE~loop-corrected) masses, as \EG~in the green curves of
\fig{fig:xidepwidth}. The usual prejudice against this procedure arises
from the fact that the thus shifted argument of the loop functions
displaces or even obstructs internal effects, \EG~thresholds. This is
especially true in the case of a light Higgs state (with mass at or
below the electroweak scale), since radiative effects can compete with
the tree level. Nevertheless, as we argued above, it is unlikely that
directly inputting a loop-corrected mass in the loop functions
actually improves the reliability of the calculation, since it then
introduces gauge-violating pieces and explicit dependence on the
field~renormalization.\footnote{Of course, the terms of two-loop order
are legitimate if the considered order is fully under control.}

Thus, if one chooses to inject a loop-corrected mass in the decay
amplitudes, gauge~dependence should be carefully analyzed, either for
an estimate of the associated uncertainties or for an attempt at
restoring the symmetry. The latter obviously requires the addition of
a two-loop order piece absorbing the $\xi$-dependent terms
in \refeqs{eq:massrecursion}, \refeqs{eq:vertxi} and
\refeqs{eq:mixxi}. Below, we continue to focus on a `physical'
scheme, since it makes the analysis of gauge dependence more
convenient. Here, we observe that:
\begin{itemize}
\item the $\xi$-dependence in the three-point functions only appears
  at the level of the vertex diagrams; thus, this gauge~dependence
  must be neutralized separately;
\item the $\xi$-dependence in the two-point functions appears both in
  vertex diagrams and self-energies; therefore, both objects must be
  combined consistently in order to neutralize this form of
  gauge~dependence;
\item the $\xi$-dependence in the one-point functions appears in the
  vertex and self-energy diagrams, as well as counterterms
  (\EG\ tadpoles);
\item the $\xi$-dependence in the two- and three-point functions
  originates in the mismatch between the kinematical mass and the
  tree-level mass appearing in the Higgs--Goldstone couplings, while
  additional sources intervene at the level of the $A_0$~functions.
\end{itemize}
A strategy outlined in \EG\ \citeres{Goodsell:2017pdq,Domingo:2018uim}
would simply upgrade the Higgs--Goldstone couplings---see
\refeq{eq:gaugecoup}---by substituting a kinematical mass to the
tree-level one. This solution works at the level of the three-point
functions and sets IR~divergences under control. However, applied to
the two-point functions of \refeqs{eq:massrecursion} or
\refeqs{eq:mixxi}, it shifts the amplitude by an
ultraviolet~(UV)-divergent piece (as already noticed in
\citere{Dao:2019nxi}). Such UV-divergences can admittedly be regularized
in an ad-hoc fashion, but this means that UV-logarithms are thus
arbitrarily introduced. In addition, the gauge-dependent $A_0$~term is
not removed by this method. We detail below how far one has to extend
this procedure to fully restore gauge invariance in the determination
of masses through an iterated pole search and in decay amplitudes
evaluated at a loop-corrected mass. We do not believe these methods to
be competitive with the simple `truncation' approach, but we expose
them for the sake of closing on the `generalized-coupling' procedure
considered in earlier works.

In order to simultaneously work with loop-corrected masses, preserve
gauge invariance, and keep control over the UV-divergences, a more
elaborate and consistent procedure needs to be constructed than just
shifting the Higgs--Goldstone couplings. The cancellation of
the gauge-dependent
terms in the two- and three-point functions makes it clear
that the promotion of the Higgs--Goldstone couplings is a necessary
step if one aims at restoring $\xi$-independence. However, we note
that this `upgrade' is just a subset, restricted to these specific
couplings, of a larger transformation of the Higgs potential that
would impose the loop-corrected Higgs mass as a tree-level value of
the new potential. Such a reshaping of the Higgs potential is
straightforward to implement as the THDM~parameters (or at least
specific linear combinations) can be expressed in terms of the masses
and mixing angles\,\cite{Boudjema:2001ii}---see
\citere{Chalons:2012qe} for such a reconstruction in the context of
a~THDM with an additional singlet. In addition, it is not possible in
general to preserve the properties of a SUSY tree-level Higgs sector
when upgrading the Higgs masses to their loop-corrected value, because
this operation spoils the connection of the quartic Higgs parameters
with the electroweak gauge couplings. Properties of the tree-level
spectrum, such as~\mbox{$m^2_{H^{\pm}}=m^2_A+M_W^2$}, are violated at
the radiative level (though still constrained by the electroweak
symmetry). Relaxing these relations appears as a necessary sacrifice
in order to restore gauge~invariance in a controlled fashion. As a
consequence, the gauge counterterms appearing in the Higgs
self-energies in the SUSY context lose any sort of meaning in the
generalized framework, and are insufficient in order to absorb the
UV-divergences. Therefore, the calculation of the Higgs self-energies
needs to be performed in the new framework (after re-definition of the
Higgs couplings), that of a THDM (with singlet) with SUSY matter
content.

A detailed procedure allowing to map the MSSM onto a THDM+SUSY
framework is provided in appendix\,\ref{ap:mapping} and can be applied
to the recursive determination of the Higgs masses
of \refeqs{eq:massrecursion}. Then, the Feynman amplitudes employ the
value of the loop-corrected Higgs masses as tree-level input
parameters both explicitly---when the Higgs states appear in
propagators---and implicitly---in the cubic and quartic Higgs
couplings. The generated shift is still formally of two-loop order
with respect to the original calculation in the SUSY context and,
indeed, allows one to restore gauge invariance. Nevertheless, as
explained in appendix\,\ref{ap:mapping}, the extension of the
renormalization conditions of the~(N)MSSM to the THDM (with singlet)
framework for the parameters of the Higgs potential is not unambiguous
in general. Several choices may appear as `natural', such as restoring
the logarithms of the SUSY self-energies at tree-level on-shell
external momenta, or employing logarithms of the same form as those
appearing in the gauge~counterterms. Failing to identify a physical
principle determining these logarithms, we must concede that, while
the UV-divergences are now under control, the added UV-logarithms are
still largely arbitrary. This arbitrariness can be exploited in the
form of a scale~dependence (which we denote as~$\mu_{\text{map}}$
below) as a measure of the uncertainty introduced in the mapping. This
form of uncertainty replaces that of the field~renormalization in the
original SUSY~calculation with off-shell external momentum, while the
$\xi$-dependence has been neutralized: the result satisfies the
symmetry principle.

\begin{figure}[p!]
\centering
\includegraphics[width=0.49\textwidth]{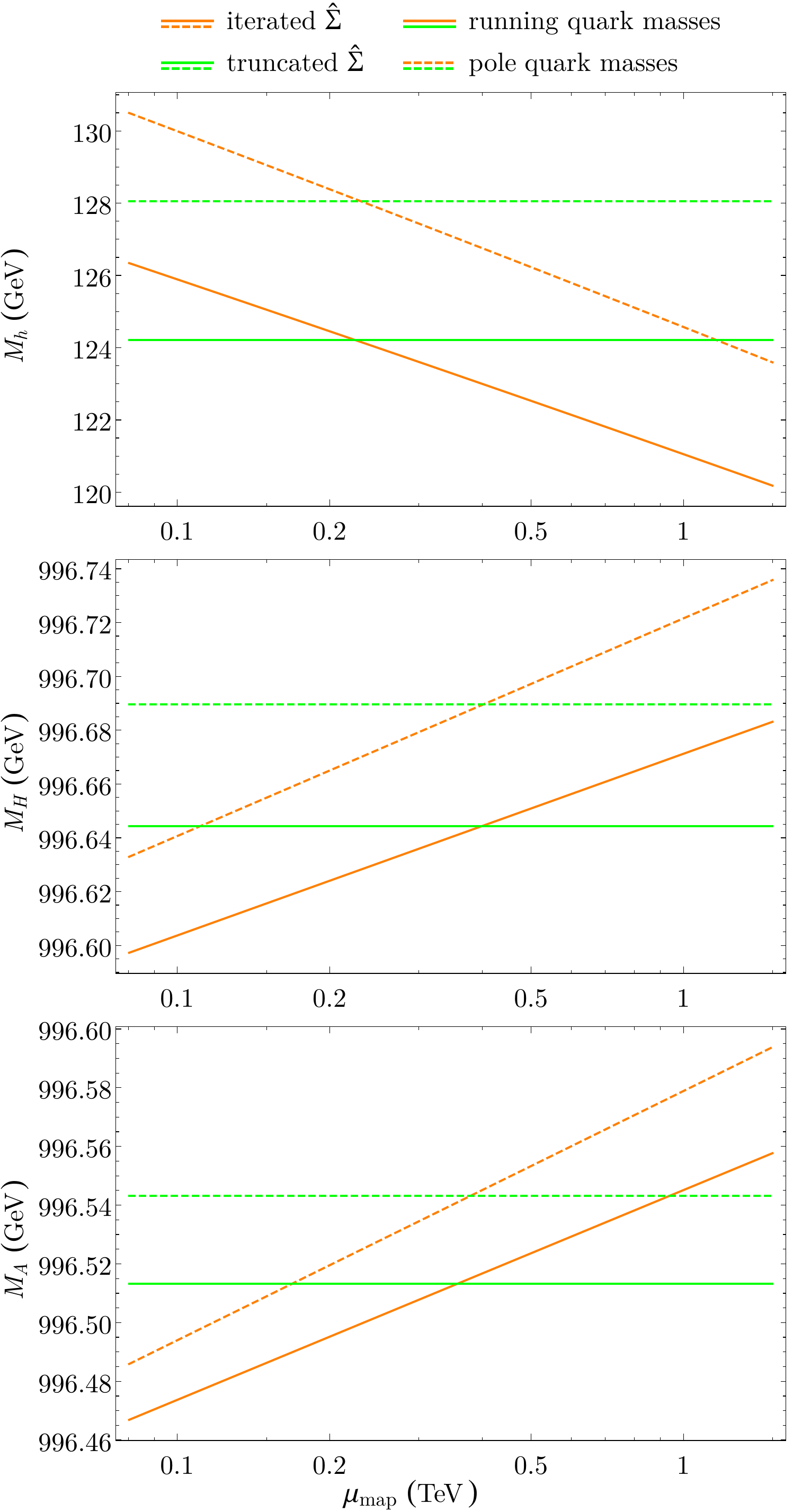}
\includegraphics[width=0.49\textwidth]{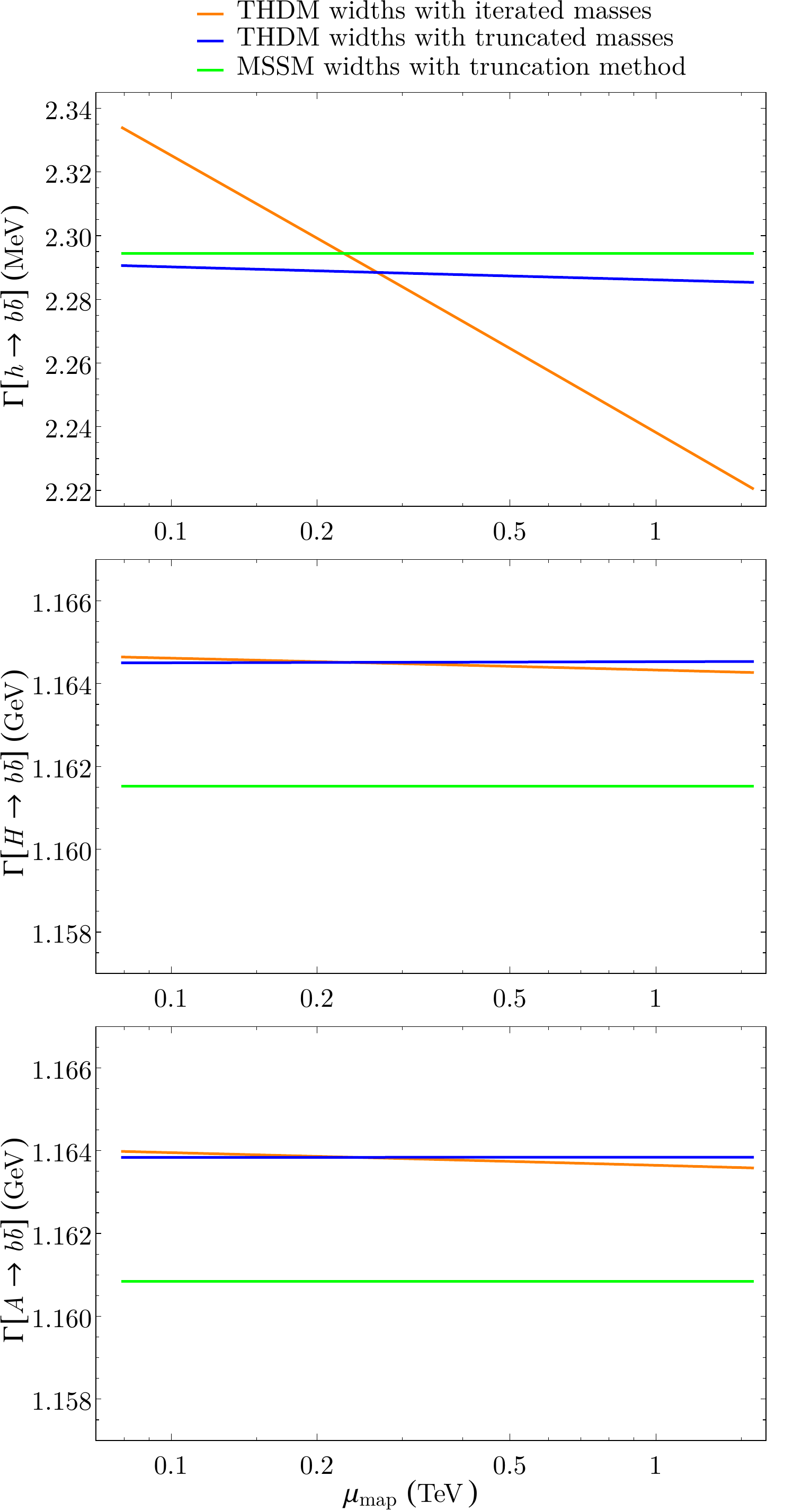}
\caption{{\em Left:} The scale~dependence associated with the mapping
  procedure is shown for the neutral Higgs masses derived by the
  recursive condition of \refeqs{eq:massrecursion} with
  $\xi$-independent self-energies in a~THDM for the scenario of
  Fig.\,\ref{fig:xidep_MSSM} (orange curves). The mapping scale is
  varied between~$M_W$ and $1.5$\,TeV. The masses obtained in the MSSM
  from the expansion of \refeq{eq:massbasic} (`truncation' method) are
  shown in green for reference. The solid and dashed curves employ
  different definitions of the fermion masses, offering an estimate of
  the magnitude of the $\mathcal{O}{\left(\alpha_q\alpha_s\right)}$
  corrections.
  \newline {\em Right:} The scale~dependence associated
  with the matching procedure is shown for the Higgs decays into
  bottom quarks at the kinematic Higgs masses given by the
  loop-corrected value from the expansion of \refeq{eq:massbasic}. The
  green curves display the $\xi$-independent widths in the MSSM, \IE\
  with an evaluation of the decay amplitude at the tree-level Higgs
  masses (`truncation'); the other curves are obtained in the
  THDM~framework (`matching'), \IE\ they are free of gauge-violating
  terms by construction. The blue curves are derived using the
  `truncated' Higgs masses of the MSSM (with \refeq{eq:massbasic}) as
  matching input. They illustrate the (small) explicit scale
  dependence of the widths. The orange curves characterize the
  parametric dependence, originating in the mass determination via a
  pole search in a THDM framework (`mapping'), when the latter is
  chosen as input instead. The quark masses are set to running masses
  at the scale of the decaying Higgs. \label{fig:mu2dep_MSSM}}
\end{figure}

On the left-hand side of \fig{fig:mu2dep_MSSM}, we show the
mapping-scale dependence in the recursive mass~determination for the
same scenario as in \fig{fig:xidep_MSSM}. A comparison of the range of
variation with that of \fig{fig:xidep_MSSM} is not meaningful, since
the $\xi$-dependence in the latter is polynomial, hence problematic
as~\mbox{$\xi\to\infty$}, while the scale~dependence
in \fig{fig:mu2dep_MSSM} is logarithmic and has been freed of
symmetry-violating effects. In terms of predictivity, there is no
obvious gain with respect to the mass~determination
of \refeq{eq:massbasic}, except perhaps in the control of the
uncertainty associated with electroweak corrections~\mbox{of higher order}.

Let us now assume that the physical Higgs masses are calculated in a
consistent fashion (either via the truncation method
or through the mapping procedure) and are gauge-independent
objects. We turn to the question of the Higgs decays. In order to
define $\xi$-independent transition amplitudes, we can employ the same
strategy as for the mass~determination, \IE~work in a
THDM+SUSY~framework where the physical masses are tree-level
masses. However, such a framework where tree-level and loop-corrected
masses coincide is that of a THDM with on-shell renormalization
conditions for the Higgs masses: indeed, the Higgs potential of the
THDM possesses enough degrees of freedom to allow for an on-shell
definition of the masses. In contrast to the mass~calculation, there
is no arbitrariness in the renormalization conditions for the
definition of the decay amplitude (except in the case of
Higgs-to-Higgs transitions). The precise implementation of this
on-shell THDM+SUSY is discussed in appendix\,\ref{ap:OSTHDMmatching}:
it can be viewed as a simple switch of renormalization scheme with
respect to the~MSSM. The arbitrariness of renormalization that we
mentioned at the level of the mass~determination is now lifted (though
the associated uncertainty is still hidden in the input values for the
Higgs masses). Corresponding results are shown on the right-hand side
of \fig{fig:mu2dep_MSSM} in the case of neutral Higgs decays into
bottom quarks. The explicit uncertainty associated with the matching
procedure (variations of the blue curves) is very small. The implicit
dependence on the mapping scale (orange curves) when the input values
of the Higgs masses are defined by the recursive condition in the~THDM
mostly matters for the SM-like state. Finally, we note that the
predictions for the decay widths obtained with this matching procedure
are very close to that of the truncation method in the~MSSM (green
curve) and spread far less than the $\xi$-dependent results
of \fig{fig:xidepwidth}.

Nevertheless, there is another source of uncertainty in this
mapping/matching procedure of the MSSM onto a~THDM, because the Higgs
potential of the~THDM is not fully determined by the identification of
the (four)~Higgs~masses, leading to an arbitrariness in the choice of
the (seven)~$\lambda_i$~parameters (out of which three are
complex). The simplest choice consists in
applying~\mbox{$\lambda_{5,6,7}\stackrel{!}{=}0$}, as in the~MSSM, and
can be justified formally. However, if the radiative corrections to
the Higgs masses involve large effects of~$\lambda_{5,6,7}$-type, such
as a large splitting between the two neutral heavy-doublet Higgs
bosons, the inadequate mapping of these effects
onto~$\lambda_{1,2,3,4}$ can lead to possibly large spurious
Higgs-to-Higgs corrections: this means that the exact form of the~THDM
must be carefully chosen in order to be consistent with the Higgs
spectrum. As yet, we do not have a systematic recipe to optimize this
selection. The latter would require assessing several Higgs-to-Higgs
transitions in order to shape the radiative Higgs potential in a more
realistic way.

\tocsubsection{Gauge-dependence in the (near-)degenerate case\label{sec:neardeg}}

The procedures that we discussed above, restoring explicit
gauge~invariance in the renormalized diagonal self-energies or the
transition amplitudes, are well-defined only in scenarios where Higgs
states do not receive large mixing effects at the radiative level
(\IE~in scenarios where the LSZ~expansion applies). Now we consider
the near-degenerate case. The main difficulty consists in defining the
mixed state in a gauge-invariant way, because of the gauge~dependence
present in the off-diagonal self-energies. Let us focus on a
two-dimensional near-degenerate subspace, generated by the tree-level
fields~$h_i$ and~$h_j$ (larger degenerate sectors follow the same
logic). Then, the pole equation reads
\begin{align}\label{eq:2statemix}
  \left[p^2-m_{h_i}^2+\hat{\Sigma}_{h_ih_i}{\left(p^2\right)}\right]
  \left[p^2-m_{h_j}^2+\hat{\Sigma}_{h_jh_j}{\left(p^2\right)}\right]=
  \hat{\Sigma}_{h_ih_j}{\left(p^2\right)}\,\hat{\Sigma}_{h_jh_i}{\left(p^2\right)}\,.
\end{align}
The right-hand side---of two-loop order---is not neglected because
both factors in the left-hand side are of one-loop order each
(since~\mbox{$\big\lvert
m^2_{h_i}-m^2_{h_j}\big\rvert\sim\big\lvert\hat{\Sigma}_{h_ih_j}{\left(p^2\right)}\big\rvert$}
and the difference between $p^2$, $m^2_{h_i}$ and $m^2_{h_j}$ is of
one-loop order when $p^2$ coincides with the pole mass). At this
leading (non-trivial) order, one can freeze the momentum in the
self-energies,
\EG~\mbox{$\hat{\Sigma}_{h_kh_l}(p^2)\to\hat{\Sigma}_{h_kh_l}{\left(\frac{1}{2}(m^2_{h_k}+m^2_{h_l})\right)}$}
for~\mbox{$k,l\in\{i,j\}$}. Then, the only gauge-dependent pieces in
\refeq{eq:2statemix} appear in the off-diagonal self-energies and are
due to the
term~\mbox{${\propto}\,(p^4-m^2_{h_i}\,m^2_{h_j})\to(m_{h_i}^2-m_{h_j}^2)^2/4$}
of \refeq{eq:selfxidep}. However,
as~\mbox{$m_{h_i}^2-m_{h_j}^2=\mathcal{O}(\text{1L})$} (by
assumption), such terms are formally of three-loop order and could be
explicitly set to~$0$ in the self-energy. This would again imply an
arbitrary regularization of the associated UV-divergence, which, as
before, may be translated into a scale uncertainty---equivalently, the
$\xi$-dependence could be kept and varied in order to estimate the
associated uncertainties. The corresponding object (with possibly
neutralized $\xi$-dependence) is denoted
as~$\tilde{\Sigma}_{h_ih_j}{\big(\frac{1}{2}(m^2_{h_i}+m^2_{h_j})\big)}$.
Another feature in the choice of momenta as presented above is the
independence of the thus defined Higgs masses from field
renormalization, hence the absence of corresponding uncertainties. We
can then consider the effective mass~matrix in the degenerate sector,
\begin{align}\label{eq:effmass}
  \mathcal{M}^{2\,\text{eff}}&\equiv
  \mathrm{diag}{\left(m^2_{h_i},m^2_{h_j}\right)}
  - \hat{\Sigma}^{\text{eff}}\,,\quad
  \hat{\Sigma}^{\text{eff}}\equiv
  \begin{pmatrix}
    \hat{\Sigma}_{h_ih_i}{\left(m^2_{h_i}\right)} &
    \tilde{\Sigma}_{h_ih_j}{\left(\frac{1}{2}\big(m^2_{h_i} +\, m^2_{h_j}\big)\!\right)}\\
    \tilde{\Sigma}_{h_jh_i}{\left(\frac{1}{2}\big(m^2_{h_i} +\, m^2_{h_j}\big)\!\right)} &
    \hat{\Sigma}_{h_jh_j}{\left(m^2_{h_j}\right)}
\end{pmatrix},
\end{align}
which is symmetric, hence diagonalizable in an orthogonal basis with
eigenvalues~$\tilde{m}^2_{H_k}$, and
eigenvectors~\mbox{$H_k=S^*_{ki}\,h_i+S^*_{kj}\,h_j$}. This definition
of the masses and fields at the one-loop order essentially extends the
truncation procedure to the degenerate case. We stress that the
inclusion of the off-diagonal elements is only meaningful because
$m_{h_i}^2-m^2_{h_j}=\mathcal{O}(\text{1L})$. Indeed, in the
non-degenerate case, the off-diagonal element is a piece contributing
at an incomplete higher order, hence it only increases the uncertainty
in the determination of the Higgs properties. Setting its argument to
the average squared mass, or the explicit neutralization of its
$\xi$-dependence also become transformations of this object by
one-loop effects, depriving it of any quantitative meaning. Of course,
assessing the exact point of transition between the near- and
non-degenerate regimes remains largely arbitrary.

Further momentum-dependent corrections (though formally subleading)
may be included as `diagonal' effects according to
\mbox{$\mathcal{M}_{H_k}^2=\tilde{m}^2_{H_k}-S_{km}\,S_{kn}\left[\hat{\Sigma}{\left(\mathcal{M}_{H_k}^2\right)}-\hat{\Sigma}^{\text{eff}}\right]_{h_mh_n}$},
but explicit $\xi$-dependence would be re-introduced
unless~$S_{km}\,S_{kn}\,\hat{\Sigma}_{h_mh_n}(\mathcal{M}_{H_k}^2)$ is
again re-defined by a shift of two-loop order absorbing such
dependence. This can be achieved in the context of a~THDM+SUSY with
tree-level fields $\bar{H}_k$ absorbing both the tree-level rotation
by~$\mathbf{U}_n$ and the loop-level rotation by~$\mathbf{S}$ of
the~MSSM. Still, the imaginary parts in~$\mathbf{S}$ induce further
complications for the mapping, so that the complex
rotation~$\mathbf{S}$ is conveniently replaced by a real
rotation~$\mathbf{S}'$ minimizing the size of the off-diagonal term in
$\mathbf{S}'\cdot\mathcal{M}^{2\,\text{eff}}\cdot\mathbf{S}'^T$. Again,
the form of the Higgs potential should be carefully chosen so that
mass~corrections are appropriately mapped.

Having defined the external states in an explicitly gauge-invariant
way, we may now consider the decay amplitudes. In the `truncation'
approach, one can define the gauge-independent object as
\begin{align}\label{eq:decaymix}
  &\Amp[\tilde{\mathcal{A}}]{}{}{\AtoB{H_k}{XX}} =\notag\\
  &S_{ki}\left\{\Amp{tree}{}{\AtoB{h_i}{XX}}
  \left[1-\frac{1}{2}\frac{d\hat{\Sigma}_{h_ih_i}}{dp^2}\right]
  - \Amp{tree}{}{\AtoB{h_l}{XX}}\,
  \frac{\hat{\Sigma}_{h_ih_l}-\tilde{\Sigma}^{\text{eff}}_{h_ih_l}}{m^2_{h_i}-m^2_{h_l}}
  + \Amp{vert}{}{\AtoB{h_i}{XX}}\right\}_{p^2=m^2_{h_i}}\hspace{-1em}
\end{align}
where
\mbox{$\tilde{\Sigma}^{\text{eff}}_{h_ih_l}=\tilde{\Sigma}_{h_ih_l}{\left(\frac{1}{2}(m^2_{h_i}+m^2_{h_l})\right)}$}
if both $h_l$ and $h_i$ belong to the same degenerate subsystem, and
\mbox{$\tilde{\Sigma}^{\text{eff}}_{h_ih_l}=0$} otherwise.
Alternatively, it is again possible to embed the Higgs masses in an
on-shell model with SUSY matter content. The only difference with
respect to the non-degenerate case is that in the on-shell model, the
$\mathbf{S}^{(\prime)}$-rotation must be included in the definition of
the mixing matrix
$\mathbf{U}_n\to\mathbf{S}^{(\prime)}\cdot\mathbf{U}_n$, so that the
tree-level fields differ in the SUSY model and its on-shell
counterpart.

\begin{figure}[p!]
\centering
\includegraphics[width=\textwidth]{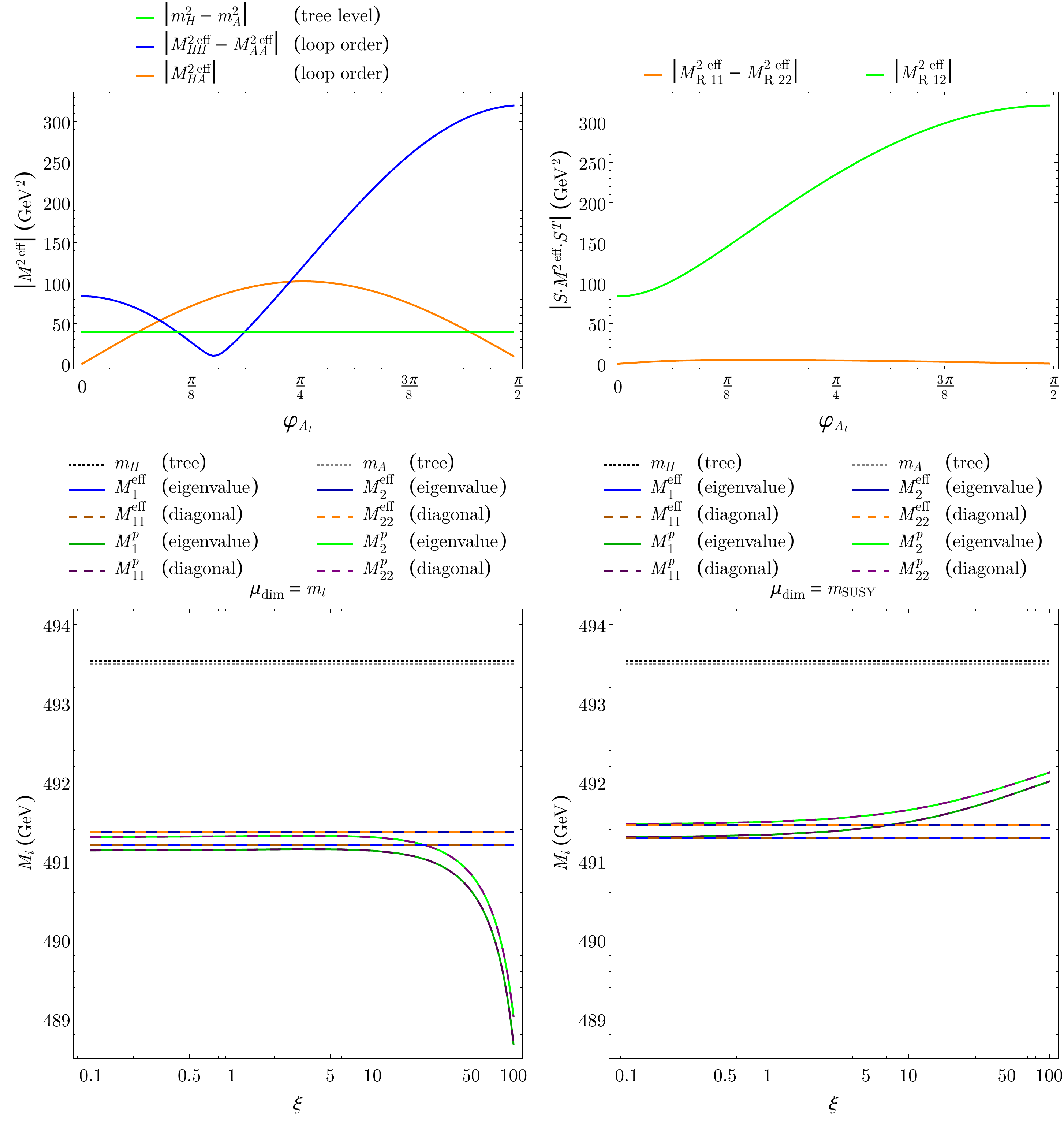}
\caption{The mass determination is illustrated in the MSSM for the
  degenerate $(H, A)$~pair
  with \CP-violation. \mbox{$t_{\beta}=30$}, \mbox{$M_{H^{\pm}}=0.5$\,TeV}, \mbox{$m_{\text{SUSY}}=1$\,TeV}, \mbox{$A_t=3$\,TeV}, \mbox{$\varphi_{A_t}\in\left[0,\frac{\pi}{2}\right]$}
  (upper row), then \mbox{$\varphi_{A_t}=\frac{3\,\pi}{20}\approx
  0.47$}.
\newline{\em Upper left}: the mass-splitting vs.\ mixing is shown:
in green, the mass-splitting between the \CP-even and \CP-odd heavy
scalars at the tree level; in blue, the mass-splitting between the two
diagonal entries of the effective mass~matrix, as defined in
\refeq{eq:effmass}; in orange, the mixing term of \refeq{eq:effmass}
(in absolute value).
\newline{\em Upper right}: the mass-splitting vs.\ mixing is shown
after rotation by an appropriate real orthogonal matrix: in green, the
splitting between the diagonal elements of the rotated effective
mass~matrix \mbox{$M^{2\,\text{eff}}_{\mathrm{R}}=S\!\cdot
M^{2\,\text{eff}}\!\cdot S^T$}; in orange, the magnitude of the
subsisting off-diagonal element---the effective mass~matrix being
complex, it cannot be fully diagonalized by a real orthogonal matrix,
but the off-diagonal element can be minimized.
\newline{\em Lower left}: the mass determination is shown for
\mbox{$\varphi_{A_t}=\frac{3\,\pi}{20}$}, \mbox{$\mu_{\text{dim}}=m_t$}:
the tree-level masses are shown in dotted gray and black; the
eigenvalues of the effective mass~matrix are shown in shades of blue;
the diagonal elements of the rotated effective mass~matrix are shown
in dashed brown/orange; the poles retaining full momentum~dependence
in the MSSM~self-energies are shown in green; the corresponding
diagonal elements after rotation are shown in purple. Eigenvalues and
diagonal elements after rotation essentially coincide, resulting in
pairs of curves overlapping each other.
\newline{\em Lower right}: the mass determination is shown for
\mbox{$\varphi_{A_t}=\frac{3\,\pi}{20}$}, \mbox{$\mu_{\text{dim}}=m_{\text{SUSY}}$}
(identical color code).\label{fig:degen_MSSM}}
\end{figure}

\begin{figure}[tb!]
\centering
\includegraphics[width=\textwidth/2]{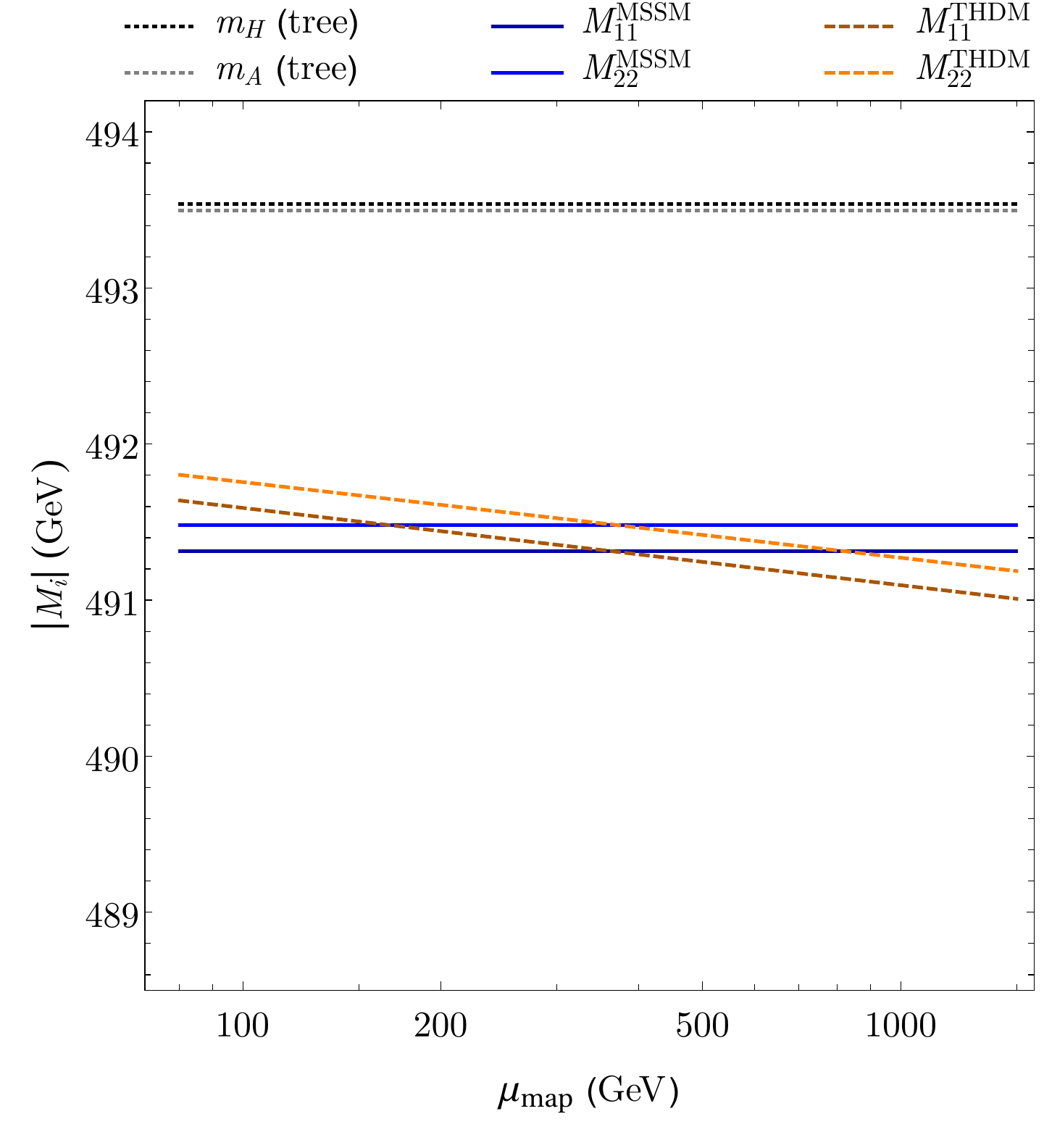}
\caption{The mass determination is shown for the degenerate
$(H,A)$~system of \fig{fig:degen_MSSM} with
\mbox{$\varphi_{A_t}=\frac{3\,\pi}{20}$} through a mapping of the Higgs
self-energies onto a THDM+SUSY model. The dotted lines correspond to
the (MSSM) tree-level values and the horizontal blue lines to the MSSM
loop-corrected evaluation employing \refeq{eq:effmass}. The dashed
lines are obtained with self-energies that are calculated in a THDM
where the Higgs potential takes tree-level masses that coincide with
the loop-corrected mass value that are injected as external momenta
and tree-level fields including the $\mathbf{S}$ rotation. The
parameters of this THDM are determined from the conditions
of \refeq{eq:lambdavsmass} together with the (arbitrary)
constraints \mbox{$\lambda_1\equiv\lambda_2$}, \mbox{$\lambda_6^{r,i}\equiv0$},
\mbox{$\lambda_7^{r}\equiv0$}. The dependence on
$\mu_{\text{map}}$ originates in the freedom of scheme for the fixing
of the renormalization conditions of these Higgs-potential parameters
and can be seen as a symmetry-conserving uncertainty on the
mass-determination.\label{fig:THDMmap}}
\end{figure}

We illustrate this discussion for the degenerate case with
\fig{fig:degen_MSSM}. We consider a \CP-violating scenario with
\mbox{$t_{\beta}=30$}, \mbox{$m_{H^{\pm}}=0.5$\,TeV},
\mbox{$m_{\text{SUSY}}=1$\,TeV}, \mbox{$A_t=3$\,TeV},
\mbox{$\varphi_{A_t}\in\left[0,\frac{\pi}{2}\right]$}---likely
to exhibit several phenomenological shortcomings (\EG~direct searches
for heavy Higgs bosons\,\cite{Aad:2020zxo}, $B$-meson
decays\,\cite{Misiak:2017zan}, electric dipole
moments\,\cite{Arbey:2014msa}), but solely presented here in order to
exemplify the impact of gauge~dependence in degenerate systems. The
phase of the trilinear stop couplings induces a mixing between the
(tree-level) \CP-even and \CP-odd neutral heavy-doublet components,
which are close in mass due to the
$SU(2)_{\mathrm{L}}$-symmetry. This \CP-violating mixing strictly
appears at the radiative level (the tree-level MSSM~Higgs~sector is
still \CP-conserving) and does not involve the gauge sector: the
mixing term is thus automatically $\xi$-independent, without need of
further manipulations. We then consider the effective mass matrix
$\mathcal{M}^{2\,\text{eff}}$ of \refeq{eq:effmass} and compare the
mixing with the mass-splitting between the diagonal entries: this is
depicted in the plot in the upper left-hand quadrant. The degeneracy
at the tree level (green curve) is partially lifted by the diagonal
loop corrections (blue curve), but the mixing (orange curve) still
competes in the vicinity of~\mbox{$\varphi_{A_t}\sim0.5$}. We then
introduce the rotation matrix~$\mathbf{S}$ (we omit~$'$) that
minimizes the size of the off-diagonal entry
of~$\mathcal{M}^{2\,\text{eff}}$. It is clear that a complex
matrix~$\mathbf{S}$ could fully
diagonalize~$\mathcal{M}^{2\,\text{eff}}$. However, in order to recast
the system onto an effective~THDM (see \fig{fig:THDMmap}), it is more
convenient to introduce a real orthogonal~$\mathbf{S}$, so that a
subsidiary off-diagonal piece remains as a tribute to the imaginary
parts in~$\mathcal{M}^{2\,\text{eff}}$. As is shown on the upper
right-hand side of \fig{fig:degen_MSSM}, this term is subleading and
can be neglected in the mass~determination. Next, we focus on the
point~\mbox{$\varphi_{A_t}=\frac{3\,\pi}{20}$}, where~$\mathbf{S}$
shows a mixing angle of about~$\frac{\pi}{4}$. The plots in the lower
row of \fig{fig:degen_MSSM} illustrate the $\xi$-dependence in the
mass~determination. The dotted lines correspond to the tree-level
masses. The solid lines correspond to the eigenvalues of the
$(2\times2)$~mass~system while the dashed lines are obtained from the
diagonal entries after rotation by the (gauge-independent)
matrix~$\mathbf{S}$: the good agreement between both approaches proves
the reliability of the second one. Then, the horizontal lines are
obtained from the effective mass~matrix~$\mathcal{M}^{2\,\text{eff}}$
and are, by construction, $\xi$-independent. The curves showing a
$\xi$-variation are obtained by retaining full momentum~dependence in
the MSSM~self-energies: as before, this approach is both
gauge~dependent and renormalization-scheme dependent---the latter is
made obvious by the impact of the renormalization
scale~$\mu_{\text{dim}}$. If the amplitude of the variations in~$\xi$
and~$\mu_{\text{dim}}$ is interpreted as the uncertainty on the
electroweak corrections, we see that the latter represents a
non-negligible fraction of the mass~shift between tree-level and
one-loop result, which is problematic. At low~$\xi$, the scale
variation of the masses still reaches~$\simord50\%$ of the size of the
loop-corrected mass-splitting.

In \fig{fig:THDMmap}, we extend the calculation to a THDM+SUSY
framework as explained in appendix\,\ref{ap:mapping}, so that the
self-energies can receive loop-corrected mass values as external
momenta without violating the gauge symmetry. Due to the degeneracy in
the MSSM, the $\mathbf{S}$ rotation is included in the definition of
the tree-level Higgs fields of the THDM. The parameters of the
tree-level Higgs potential are defined from the loop-corrected mass
values according to \refeq{eq:lambdavsmass}, together with the
(arbitrary)
constraints \mbox{$\lambda_1\equiv\lambda_2$}, \mbox{$\lambda_6^{r,i}\equiv0$}
and \mbox{$\lambda_7^{r}\equiv0$}. The scale~$\mu_{\text{map}}$
encodes the arbitrariness in the associated renormalization conditions
and can be seen as a measure of the uncertainty associated with the
mapping procedure. The resulting self-energies in the THDM are
$\xi$-independent and differ from the MSSM ones by a shift of two-loop
order. The associated masses are obtained at the strict one-loop order
from the THDM self-energies projecting onto the tree-level Higgs-field
directions of the THDM (the off-diagonal term is subleading by
construction): they are displayed in dashed orange/brown and show a
variation of $\simord 0.5$\,GeV
for \mbox{$\mu_{\text{map}}\in[M_{W},m_{\text{SUSY}}]$}. This
uncertainty, from which electroweak-violating effects have been
cleansed, is roughly comparable to the one observed
in \fig{fig:degen_MSSM} which is triggered by $\xi$-dependent
contributions. The iteration procedure is stable (the shift is hardly
noticeable after one iteration). Finally we note that the mass values
are consistent with those of the truncation method (solid blue lines).

\begin{figure}[tb!]
\centering
\includegraphics[width=\textwidth]{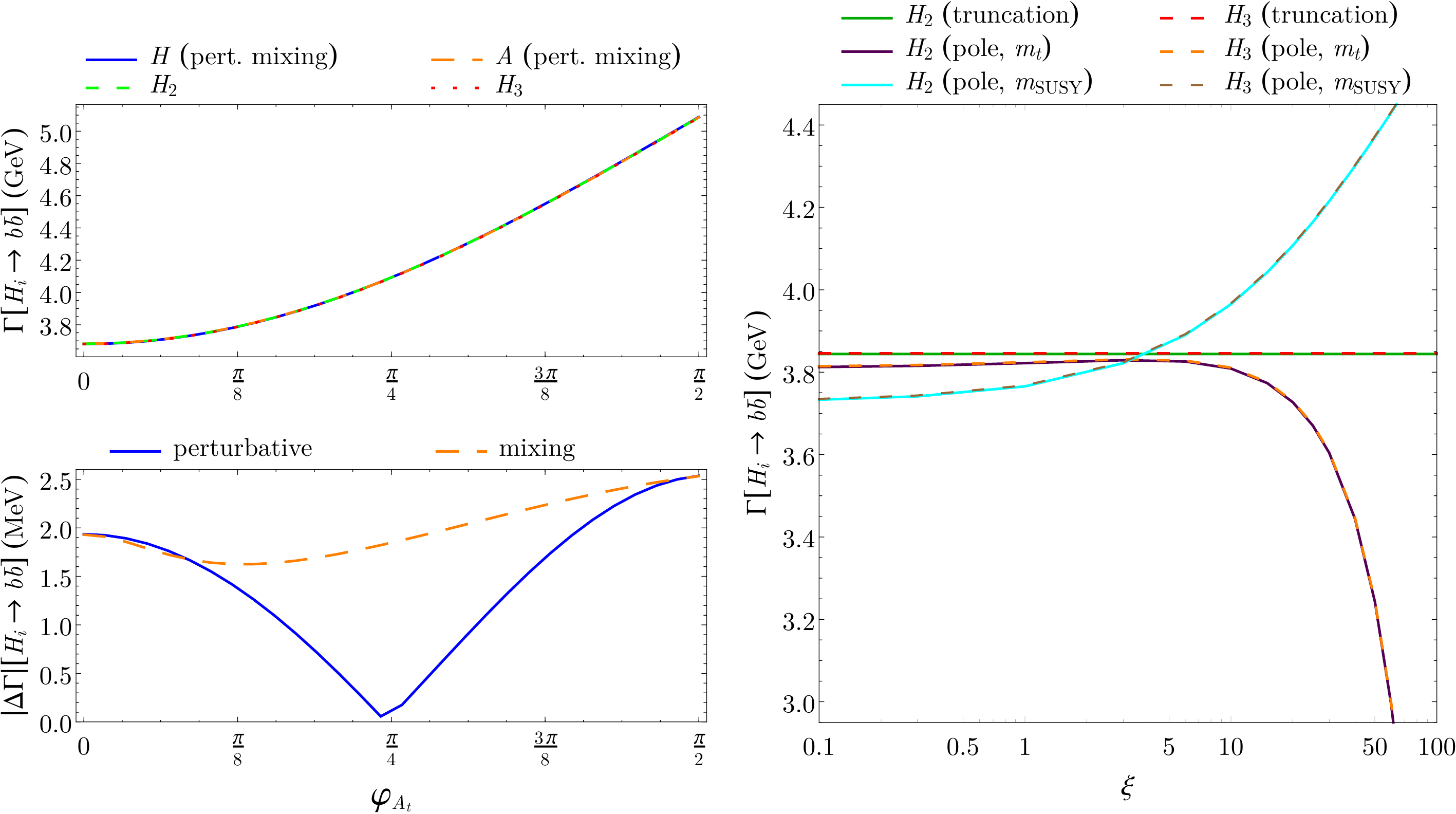}
\caption{The decay widths for the $b\bar{b}$ channel in the scenario
of \fig{fig:degen_MSSM} are shown.
\newline {\em Upper left}: the predictions of \refeq{eq:decaymix}
for the decays of the \CP-admixtures $H_{2,3}$ (dashed green and
dotted red) are compared to the widths obtained with a perturbative
treatment of the \CP-violating mixing (blue and dashed orange). All
are almost identical.
\newline {\em Lower left}: the difference between the predicted decay
widths in the perturbative (solid blue) and mixing (dashed orange)
description is depicted.
\newline {\em Right}: the gauge- and field-renormalization dependence
of the decay widths obtained with \refeq{eq:decaymix} (green and
dashed red) and with the mixing formalism associated to an iterative
pole search (purple and dashed orange for field counterterms at the
scale $m_t$; blue and dashed brown for field counterterms at the scale
$m_{\text{SUSY}}$) are shown. Due to the $SU(2)_{\text{L}}$ symmetry,
the curves appear in overlapping pairs.\label{fig:degMSSM_dec}}
\end{figure}

Finally, in \fig{fig:degMSSM_dec}, we present the Higgs decay widths
into the $b\bar{b}$ final state. In the plot on the left-hand side, we
compare the decay widths obtained in a purely perturbative
expansion---\IE~without resumming the \CP-violating mixing---and those
derived with the mixing formalism of \refeq{eq:decaymix}. The latter
are depicted in dashed green and dotted red for the two mixed states
$H_2$ and $H_3$, while the former are shown in blue and dashed orange
for the \CP-even and the \CP-odd Higgs respectively. The
$SU(2)$-symmetry (albeit $m_{H^{\pm}}/v$ is not so large) results in
very close predictions for all these widths. In fact, one needs to
consider the differences of widths---\IE~measurements of the
$SU(2)_{\text{L}}$-violation, a factor~$10^3$ smaller---in order to
see the deviation between the perturbative and mixing descriptions,
see the lower plot. The impact of the mixing on the decay widths thus
remains very mild in the whole parameter space considered in this
scenario. The reason is that the scalar and pseudoscalar $Hb\bar{b}$
operators hardly interfere. While, strictly speaking, the mixing
formalism is only needed in the middle region---where the off-diagonal
self-energy is comparable to the diagonal mass-splitting---it remains
legitimate in the whole range of~$\varphi_{A_t}$ because \mbox{$\lvert
m_H^2-m_A^2\rvert\sim\lvert\hat{\Sigma}_{HH}(m^2_H)-\hat{\Sigma}_{AA}(m^2_A)\rvert=\mathcal{O}(\text{1L})$}. On
the other hand, the strict perturbative approach is \textit{a priori}
insufficient in the middle range of~$\varphi_{A_t}$, but performs
quite well numerically in this example. On the right-hand side, we
compare the predictions of \refeq{eq:decaymix} with those of the
mixing formalism associated with an iterative pole search (as
in \EG~\citeres{Williams:2011bu,Domingo:2017rhb}), for the
point \mbox{$\varphi_{A_t}=\frac{3\,\pi}{20}$} with near-maximal
mixing. Again the $SU(2)_{\text{L}}$-symmetry results in very close
predictions between the widths for the two states $H_2$ (in solid
lines) and $H_3$ (in dashed lines). Similarly to the masses, the
widths obtained with the iterative pole-search procedure are sensitive
to gauge and field-renormalization variations, these causing an
uncertainty of~$\mathcal{O}(5\%)$ for \mbox{$\xi\lesssim5$},
comparable to the magnitude of electroweak (non-SUSY) 1L
contributions. The width obtained with \refeq{eq:decaymix} is by
construction independent from these computational artifacts. We
observe an agreement within a few percent with the predictions of the
pole-search formalism at~\mbox{$\xi=1$}, within the uncertainties from
gauge and scale variations of this method. In any case, in such a
strong mixing configuration where the two diagonal entries are almost
degenerate at the 1L order, 2L corrections may significantly alter the
quantitative features of the mixing.

Another mixing scenario in the NMSSM is presented in
appendix\,\ref{ap:SHmix}, in a less symmetric configuration involving
a singlet and a doublet \CP-even Higgs.

As a concluding remark on the gauge~dependence, we have seen that the
higher-order terms introduced in the determination of observable
quantities, via \EG~not expanding and truncating the momenta in the
radiative corrections, are liable to spoil the quality of these
calculations. We note, however, that sizable deviations from
gauge-independent definitions only appear for large values of the
gauge-fixing parameter. In particular, the choice~\mbox{$\xi=1$}
always seems to produce results that are `accidentally' close to the
gauge-independent definitions, so that the impact of the
gauge~dependence should remain small at the numerical level for
calculations performed in the 't\,Hooft--Feynman~gauge. Nevertheless,
the explicit dependence on the field renormalization---associated with
the UV-regularization of self-energies away from their
mass~shell---obviously makes noticeable contributions to the
theoretical uncertainty.

\tocsection{Logarithms in loop-order Higgs mixing\label{sec:log}}

The analysis of the gauge dependence in Higgs decays has shown that
splitting the loop corrections between mixing and vertex contributions
to the decay amplitudes is a purely artificial procedure---which was
already clear from the fact that it is scheme dependent. While this
separation is not a fundamental problem \textit{per se}, it could lead
to misleading results through the inclusion of incomplete higher
orders. We aim to illustrate this below by comparing decay amplitudes
in the limit of heavy-doublet states, at a scale where the
electroweak-violating effects are subleading, hence the global
$SU(2)_{\mathrm{L}}$-symmetry should be approximately satisfied (up to
breaking terms scaling with~$v$ and suppressed by the heavy Higgs
mass). While very large values
of \mbox{$m_{H^{\pm}}\sim10$--$100$}\,TeV are currently of limited
phenomenological interest, the purpose in investigating this regime
consists in identifying spurious $SU(2)_{\text{L}}$-violating
artifacts induced by the formalism---\IE~not controlled by the
spontaneous breaking of the electroweak symmetry. Once these
unphysical effects are exposed and corrected, the impact on the
predictions at the TeV-scale is found to be
non-negligible, \IE~comparable to the size of the EW corrections
themselves or of 2L corrections.

\tocsubsection{Fermion-loop contributions to the Higgs mixing\label{sec:1Lfermion}}

For simplicity, we consider the MSSM with decoupled SUSY particles and
$m_{H^{\pm}}$ much above the electroweak scale. This is akin to a
THDM~framework of type\,II in the decoupling limit. In particular, the
mixing in the tree-level Higgs sector can be approximated by a
$\beta$-angle rotation:
\begin{align}
\begin{pmatrix} h\\ H\end{pmatrix} &\approx
\begin{pmatrix} c_{\beta} & s_{\beta}\\ s_{\beta} & -c_{\beta}\end{pmatrix}
\begin{pmatrix} h_d^0\\h_u^0\end{pmatrix}, &
\begin{pmatrix} G^0\\ A\end{pmatrix} &=
\begin{pmatrix} -c_{\beta} & s_{\beta}\\ s_{\beta} & c_{\beta}\end{pmatrix}
\begin{pmatrix} a_d^0\\ a_u^0\end{pmatrix}, &
\begin{pmatrix} G^{\pm}\\ H^{\pm}\end{pmatrix} &=
\begin{pmatrix} -c_{\beta} & s_{\beta}\\ s_{\beta} & c_{\beta}\end{pmatrix}
\begin{pmatrix} H_d^{\pm}\\ H_u^{\pm}\end{pmatrix},
\end{align}
where $h_{u,d}^0$, $a_{u,d}^0$ and $H_{u,d}^{\pm}$ are the \CP-even,
\CP-odd and charged Higgs fields (respectively) in the
gauge-eigenstate basis. In addition, \mbox{$m_{H^{\pm}}\approx
m_{H}\approx m_A\gg m_h\approx M_Z$}. We will exploit this
approximation below for the purpose of deriving simple analytical
formulae capturing the main features of the calculation, although our
numerical results still consist in a full calculation of one-loop
order.

We further target corrections associated with quark Yukawa couplings
of the third generation~$Y_{t,b}$, \IE\ top and bottom loops. For a
heavy SUSY sector, the sfermion loops contributing at the same order
can be regarded as constant with respect to variations of the Higgs
external momentum $p$, and we do not document them further (though
they may involve large logarithms of the form $\ln
m^2_{\text{SUSY}}/M_{\text{EW}}^2$). As before, we work in a scheme
where the charged-Higgs mass, as well as the $W$\nobreakdash-,
$Z$\nobreakdash-, and fermion masses are renormalized on-shell, while
other parameters receive a $\overline{\text{DR}}$-renormalization with
the ultraviolet regulator set to $m_t$.

The leading contributions to the renormalized Higgs self-energies can
be summarized as follows:
\begin{subequations}\label{eq:mixHferm}
\begin{align}
  \hat{\Sigma}_{hh}{\left(p^2\right)} &\simeq -\frac{3}{16\pi^2}\left(
    Y_t^2\,s_{\beta}^2 + Y_b^2\,c_{\beta}^2\right) p^2\,
    \ln\frac{p^2}{M_{\text{EW}}^2}\,,\\
  \hat{\Sigma}_{HH}{\left(p^2\right)} &\simeq
    \hat{\Sigma}_{AA}{\left(p^2\right)}\simeq
    \hat{\Sigma}_{H^+H^-}{\left(p^2\right)} \simeq -\frac{3}{16\pi^2}\left(
    Y_t^2\,c_{\beta}^2 + Y_b^2\,s_{\beta}^2\right)\left(p^2-m_{H^{\pm}}^2\right)
    \ln\frac{p^2}{M_{\text{EW}}^2}\,,\\
  \hat{\Sigma}_{Hh}{\left(p^2\right)} &\simeq
    -\hat{\Sigma}_{AG^0}{\left(p^2\right)}\simeq
    -\hat{\Sigma}_{H^+G^-}{\left(p^2\right)} \simeq \frac{3}{16\pi^2}\left(
    Y_t^2 - Y_b^2\right) s_{\beta}\,c_{\beta}\,p^2\,
    \ln\frac{p^2}{M_{\text{EW}}^2}\,.
\end{align}
\end{subequations}
Here, terms ${\propto}\,v^2$ are neglected, unless $p^2$ is set to a
value of electroweak magnitude (like in
$\hat{\Sigma}_{hh}{\left(p^2\right)}$ for the determination of
$M_h$). The scale \mbox{$M_{\text{EW}}^2\sim M_W^2\sim M_Z^2\sim
m_t^2$} is characteristic of the scheme.

From \refeqs{eq:mixHferm}, we can extract the leading contributions to
the heavy Higgs masses and mixing. In the considered scheme (with the
charged Higgs renormalized on-shell), these one-loop corrections to
the masses are of subleading order
$\mathcal{O}{\left(Y_{t,b}^2\,M_{\text{EW}}^2/(16\,\pi^2)\right)}$ in
the limit \mbox{$M_{\text{EW}}^2\ll m_{H^\pm}^2$}. On the other hand,
the contributions to the wave-functions and mixing of the
heavy-doublet states can be sizable and matter at the level of the
Higgs decays:
\begin{subequations}\label{eq:LSZmix}
\begin{align}
  -\frac{1}{2}\frac{d\hat{\Sigma}_{HH}}{dp^2}{\left(p^2\sim m_{H^{\pm}}^2\right)}
  &\simeq -\frac{1}{2}\frac{d\hat{\Sigma}_{AA}}{dp^2}{\left(p^2\sim m_{H^{\pm}}^2\right)} \simeq -\frac{1}{2}\frac{d\hat{\Sigma}_{H^+H^-}}{dp^2}{\left(p^2\sim m_{H^{\pm}}^2\right)}\notag\\
  &\simeq
  \frac{3}{32\,\pi^2}\left(Y_t^2\,c_{\beta}^2 + Y_b^2\,s_{\beta}^2\right)
  \ln\frac{m_{H^{\pm}}^2}{M_{\text{EW}}^2}\,,\\[1ex]
  \left.-\frac{\hat{\Sigma}_{Hh}{\left(p^2\right)}}{p^2-m_{h}^2}\right|_{p^2\sim m_{H^{\pm}}^2} &\simeq
  \left.\frac{1}{p^2}\,\hat{\Sigma}_{AG^0}{\left(p^2\right)}\right|_{p^2\sim m_{H^{\pm}}^2} \simeq
  \left.\frac{1}{p^2}\,\hat{\Sigma}_{H^+G^-}{\left(p^2\right)}\right|_{p^2\sim m_{H^{\pm}}^2}\nonumber\\
  &\simeq -\frac{3}{16\,\pi^2}\left(Y_t^2 - Y_b^2\right) s_{\beta}\,c_{\beta}\,
  \ln\frac{m_{H^{\pm}}^2}{M_{\text{EW}}^2}\,.
\end{align}
\end{subequations}
It is common practice\,\cite{Williams:2011bu,Domingo:2018uim} to
include the wave-function corrections and the mixing in the \CP-even
sector within a loop-corrected mixing matrix, while the mixing with
the Goldstone bosons is kept separately at the diagrammatic level. One
could regret that the $SU(2)_{\mathrm{L}}$-symmetry is thus explicitly
broken by the formalism, but this should not be cause for any
inconsistency, in principle. The motivation behind absorbing the Higgs
mixing within the definition of the external field rests with a
resummation of mixing effects in the case where tree-level states are
almost degenerate. The physical heavy-doublet field is then defined
as \mbox{$h_2 = Z_{21}\,h + Z_{22}\,H = Z_H\left[\zeta_{Hh}\,h +
H\right]$} with \mbox{$\mathbf{Z}\equiv(Z_{21},Z_{22})^T$} being an
eigenvector of the effective mass~matrix
\mbox{$[m^2_{h_j}\delta_{jk}-\hat{\Sigma}_{h_jh_k}(p^2)],\,j,k\in\{1,2\},$}
for the eigenvalue \mbox{$p^2=M_H^2$} and satisfying the normalization
condition
\mbox{$\mathbf{Z}^T\cdot\!\big[\mathbf{1}+\mathbf{\hat{\Sigma}}_{h}^\prime(M_H^2)\big]\!\cdot\mathbf{Z}=1$}. At
the order considered in \refeqs{eq:mixHferm}, we have
\begin{align}
  Z_H &\approx 1 + \frac{3}{32\,\pi^2}\left(
  Y_t^2\,c_{\beta}^2 + Y_b^2\,s_{\beta}^2\right)
  \ln\frac{m_{H^{\pm}}^2}{M_{\text{EW}}^2}\,, &
  \zeta_{Hh} &\approx -\frac{3}{16\,\pi^2}\left(Y_t^2 - Y_b^2\right)
  s_{\beta}\,c_{\beta}\,\ln\frac{m_{H^{\pm}}^2}{M_{\text{EW}}^2}\,.
\end{align}
Similarly, the fields of the \CP-odd and charged Higgs receive loop
corrections according to \AtoB{A}{Z_A\,A} and
\AtoB{H^{\pm}}{Z_{H^{\pm}}\,H^{\pm}} with \mbox{$Z_A\simeq Z_H\simeq
  Z_H^{\pm}$}, while the mixing with the Goldstone bosons is kept
apart. Below, we continue to use the notations \mbox{$h\sim h_1$}
or \mbox{$H\sim h_2$} indifferently.

\tocsubsection{Two-loop \texorpdfstring{$\mathcal{O}{\left(Y_q^4\right)}$}{\unicodescriptO(\unicodeitalicY\unicodesubt\unicodesupfour)} corrections to the Higgs masses}

The choice of processing the mixing in the \CP-even sector differently
than in the \CP-odd or charged sectors intervenes first at the level
of the mass~calculation, when corrections of two-loop order are
considered. Indeed, the off-diagonal corrections of one-loop order
contribute only from this order on, but the two-loop effects of
$\mathcal{O}{\left(Y_q^4\right)}$ belong to those that are commonly
included in the mass~determination. From the perspective of an
expansion up to the two-loop order one has
\begin{align}\label{eq:expmassHferm}
  M^2_{h_i} &= m^2_{h_i} - \Real{\hat{\Sigma}_{ii}^{\text{1L}}{\left(m^2_{h_i}\right)}
    + \hat{\Sigma}_{ii}^{\text{2L}}{\left(m^2_{h_i}\right)}
    - \hat{\Sigma}_{ii}^{\text{1L}}{\left(m^2_{h_i}\right)}\,
    \frac{d\hat{\Sigma}_{ii}^{\text{1L}}}{dp^2}{\left(m^2_{h_i}\right)}
    - \sum_{j\neq i}\frac{\hat{\Sigma}_{ij}^{\text{1L}}{\left(m^2_{h_i}\right)}\,
      \hat{\Sigma}_{ji}^{\text{1L}}{\left(m^2_{h_i}\right)}}{m^2_{h_i}-m^2_{h_j}}}\,.
\end{align}

We expect the mass-splitting between heavy-doublet states to be
protected by the $SU(2)_{\mathrm{L}}$-symme\-try, \IE~contributions of
order \mbox{$m^2_{H^{\pm}}\left\{\ln^2{\left(m^2_{H^{\pm}}/M^2_{\text{EW}}\right)},\,\ln{\left(m^2_{H^{\pm}}/M^2_{\text{EW}}\right)},\,1\right\}$}
should disappear in the mass difference, leaving only terms of the
form \mbox{$M^2_{\text{EW}}\left\{\ln^2{\left(m^2_{H^{\pm}}/M^2_{\text{EW}}\right)},\,\ln{\left(m^2_{H^{\pm}}/M^2_{\text{EW}}\right)},\,1\right\}$}.\footnote{Terms
scaling like $m_{H^\pm}\,M_{\text{EW}}$ do not appear from
contributions of Yukawa type.} As the charged Higgs mass is
renormalized on-shell in our scheme, the radiative corrections should
satisfy the above property directly at the level of the renormalized
self-energies---\IE~the counterterms should automatically remove the
$SU(2)_{\mathrm{L}}$-symmetry-violating terms.

\begin{figure}[t!]
\centering
\includegraphics[width=\linewidth]{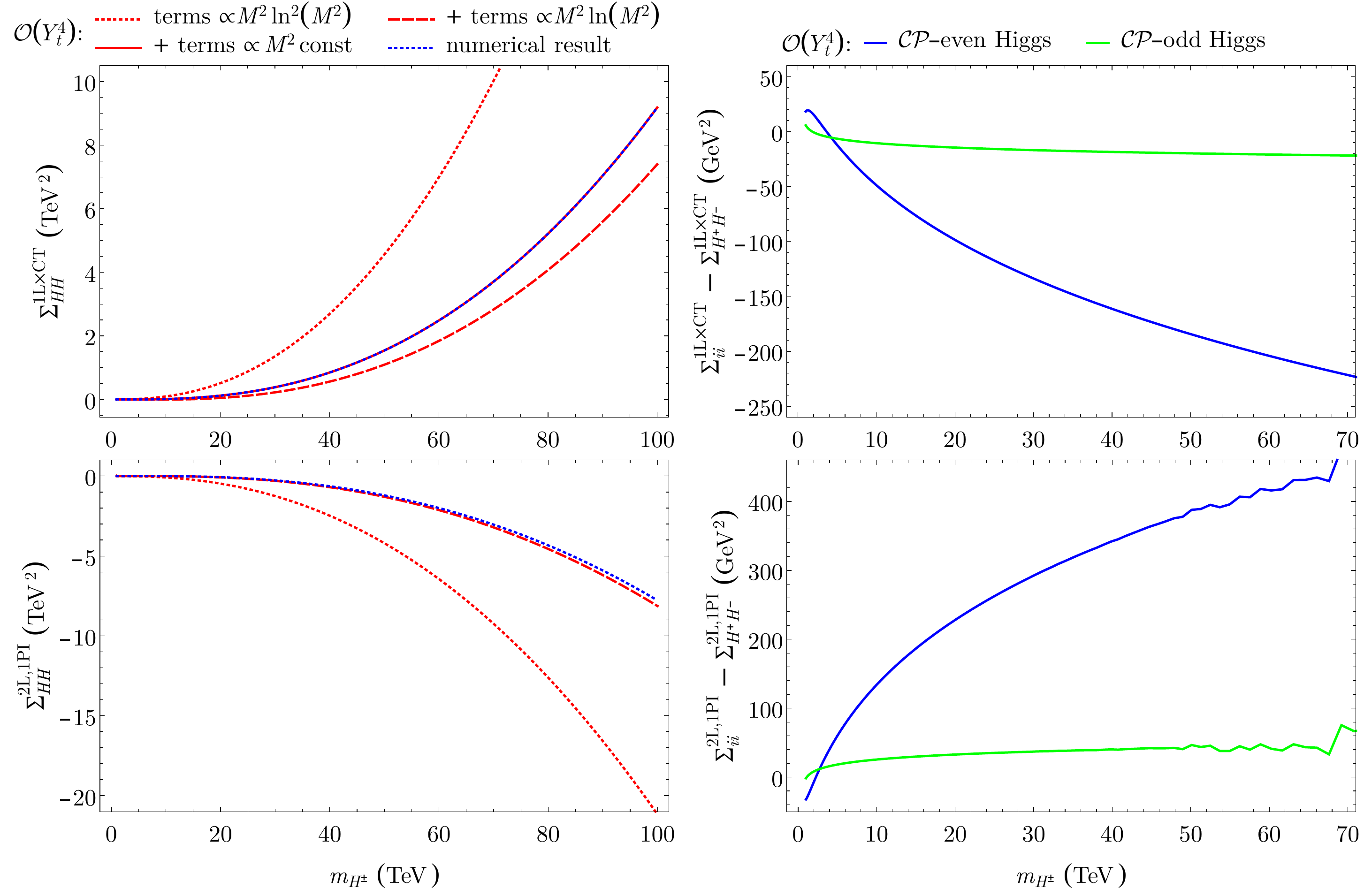}
\caption{Contributions of order $Y_t^4$ (by fermion loops) to the
  diagonal self-energies of heavy-doublet Higgs states are
  displayed. \mbox{$t_{\beta}=3$} and the SUSY sector is decoupled (at
  a scale of $200$\,TeV). Plots on the left-hand side show the
  convergence of the expansion in $M^2_{\text{EW}}/p^2$,
  with \mbox{$p^2\sim M^2_{H^{\pm}}$}. For the case of two-loop
  integrals, we only extracted analytically the logarithmic terms,
  which is why there is no solid red line in the lower plot. Plots on
  the right-hand side show the cancellation with corresponding
  charged-Higgs self-energies, intervening in the
  counterterm. Resulting diagonal contributions to the \CP-even (blue)
  or \CP-odd (green) self-energies are of electroweak size. The
  numerical evaluation of the two-loop self-energies has been achieved
  with \texttt{TSIL}. The wiggles for \mbox{$M_{H^{\pm}}\geq50$}\,TeV
  in the plot in the lower right-hand quadrant shows
  that \texttt{TSIL} becomes unstable in this regime. The
  corresponding instabilities appear at much lower mass~values
  ($\simord$\,TeV) for the contributions of the bottom quark,
  justifying our choice of displaying only the order $Y_t^4$ and
  considering the low value of \mbox{$t_{\beta}=3$}, where the bottom
  contributions are suppressed.\label{fig:2Lself}}
\end{figure}

Let us first examine the genuine diagonal two-loop piece
$\hat{\Sigma}_{ii}^{\text{2L}}$.\footnote{In the following we refer to
the momentum-dependent contributions by top and bottom quarks; in the
collection of two-loop contributions, we also display the
contributions by stops and sbottoms that have been derived
in\,\cite{Hollik:2014bua,Hollik:2014wea,Passehr:2017ufr}.} As the
two-loop electroweak calculation is incomplete in
the~MSSM,\footnote{The full two-loop gauge contributions are known for
scalar particles\,\cite{Goodsell:2019zfs}, but the two-loop
self-energies for the gauge bosons that are needed for the
counterterms in the Higgs sector to connect to observables are not yet
available.} we work in the gaugeless limit in this subsection. The
full two-loop self-energy $\hat{\Sigma}_{ii}^{\text{2L}}$ can be
decomposed into several contributions, starting with genuine
one-particle irreducible~(1PI) two-loop self-energy diagrams
$\Sigma_{ii}^{\text{2L,1PI}}$, 1PI~one-loop self-energy diagrams with
counterterm insertion $\Sigma_{ii}^{\text{1L$\times$CT}}$, two-loop
counterterm diagrams involving a pair of one-loop counterterms
$\Sigma_{ii}^{\text{CT$\times$CT}}$ and the genuine two-loop
counterterm contribution $\Sigma_{ii}^{\text{2L,CT}}$. In the
gaugeless limit and in our renormalization scheme, the counterterms to
the non-expanded heavy neutral self-energies are exactly equal to the
on-shell charged self-energy that can be decomposed in the same way;
thus, the renormalized self-energies are given by
$\hat{\Sigma}_{ii}^{\text{2L}}=\Sigma_{ii}^{\text{2L,1PI}}-\Sigma_{H^+H^-}^{\text{2L,1PI}}+\Sigma_{ii}^{\text{1L$\times$CT}}-\Sigma_{H^+H^-}^{\text{1L$\times$CT}}+\Sigma_{ii}^{\text{CT$\times$CT}}-\Sigma_{H^+H^-}^{\text{CT$\times$CT}}$
(see below for additional counterterms once the one-loop squared terms
are included). Each piece can be expanded in the heavy-mass limit,
providing the following leading terms
of~$\mathcal{O}{\left(Y_q^4\right)}$:\footnote{The coefficients in the
expansion of the occurring two-loop integrals have mostly been
determined by numerical methods with the help
of \texttt{TSIL}\,\cite{Martin:2005qm}; some analytic expansions were
performed with methods of \citeres{Scharf:1993ds,Davydychev:1993pg}.}
\begin{subequations}\allowdisplaybreaks
\begin{align}
  \Sigma_{ii, H^+H^-}^{\text{2L,1PI}}{\left(M^2\right)} &\sim
  \frac{3}{512\,\pi^4} \left[
    3\,Y_t^4\,c_{\beta}^2\left(4\,c_{\beta}^2-1\right)
    + 3\,Y_b^4\,s_{\beta}^2 \left(4\,s_{\beta}^2 - 1\right) - Y_t^2\,Y_b^2\right]
  M^2\,\ln^2{\frac{M^2}{M^2_{\text{EW}}}}\,,\\[1ex]
  \Sigma_{ii,H^+H^-}^{\text{1L$\times$CT}}{\left(M^2\right)} &\sim
  \frac{3}{1024\,\pi^4}\, \Big\{\begin{aligned}[t]
    & 3\,Y_t^4\,c_{\beta}^2\left(1 + 3\,\delta_{m_t}^{\text{OS},1}\,c_{\beta}^2\right)
    + 3\,Y_b^4\,s_{\beta}^2\left(1 + 3\,\delta_{m_b}^{\text{OS},1}\,s_{\beta}^2\right)\\
    &{+}\, Y_t^2\,Y_b^2 \left[3\left(1+3\,c^2_{\beta}\right)\delta_{m_t}^{\text{OS},1}
      +3\left(1+3\,s^2_{\beta}\right)\delta_{m_b}^{\text{OS},1}+1\right]\!\!\Big\}\,
    M^2\,\ln^2{\frac{M^2}{M^2_{\text{EW}}}}\,,\taghere\end{aligned}
    \hspace{1.5em}\\[1ex]
  \Sigma_{ii}^{\text{CT$\times$CT}}{\left(M^2\right)} &=
  \Sigma_{H^+H^-}^{\text{CT$\times$CT}}{\left(M^2\right)} \sim
  -\frac{9}{512\,\pi^4}\left(Y_t^2\,c^2_{\beta} + Y_b^2\,s^2_{\beta}\right)
  M^2\,\ln^2{\frac{M^2}{M^2_{\text{EW}}}}\,,
\end{align}
\end{subequations}
where $\delta_{m_t}^{\text{OS},1}=1=\delta_{m_b}^{\text{OS},1}$ are
related to the $\overline{\text{DR}}$/on-shell conversion of fermion
masses. The convergence of this expansion is illustrated on the
left-hand side of \fig{fig:2Lself} for both
$\Sigma_{ii}^{\text{2L,1PI}}$ and $\Sigma_{ii}^{\text{1L$\times$CT}}$,
considering only the terms of order $Y_t^4$, at
\mbox{$t_{\beta}=3$}. \texttt{TwoCalc}\,\cite{Weiglein:1993hd} and
\texttt{TLDR}\,\cite{Goodsell:2019zfs} have been utilized for
analytical manipulations of the two-loop diagrams, and numerical
results for the two-loop integrals are obtained
via \texttt{TSIL}\,\cite{Martin:2005qm}. These terms are exactly
mirrored by the corresponding contributions to the charged Higgs
self-energy which, in our renormalization scheme, appears in the
two-loop counterterms to the neutral self-energies. The resulting
cancellation is shown on the right-hand side of \fig{fig:2Lself},
resizing the diagonal contributions to the renormalized Higgs
self-energies to electroweak magnitude (as we expected).

Now, let us turn to the one-loop squared terms of
\refeq{eq:expmassHferm}. The diagonal one,
$\hat{\Sigma}_{ii}^{\text{1L}}\,d\hat{\Sigma}_{ii}^{\text{1L}}/dp^2$,
obviously satisfies the property that we stated before: when replacing
the renormalized one-loop self-energies by their approximate
expressions of \refeqs{eq:mixHferm}, we find only terms of electroweak
size. The off-diagonal term (last term of \refeq{eq:expmassHferm})
behaves differently, however, and generates a leading contribution of
order
$m^2_{H^{\pm}}\,\ln^2{\left(m^2_{H^{\pm}}/M^2_{\text{EW}}\right)}$:
{\setlength{\abovedisplayskip}{1.8ex}
\setlength{\belowdisplayskip}{1.8ex}
\begin{align}
  \sum_{j\neq i}\frac{\hat{\Sigma}_{ij}^{\text{1L}}{\left(m^2_{h_i}\right)}\,
    \hat{\Sigma}_{ji}^{\text{1L}}{\left(m^2_{h_i}\right)}}{m^2_{h_i}-m^2_{h_j}}&\approx
  \left[\frac{3}{16\,\pi^2}\left(Y_t^2 - Y_b^2\right) s_{\beta}\,c_{\beta}\right]^2
  m^2_{H^{\pm}}\,\ln^2\frac{m^2_{H^{\pm}}}{M^2_{\text{EW}}}\,.
\end{align}}%
This applies to the \CP-even sector \mbox{($h_i\equiv H$, $h_{j\neq
i}\equiv h$)}, but also to the \CP-odd \mbox{($h_i\equiv A$, $h_{j\neq
i}\equiv G^0$)} and charged \mbox{($h_i\equiv H^{\pm}$, $h_{j\neq
i}\equiv G^{\pm}$)} ones, so that the $SU(2)_{\mathrm{L}}$-symmetry is
still not (strongly) violated. In the renormalization scheme under
consideration, with an on-shell charged Higgs, this also means that
the two-loop counterterm contribution $\Sigma_{ii}^{\text{2L,CT}}$
should contain the term
$-\hat{\Sigma}_{H^+G^-}^{\text{1L}}\,\hat{\Sigma}_{G^+H^-}^{\text{1L}}/m^2_{H^{\pm}}$,
and balance the one-loop squared term directly at the level of the
renormalized self-energy. To be explicit, the renormalization
condition \mbox{$\Real{\hat{\Sigma}_{H^+H^-}{\left(m_{H^\pm}^2\right)}}=0$}
should be applied to the full expression, including off-diagonal
contributions, at each order of the loop expansion. At the two-loop
order, this fixes the on-shell counterterm to
{\setlength{\abovedisplayskip}{1.8ex}
\setlength{\belowdisplayskip}{1.8ex}
\begin{align}\label{eq:dMHp2}
  \delta^{\mathrm{2L}} m_{H^\pm}^2 &=
  \Real{\Sigma^{\mathrm{2L}}_{H^+H^-}{\left(m_{H^\pm}^2\right)}
    - \hat{\Sigma}^{\mathrm{1L}}_{H^+H^-}{\left(m_{H^\pm}^2\right)}\,
    \frac{d\hat{\Sigma}^{\mathrm{1L}}_{H^+H^-}}{dp^2}{\left(m_{H^\pm}^2\right)}
    - \frac{\hat{\Sigma}^{\mathrm{1L}}_{H^+G^-}{\left(m_{H^\pm}^2\right)}\,
      \hat{\Sigma}^{\mathrm{1L}}_{G^+H^-}{\left(m_{H^\pm}^2\right)}}
    {m_{H^\pm}^2 - m_{G^\pm}^2}}\,.\hspace{2em}\notag\\[-3ex]
\end{align}}%
The renormalization scheme at the one-loop order is not repeated here
(see \EG~\citere{Frank:2006yh}). We note that---although
$\Real{\hat{\Sigma}_{H^+H^-}^{\mathrm{1L}}{\left(m_{H^\pm}^2\right)}}$
is requested to vanish---the term of the momentum expansion,
$\Real{\hat{\Sigma}^{\mathrm{1L}}_{H^+H^-}{\left(m_{H^\pm}^2\right)}\,
d\hat{\Sigma}^{\mathrm{1L}}_{H^+H^-}/dp^2{\left(m_{H^\pm}^2\right)}}$,
still contributes a finite shift due to the imaginary parts (see
\EG~\citere{Bahl:2018qog} for details). In addition, \mbox{$m_{G^\pm}^2=0$}
in the gaugeless limit (we remind the reader that
$\hat{\Sigma}^{\text{1L}}_{H^{\pm}W^{\mp}}$ and
$\hat{\Sigma}^{\text{1L}}_{AZ}$ also vanish in this approximation),
while in the presence of gauge contributions the combination
of the mixings with gauge and
Goldstone bosons also restores the
denominator~$1/m_{H^\pm}^2$, as explained
in \citeres{Williams:2011bu,Domingo:2018uim} for the neutral
case.\footnote{In the MSSM with real parameters, another popular
renormalization condition consists in requiring an on-shell \CP-odd
Higgs instead of an on-shell charged Higgs; in that case, the
renormalization condition implies similar contributions to the mass
counterterm, containing the off-diagonal
self-energy~$\hat{\Sigma}_{AG^0}$ squared. If neither the charged
nor \CP-odd Higgs are renormalized on-shell, the off-diagonal
self-energy squared terms continue to explicitly appear in the
radiative corrections to each individual propagator, still subtracting
one another at the level of the mass-splitting.}

\begin{figure}[p!]
\centering
\includegraphics[width=\linewidth]{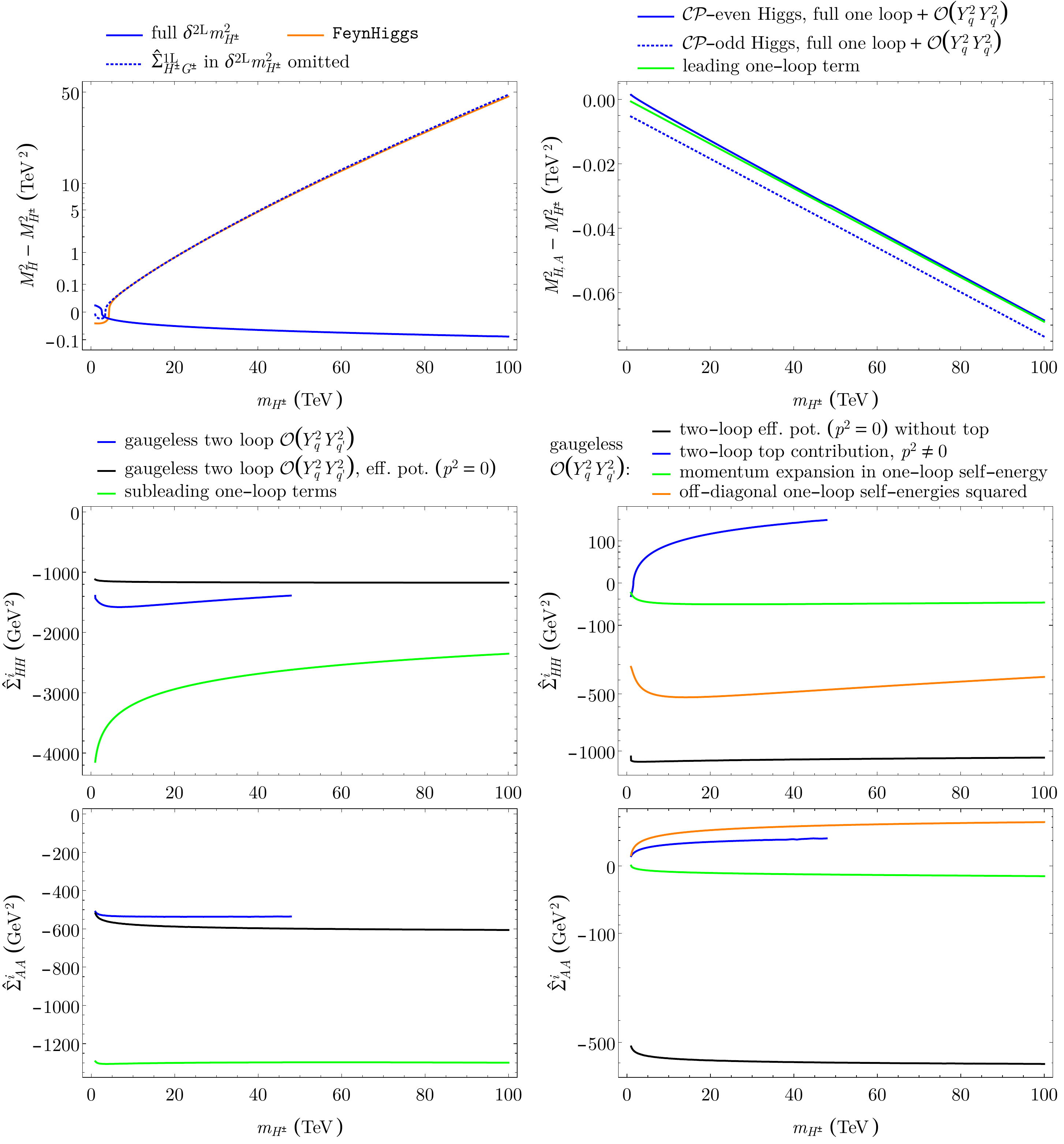}
\caption{Contributions to the squared mass-splitting for the
heavy-doublet states in the scenario of \fig{fig:2Lself}.
\newline{\em Upper left}: one-loop and $\mathcal{O}(Y_q^4)$ two-loop
contributions to the squared mass-splitting between the \CP-even and
charged Higgs in \texttt{FeynHiggs-2.16.0} (orange), in our
calculation (solid blue), and omitting the off-diagonal contribution
to \refeq{eq:dMHp2} (dotted blue).
\newline{\em Upper right}: one-loop and $\mathcal{O}(Y_q^4)$ two-loop
contribution to the squared mass-splitting between the neutral
(\CP-even: solid blue; \CP-odd: dotted blue) and charged Higgs; the
leading one-loop contribution from gauge corrections is shown in
green;
\newline{\em Middle left}: Contributions to the \CP-even self-energy
beyond the leading one-loop term from gauge corrections; subleading
one-loop contributions in green; two-loop $\mathcal{O}(Y_q^4)$
contributions in the effective potential approximation (black), and in
the gaugeless limit (blue);
\newline{\em Middle right}: Contributions of order $\mathcal{O}(Y_q^4)$
to the \CP-even self-energy; SUSY and bottom (black), and top (blue)
diagonal two-loop pieces; diagonal one-loop squared piece (green);
off-diagonal one-loop squared piece (orange);
\newline{\em Bottom left and right}: as middle left and right but applied
to the \CP-odd state.
\newline The two-loop contribution to the diagonal self-energy in the
gaugeless limit is only displayed up to $50$\,TeV in order to avoid
numerical instabilities.
\label{fig:2Lmass}}
\end{figure}

A problem arises when the off-diagonal contribution to the
charged-Higgs mass counterterm is overlooked, as seems to have been
the case in some earlier calculations---see \EG~Eq.\,(11)
in\,\cite{Bahl:2018qog}---, leading to an explicit violation of the
$SU(2)_{\mathrm{L}}$-symmetry and an inconsistent order
$\mathcal{O}{\left(Y_q^4\right)}$. While this issue is made obvious
for heavy-doublet states by the analysis of logarithms, it actually
points at a conceptual shortcoming in the mass~calculation for all
Higgs states. Even in a calculation of one-loop order, the partial
inclusion of the \CP-even off-diagonal contribution, but not of the
corresponding Higgs--Goldstone mixing in the charged and/or \CP-odd
sector, in fact worsens the quality of the mass~prediction for the
heavy states with respect to a simple truncation at strict one-loop
order, since it introduces unwarranted $SU(2)_{\mathrm{L}}$-violating
effects. This is what causes the shift between the green and orange
lines in the middle plot of \fig{fig:xidep_MSSM}
for \mbox{$m_{H^{\pm}}=1$}\,TeV, as we already commented, an effect
competing with the $\mathcal{O}(\alpha_t\alpha_s)$
corrections to $M_H$ and which is usually `hidden' in the
pole-search procedure. Nevertheless, the charged Higgs counterterm has
a limited impact on the mass prediction for the SM-like Higgs in
general, due to the decoupling of this latter state from the heavy
doublet. Therefore, only in scenarios where~\mbox{$m_{H^{\pm}}\sim v$}
and a large $h$--$H$~mixing develops could this issue have any
numerical consequences for $M_{h}$, but we note that there is no
longer any logarithmic enhancement in such scenarios either. In
addition, the partial result obtained in the gaugeless limit is not
really suited to explore such configurations.

In \fig{fig:2Lmass}, we first consider the squared mass-splitting
between the heavy-doublet states of the scenario
of \fig{fig:2Lself}. The plot on the left-hand side of the first row
shows this squared mass-splitting for the heavy \CP-even and charged
Higgs. The result
of \texttt{FeynHiggs-2.16.0}\,\cite{Heinemeyer:1998yj,Heinemeyer:1998np,Degrassi:2002fi,Frank:2006yh,Hahn:2013ria,Bahl:2016brp,Bahl:2017aev,Bahl:2018qog}
with the setting \texttt{FHSetFlags[4,4,0,1,0,0,0,3]} is displayed in
orange and shows a comparatively large and growing squared
mass-splitting. Our result is the solid blue curve, corresponding to
considerably smaller values. We are able to roughly reproduce the
result of \texttt{FeynHiggs} when omitting the off-diagonal
self-energy-squared term in the mass counterterm of \refeq{eq:dMHp2}
(dotted blue line; also the \CP-even mixing is then included in the
pole search). This mismatch is thus explained by the large
$SU(2)_{\text{L}}$-violating terms of
order~$\mathcal{O}{\left(Y_q^4\right)}$ scaling
like \mbox{$m^2_{H^{\pm}}\left\{\ln^2{\left(m^2_{H^{\pm}}/M^2_{\text{EW}}\right)},\,\ln{\left(m^2_{H^{\pm}}/M^2_{\text{EW}}\right)},\,1\right\}$}
that have been erroneously introduced in the calculation. It is
remarkable that these `wrong' two-loop effects then dominate the
one-loop corrections, which, in our prediction, explain most of the
squared mass-splitting. The plot on the right-hand side examines our
result more closely. The bulk of the variation both
in~\mbox{$M_H^2-M_{H^{\pm}}^2$} and~\mbox{$M_A^2-M_{H^{\pm}}^2$}
originates in one-loop effects of gauge type, dominated by a term
scaling linearly with the mass of the heavy
doublet,~$\simord\alpha\,M_Z\,m_{H^{\pm}}$ (plotted in green). The
following plots further analyze the remainder in the \CP-even (second
row) and the \CP-odd cases (third row). On the left, we compare the
magnitude of the remaining one-loop contributions (\IE~beyond the
gauge term scaling linearly; in green), the two-loop corrections of
order~$\mathcal{O}{\left(Y_q^4\right)}$ in the effective-potential
approximation (black curves) and in the gaugeless limit (blue
curves). It thus appears that the genuine effects
of~$\mathcal{O}{\left(Y_q^4\right)}$ are considerably smaller than
what the off-diagonal contribution in the \CP-even sector hinted. In
addition, we note that the two-loop terms obtained in the effective
potential approximation are~$\simord30\%$ away from those obtained in
the \textit{a priori} more complete gaugeless limit.\footnote{We
stress that these two-loop contributions contain large logarithms of
the type $\ln(m_{\tilde{q}}^2/M_{\text{EW}}^2)$ in our scenario with a
heavy SUSY spectrum; the threshold corrections for the heavy SUSY
sector are explained for a SM-EFT matching
in \citeres{Draper:2013oza,Bagnaschi:2014rsa,Vega:2015fna,Bagnaschi:2017xid,Harlander:2018yhj,Bagnaschi:2019esc,Bahl:2019wzx},
and for a THDM-EFT matching
in \citeres{Haber:1993an,Gorbahn:2009pp,Carena:2015uoe,Lee:2015uza,Bahl:2018jom,Murphy:2019qpm},
but this feature goes beyond the scope of this article.} Finally, the
plots on the right-hand side of \fig{fig:2Lmass}, second and third
row, show the individual contributions to the two-loop self-energies
of order~$\mathcal{O}{\left(Y_q^4\right)}$ as listed
in \refeq{eq:expmassHferm}: each piece remains of electroweak order as
could be anticipated.

\medskip
As a closing remark, we note that, while the order $Y_q^4$ is known,
it may not be the most relevant type of two-loop contribution for
heavy-doublet states. Indeed, already at the one-loop order we
mentioned that the radiative corrections to the squared masses are
dominated by a linear term~$\simord\alpha\,M_Z\,m_{H^{\pm}}$ due to
electroweak gauge effects. It is thus likely that two-loop corrections
to the masses of the heavy doublets are dominated by contributions of
the same type, \IE~by momentum-dependent electroweak
effects. This does evidently not spoil the relevance of our previous
remarks as to a consistent treatment of the one-loop squared term,
since the `erroneous' terms of
order~\mbox{$Y_q^4\,m^2_{H^{\pm}}\left\{\ln^2{\left(m^2_{H^{\pm}}/M^2_{\text{EW}}\right)},\,\ln{\left(m^2_{H^{\pm}}/M^2_{\text{EW}}\right)},\,1\right\}$}
first need to be put under control before the linear contribution
becomes apparent.

\tocsubsection{Application to the decays into quarks}

Now, we focus on the decays \AtoB{H,\,A}{t\,\bar{t}},
\AtoB{H^+}{t_{\mathrm{R}}\,\bar{b}_{\mathrm{L}}}, processes that are
mediated through $Y_t$ at the tree level. The vertex corrections at
the same order as considered in \refeqs{eq:mixHferm} read
{\setlength{\abovedisplayskip}{1.5ex}
\setlength{\belowdisplayskip}{1.5ex}
\begin{align}
  \frac{\Amp{vert}{}{\AtoB{H,\,A}{t\,\bar{t}\,}}}
       {\Amp{tree}{}{\AtoB{H,\,A}{t\,\bar{t}\,}}} &\simeq
  \frac{\Amp{vert}{}{\AtoB{H^+}{t_{\mathrm{R}}\,\bar{b}_{\mathrm{L}}}}}
       {\Amp{tree}{}{\AtoB{H^+}{t_{\mathrm{R}}\,\bar{b}_{\mathrm{L}}}}} \simeq
  -\frac{1}{16\,\pi^2}\,Y_b^2\,c_{\beta}^2\,\ln\frac{m_{H^{\pm}}^2}{M^2_{\text{EW}}}\,.
\end{align}}%
Following the LSZ reduction, we then add the contributions from
self-energy diagrams on the Higgs leg, provided by \refeqs{eq:LSZmix},
and obtain at the one-loop order:
{\allowdisplaybreaks
\begin{subequations}\label{eq:httLSZ}
\setlength{\abovedisplayskip}{1.5ex}
\setlength{\belowdisplayskip}{1.5ex}
\begin{align}
  \Amp{1L}{}{\AtoB{H}{t\,\bar{t}\,}} &= \Amp{tree}{}{\AtoB{H}{t\,\bar{t}\,}}
  + \Amp{vert}{}{\AtoB{H}{t\,\bar{t}\,}}
  - \frac{\hat{\Sigma}_{Hh}{\left(M_H^2\right)}}{M_H^2 - m_h^2}\,
    \Amp{tree}{}{\AtoB{h}{t\,\bar{t}\,}}\,\notag\\*[-1ex]
  &\quad -\frac{1}{2}\,\frac{d\hat{\Sigma}_{HH}}{dp^2}{\left(M_H^2\right)}\,
    \Amp{tree}{}{\AtoB{H}{t\,\bar{t}\,}}\,,\\
  \Amp{1L}{}{\AtoB{A}{t\,\bar{t}\,}} &= \Amp{tree}{}{\AtoB{A}{t\,\bar{t}\,}}
  + \Amp{vert}{}{\AtoB{A}{t\,\bar{t}\,}}
  - \frac{\hat{\Sigma}_{AG^0}{\left(M_A^2\right)}}{M_A^2}\,
    \Amp{tree}{}{\AtoB{G^0}{t\,\bar{t}\,}}\notag\\*[-.8ex]
  &\quad -\frac{1}{2}\,\frac{d\hat{\Sigma}_{AA}}{dp^2}{\left(M_A^2\right)}\,
    \Amp{tree}{}{\AtoB{A}{t\,\bar{t}\,}}\,,\\
  \Amp{1L}{}{\AtoB{H^+}{t_{\mathrm{R}}\,\bar{b}_{\mathrm{L}}}} &=
  \Amp{tree}{}{\AtoB{H^+}{t_{\mathrm{R}}\,\bar{b}_{\mathrm{L}}}}
  + \Amp{vert}{}{\AtoB{H^+}{t_{\mathrm{R}}\,\bar{b}_{\mathrm{L}}}}
  - \frac{\hat{\Sigma}_{H^+G^-}{\left(m_{H^{\pm}}^2\right)}}{m_{H^{\pm}}^2}\,
    \Amp{tree}{}{\AtoB{G^+}{t_{\mathrm{R}}\,\bar{b}_{\mathrm{L}}}}\notag\\*[-.8ex]
  &\quad-\frac{1}{2}\,\frac{d\hat{\Sigma}_{H^+H^-}}{dp^2}{\left(m_{H^{\pm}}^2\right)}
    \,\Amp{tree}{}{\AtoB{H^+}{t_{\mathrm{R}}\,\bar{b}_{\mathrm{L}}}}\,,\\
  \Rightarrow\quad\frac{\Amp{1L}{}{\AtoB{H}{t\,\bar{t}\,}}}{\Amp{tree}{}{\AtoB{H}{t\,\bar{t}\,}}} &\simeq
  \frac{\Amp{1L}{}{\AtoB{A}{t\,\bar{t}\,}}}{\Amp{tree}{}{\AtoB{A}{t\,\bar{t}\,}}} \simeq
  \frac{\Amp{1L}{}{\AtoB{H^+}{t_{\mathrm{R}}\,\bar{b}_{\mathrm{L}}}}}{\Amp{tree}{}{\AtoB{H^+}{t_{\mathrm{R}}\,\bar{b}_{\mathrm{L}}}}}\notag\\*
  &\simeq 1 + \frac{1}{32\,\pi^2}\left[3\,Y_t^2\left(1+s^2_{\beta}\right)
    - Y_b^2\left(2+s^2_{\beta}\right)\right]
    \ln{\frac{m^2_{H^{\pm}}}{M^2_{\text{EW}}}}\,.
\end{align}
\end{subequations}}%
The propagator in $1/p^2$ for the Goldstone bosons is justified by the
combination with the weak gauge-boson effects (see \EG\ section\,4.3
in\,\citere{Williams:2011bu}). We observe that (at the considered
order) the radiative corrections preserve the
$SU(2)_{\mathrm{L}}$-symmetry for the heavy-doublet states: their
decay widths are identical at the tree level (up to corrections of
$\mathcal{O}{\left(M_{\text{EW}}^2/m_{H^{\pm}}^2\right)}$) and
continue to be so at the one-loop order. This is indeed what we expect
for \mbox{$M_{\text{EW}}^2\ll m_{H^{\pm}}^2$}: massive states are
hardly sensitive to the electroweak symmetry-breaking effects.

\begin{figure}[t!]
\centering
\includegraphics[width=\linewidth]{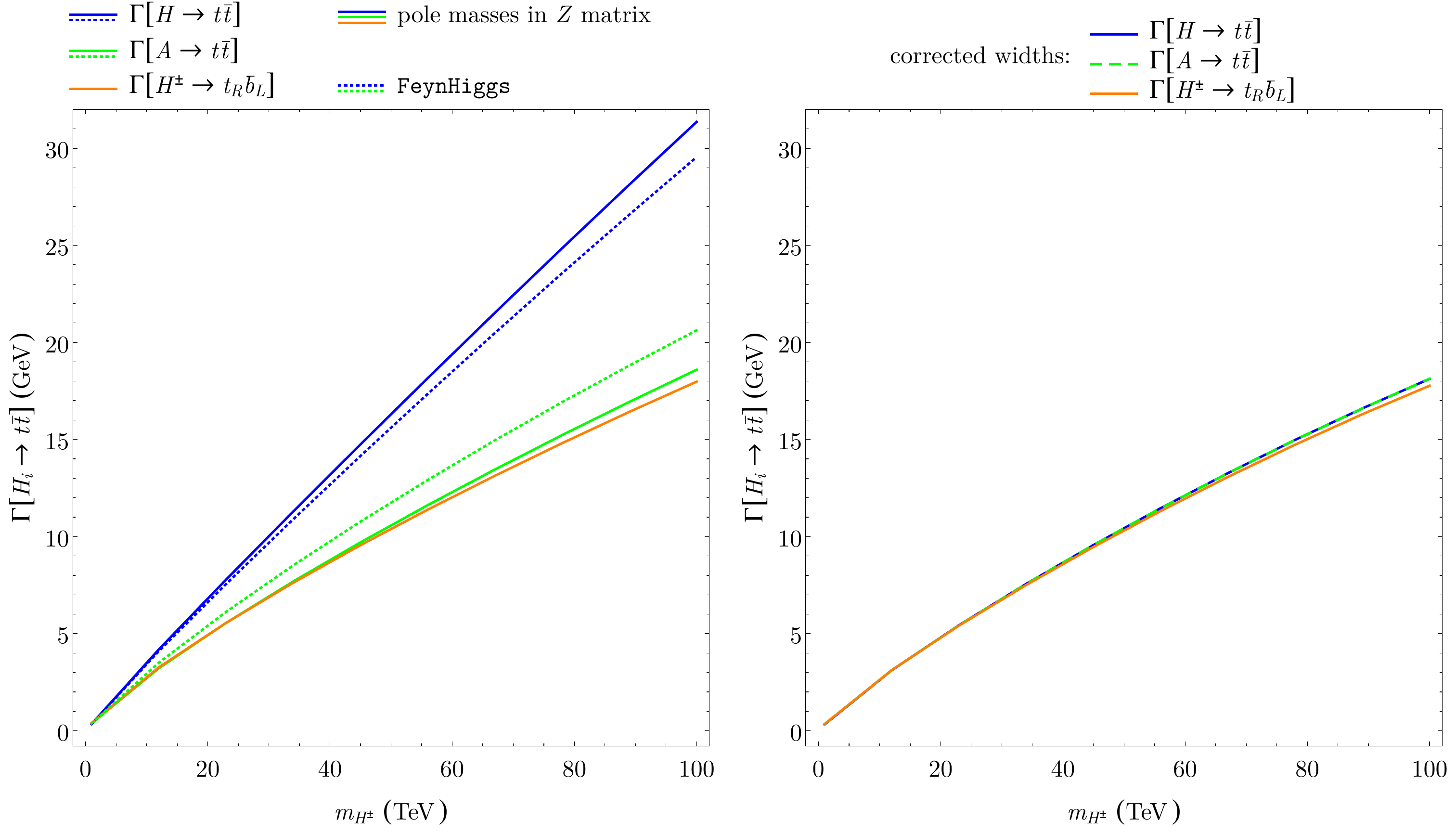}
\caption{Decay widths of heavy-doublet states into top quarks are
shown for $t_{\beta}=10$ and $m_{\text{SUSY}}\simeq100$\,TeV.
\newline{\em Left}: the mixing contributions to the \CP-even channel
are computed with Yukawa couplings at low energy (to mimic the
corresponding calculation in the mixing formalism); other terms employ
Yukawa couplings that include QCD-running up to the scale of the
decaying Higgs. Corresponding results from \texttt{FeynHiggs} are
shown as dashed curves for the neutral states---to our knowledge, the
corresponding order is not available for the charged
Higgs. Differences with our results (solid lines) originate in several
higher-order pieces (as we checked): mixing matrix vs.\ LSZ, inclusion
of a vertex$^2$ term in \texttt{FeynHiggs}, resummation of Sudakov
double logarithms in our case.
\newline{\em Right}: Mixing and vertex contributions all
consistently employ Yukawa couplings including the
QCD-running.\label{fig:httwrong}}
\end{figure}

Obviously, we would formally obtain the same expansion as that of
\refeqs{eq:httLSZ} at this order when employing the mixing formalism
instead of the LSZ reduction. Therefore, any deviation from comparable
values in the decay widths in the \CP-even, \CP-odd or charged sector
would have to originate in higher-order terms. Nevertheless, on the
left-hand side of \fig{fig:httwrong}, we observe a sizable mismatch
between the decay widths of the \CP-even state (blue curves), on the
one hand, and the \CP-odd (green) and charged Higgs (orange), on the
other hand, when employing the mixing formalism. This discrepancy
increases at large masses, where we would expect the
$SU(2)_{\mathrm{L}}$-violating effects to become smaller. The results
of \texttt{FeynHiggs} (dotted curves) show a similar behaviour but
differ from ours (solid lines) because of several higher order pieces
($\lvert\mathcal{A}^{\text{vert}}\rvert^2$ in \texttt{FeynHiggs},
resummation of Sudakov double logarithms on our side, etc.). At this
level, we are forced to consider the difference between the blue and
green curves of \fig{fig:httwrong} as setting the magnitude of the
higher-order uncertainty in the calculation, an uncertainty close
to~$100\%$ at sufficiently high mass! Of course, the high-mass regime
is currently not so interesting phenomenologically, but our purpose in
considering it is to test the robustness of the combination of vertex
and mixing pieces in a clearly $SU(2)_{\text{L}}$-conserving
regime. In fact, the origin of this issue entirely rests with the
different procedures that are employed in the \CP-even, \CP-odd and
charged sectors in order to include the loop corrections on the
external Higgs leg. As expected, these variants formally deviate by
two-loop effects of order~$\mathcal{O}{\left(Y_q^2\,\alpha_s\right)}$,
which however prove to be numerically significant. Indeed, the Higgs
mixing in the \CP-even sector has been defined at the same level as
the Higgs mass, \IE\ the Yukawa couplings are given at the electroweak
scale, as prescribed by the renormalization scheme. On the contrary,
in the \CP-odd and the charged sectors, the Higgs mixing is
incorporated at the same time as the vertex corrections,
\IE\ explicitly for the decay. At this level, the QCD~analysis (see
\EG\ \citeres{Braaten:1980yq,Drees:1990dq}) makes it clear that
QCD~logarithms can be resummed in the calculation of the inclusive
width (\IE~including gluon radiation) through the incorporation of the
QCD~running in the Yukawa couplings up to the scale of the decaying
Higgs state: QCD~logarithms are solely of ultraviolet type at the
level of the inclusive width. This justifies the use of running Yukawa
couplings defined at the high scale in the decay width, and in
particular in the \CP-odd and charged mixing contributions. As
announced above, the latter thus differ from the mixing contribution
implemented in the \CP-even sector by terms of
order~$\mathcal{O}{\left(Y_q^2\,\alpha_s\right)}$ (and higher
orders). Consequently, the resulting $SU(2)_{\mathrm{L}}$-breaking
effect is purely artificial and of higher order. However, the
QCD~analysis indicates that the recipe employed in the \CP-odd or
charged Higgs decays is more reliable in this specific case than the
mixing procedure in the \CP-even sector. 

On the right-hand side of \fig{fig:httwrong}, the same decay widths
are shown, but with a consistent combination of mixing and vertex
contributions for all the states. Then the decay widths for
the \CP-even and \CP-odd states are roughly identical. A small shift
persists between neutral and charged states. The latter has a physical
meaning: it already appears in the coefficients of the Sudakov
double-logarithms\,\cite{Domingo:2019vit}, which indeed differ at the
level of the exclusive widths (\IE~discarding $W$- and $Z$-radiation,
though inclusive with respect to~QCD and~QED~radiation), as a tribute
to the \mbox{$SU(2)_{\text{L}}\times U(1)_{\text{Y}}\to
U(1)_{\text{em}}$} breaking. Coming back to the phenomenologically
relevant regime at $M_{H^{\pm}}\sim1$\,TeV, we find that the decay
width of the heavy \CP-even Higgs has been shifted by $\simord30\%$ of
the magnitude of the electroweak corrections after purging the
formalism from artificial $SU(2)_{\text{L}}$-violating effects.

\tocsubsection{Application to the Higgs decays into weak gauge bosons}

The coupling of the heavy-doublet states to weak gauge bosons vanishes
in the decoupling limit. It is indeed difficult to build an
electroweakly invariant operator coupling exactly one doublet scalar
to two triplet or singlet vectors: this requires a breaking of the
$SU(2)_{\mathrm{L}}$-symmetry, which, as we argued before, should
appear as a subleading effect for states with a mass substantially
larger than the electroweak scale. Consequently, the decay widths
for~\AtoB{H,A}{WW,ZZ} and~\AtoB{H^{\pm}}{W^{\pm}Z} exactly or
approximately (in the \CP-even case) vanish at tree-level. On the
other hand the vertex corrections involve fermion loops that lead to
unsuppressed logarithms:
\begin{subequations}\label{eq:hvvvert}
\setlength{\abovedisplayskip}{1.0ex}
\setlength{\belowdisplayskip}{1.5ex}
\begin{align}
  \Amp{vert}{}{\AtoB{H}{VV}} &\simeq\frac{3}{16\,\pi^2}\left(Y_t^2-Y_b^2\right)
  s_{\beta}\,c_{\beta}\,\ln\frac{m_{H^{\pm}}^2}{M_{\text{EW}}^2}\,
  \Amp{tree}{}{\AtoB{h}{VV}}\,,\\[-.6ex]
  \Amp{vert}{}{\AtoB{A}{VV}} &\simeq 0\,,\\[-.6ex]
  \Amp{vert}{}{\AtoB{H^+}{W^+Z}} &\simeq
  -\frac{3}{16\,\pi^2}\left(Y_t^2-Y_b^2\right) s_{\beta}\,c_{\beta}\,
  \ln\frac{m_{H^{\pm}}^2}{M_{\text{EW}}^2}\,\Amp{tree}{}{\AtoB{G^+}{W^+Z}}\,.
\end{align}
\end{subequations}
The absence of a logarithmic contribution in the pseudoscalar case
points at an apparent deviation from the
$SU(2)_{\mathrm{L}}$-correspondence between the heavy-doublet
states. In addition, the existence of logarithmic terms for
the \CP-even and charged states contradicts our previous comment that,
in the $SU(2)_{\mathrm{L}}$-conserving limit, no operator mediating
these decays can be written. However, the result is as yet incomplete
at the considered one-loop order and should also include the mixing
contribution. Following \refeqs{eq:mixHferm}, we have
\begin{subequations}\label{eq:hvvmix}\allowdisplaybreaks
\setlength{\abovedisplayskip}{1.0ex}
\setlength{\belowdisplayskip}{1.5ex}
\begin{align}
  \Amp{mix}{}{\AtoB{H}{VV}} &\simeq -\frac{3}{16\,\pi^2}\left(Y_t^2-Y_b^2\right)
  s_{\beta}\,c_{\beta}\,\ln\frac{m_{H^{\pm}}^2}{M_{\text{EW}}^2}\,
  \Amp{tree}{}{\AtoB{h}{VV}}\,,\\[-.6ex]
  \Amp{mix}{}{\AtoB{A}{VV}} &\simeq 0\,,\\[-.6ex]
  \Amp{mix}{}{\AtoB{H^+}{W^+Z}} &\simeq
  \frac{3}{16\,\pi^2}\left(Y_t^2-Y_b^2\right)
  s_{\beta}\,c_{\beta}\,\ln\frac{m_{H^{\pm}}^2}{M_{\text{EW}}^2}\,
  \Amp{tree}{}{\AtoB{G^+}{W^+Z}}\,.
\end{align}
\end{subequations}

\begin{figure}[p!]
\centering
\includegraphics[width=\textwidth]{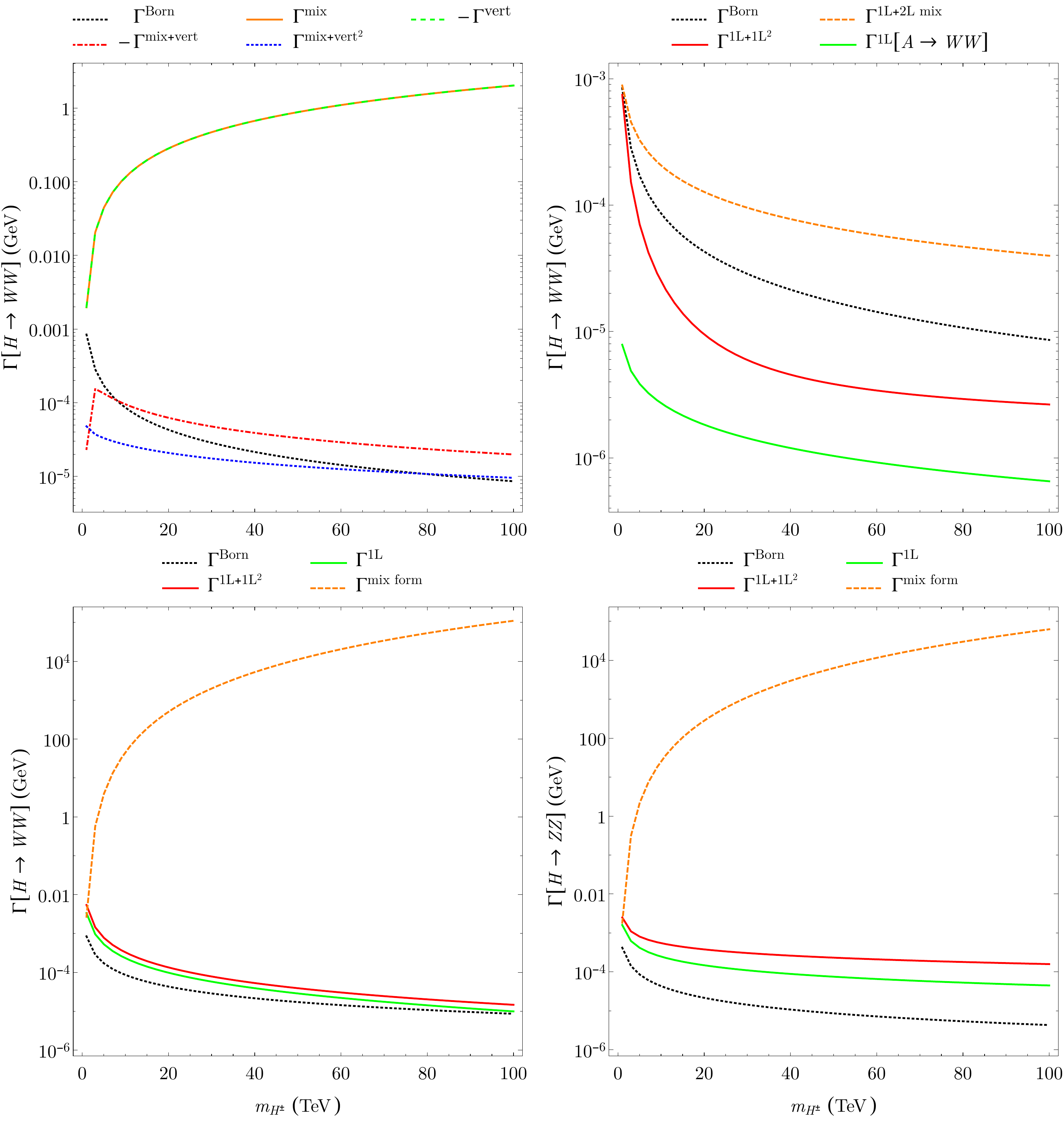}
\caption{Decay widths of heavy-doublet states into electroweak
gauge bosons are displayed for $t_{\beta}=10$ and varying
charged-Higgs mass (the SUSY sector is decoupled).
\newline{\em Upper Left}: Individual contributions of order
$\mathcal{O}{\left(Y_q^2\right)}$ to the decay of the heavy \CP-even
state; Born-level (dotted black), mixing-Born interference (orange),
vertex-born interference (absolute value, dashed green),
(mixing+vertex)-Born interference (absolute value, dot-dashed red),
1L$^2$ term (dotted blue).
\newline{\em Upper Right}: Varying definitions of the decay width with
$\mathcal{O}{\left(Y_q^2\right)}$ radiative corrections; Born-level
(dotted black), full one-loop (including 1L$^2$, red), including the
two-loop mixing effect of order
$\mathcal{O}\big(Y_{t,b}^2\,\alpha_s\big)$ (dashed orange). The green
curve corresponds to the purely radiative decay width of the \CP-odd
state.
\newline{\em Lower Left}: Full one loop corrections to the decay width
$H\to W^+W^-$; Born-level (dotted black), full one-loop (solid green)
obtained from a straightforward application of the LSZ reduction,
including also the 1L$^2$ term (solid red); the loop-corrected mixing
matrix is employed for the dashed orange curve instead.
\newline{\em Lower Right}: Similar to lower left but applied to the
channel \AtoB{H}{ZZ}.
\label{fig:hWW}}
\end{figure}

As expected, the contributions to the vertex in \refeqs{eq:hvvvert}
and to the mixing in \refeqs{eq:hvvmix} exactly cancel each other. On
the other hand, if mixing and vertex are not processed at the same
level, \EG\ due to the use of different parameters at the loop order,
then the cancellation is imperfect and large spurious effects, though
formally of higher order, develop. Such misleading effects are
exacerbated for heavy states as the fine cancellation between mixing
and vertex encompasses orders of magnitude. Unluckily, beyond the
higher orders that are inherent to the mixing formalism, explicit
effects of higher order, \EG\
$\mathcal{O}\big(Y_{t,b}^2\,\alpha_s\big)$, that are considered in the
mass calculation are also routinely included within the mixing in
the \CP-even sector. Such terms are going to cause an imbalance in the
cancellation between vertex and mixing as long as the vertex
corrections are not known to the same order. We illustrate this fact
below.

In the upper row of \fig{fig:hWW}, we present the decay width of the
heavy \CP-even Higgs into $W^+W^-$ for $t_{\beta}=10$, considering
only corrections of $\mathcal{O}{\left(Y_q^2\right)}$. Again, we scan
up to $m_{H^\pm}\sim100$\,TeV to set the calculation beyond doubt in
an $SU(2)_{\text{L}}$-conserving regime. The plot on the left-hand
side shows the magnitude of the individual contributions to the decay
width in the strict LSZ expansion. The tree-level prediction (dotted
black) is smaller by orders of magnitude than the mixing (orange) and
vertex (dashed green) contributions of one-loop order, which however
largely compensate one-another. As the sum of one-loop contributions
(dot-dashed red) is negative and larger in absolute value than the
tree level, the decay width at truncated one-loop order would be
negative at large Higgs masses. In this context, it is legitimate to
include the one-loop squared piece (dotted blue; here, the square of
the full gauge- and field-renormalization invariant amplitude of 1L
order is meant), which is expected to supersede the remaining
contributions of two-loop order. In the plot on the right-hand side,
we display the tree-level decay width (dotted black), and the width of
one-loop order including the 1L$^2$ term (solid red). In dashed
orange, the two-loop $\mathcal{O}\big(Y_{t,b}^2\,\alpha_s\big)$
corrections to the mixing have been included: the associated
prediction is in excess by a factor $10$ with respect to the one-loop
result (red): however, this enhancement is most likely not a genuine
effect, but an artifact caused by the non-cancellation of mixing and
vertex contributions of order
$\mathcal{O}\big(Y_{t,b}^2\,\alpha_s\big)$---since the corresponding
order has not been included in the vertex corrections. The purely
radiative $\Gamma[\AtoB{A}{W^+W^-}]$ (in green) is shown as a
reference. This comparison indicates that the inclusion of a partial
order $\mathcal{O}\big(Y_{t,b}^2\,\alpha_s\big)$ in the mixing in fact
worsens the quality of the prediction for the decay width by
introducing large symmetry-violating effects (though formally of
higher order). It thus appears as misleading to process vertex and
mixing contributions in a decoupled way and introduce partial higher
orders, which are liable to violate the symmetries.

However, the mixing formalism introduces further higher-order terms
due to the factorization and resummation of mixing effects in a
loop-corrected mixing matrix. These contributions are liable to blur
further the fine cancellation resulting from the symmetry
requirements. In the lower row of \fig{fig:hWW}, we no longer restrict
to the order $\mathcal{O}{\left(Y_q^2\right)}$, but include the full
electroweak and SUSY corrections---the SUSY spectrum is still at a
scale of $\simord100$\,TeV. We consider both the decay widths
for~\AtoB{H}{W^+W^-} (left) and~\AtoB{H}{ZZ} (right). The one-loop
corrections (green curves) can reach the magnitude of the Born-level
amplitudes, and the one-loop squared term may dominate the widths (red
curves). However, these decay widths obtained from the straightforward
application of the LSZ reduction all remain comparatively suppressed,
as a tribute to the electroweak symmetry. On the other hand, if vertex
and mixing contributions are not linearly added, but rather combined
via a loop-corrected mixing matrix, then the imperfect cancellation of
these contributions to the decay amplitudes generate pieces of
higher-order that break the electroweak symmetry strongly and come to
dominate the width at high masses (dashed orange curves). The dramatic
enhancement of these spurious effects is due to the kinematic
prefactor~$M_H^3/M_V^2$, that the decay amplitude needs to balance by
a careful scaling~$\mathcal{A}\propto M_V^2/M_H^2$, which is spoiled
by the separate processing/factorization of mixing contributions.
Again, this argues against an indiscriminate use of the mixing
formalism in physical transitions. Finally, we stress that, even
though the consistent combination of mixing and vertex contributions
leads to decay widths that are compatible with symmetries, the latter
still come with a sizable uncertainty for such rare processes: for,
instance, exchanging pole quark masses by QCD-running masses at the
scale of the decaying Higgs---a legitimate shift at the order
controlled in the calculation---in the scenario of \fig{fig:hWW}
typically leads to a reduction of the widths by a factor
of~$\mathcal{O}(10)$. This points at the necessity to include two-loop
contributions for a reliable assessment of these symmetry-violating
channels.

\medskip
The analysis of the one-loop radiative corrections in the decoupling
limit, in a limit where the $SU(2)_{\text{L}}$-symmetry should hold,
thus indicates that misleading large effects purely associated with
partial higher-order contributions may develop as a consequence of
considering vertex and mixing diagrams on a different footing. In
particular, the deliberate inclusion of higher-order effects in the
mixing without the corresponding terms in the vertex appears as an
unfruitful effort (as already suggested by the analysis of the gauge
dependence). This issue is self-evident in the decoupling limit, where
one may actually directly consider the model with unbroken
$SU(2)_{\text{L}}$ symmetry.\footnote{However, the
infrared behaviour
of \EG~exclusive decay widths in the unbroken electroweak description
shows that the electroweak scale still
matters in deriving the
properties of heavy Higgs states at the
radiative level.} However, it remains relevant for
\mbox{$m_{H^{\pm}}\sim1$}\,TeV, as purging the calculation from the
symmetry-violating artifacts of the formalism has a non-negligible
impact on the radiative corrections at the level \mbox{of observable
quantities.}

\vspace{-1ex}
\tocsection{Conclusions}

In this paper, we have analyzed how terms of higher order introduced
in the calculation of the radiative corrections to observables in
extended SUSY Higgs sectors could lead to spurious effects in view of
the symmetries. As noticeable numerical variations accompany these
artifacts, it appears necessary to consider them seriously in the
uncertainty estimates. On the other hand, the associated behaviour is
unphysical and needlessly burdens the error budget; therefore, we
regard it as meaningful to attempt and avoid such symmetry-violating
pieces of higher order.

We first discussed gauge dependence in Higgs-mass determinations and
decays and explained how setting the arguments of the loop functions
away from the tree-level mass values generates pieces that depend both
on the gauge-fixing parameter and the field renormalization. We then
presented two possible strategies avoiding such undesirable effects:
the first one consists in systematically expanding and truncating the
amplitudes at the relevant order controlled in the calculation; the
alternative one extends the SUSY model by a more flexible structure
where the Higgs potential can adjust to the values of the
loop-corrected masses. While the former method is more conventional
and straightforward to implement, the latter one can be interesting in
order to assess uncertainties in a gauge-conserving context, although
redundancies in the definition of the Higgs potential limit its
efficiency. We also discussed how to define a mixing matrix in a
gauge-invariant way in the case of near-degenerate states, a recipe
that fails in the non-degenerate case, where gauge invariance is most
efficiently enforced by a strict application of the LSZ reduction. In
particular, we saw that gauge invariance required a careful
combination of vertex and mixing contributions to the decay
amplitudes, so that a separation of both, \EG~through the definition
of a mixing matrix at radiative order, can be source of inconsistent
behaviours.

\needspace{3ex}
Then, we focussed on the decoupling limit of the MSSM, where the
$SU(2)_{\text{L}}$-symmetry still controls the dominant properties of
the heavy-doublet states. We illustrated this analytically, with
expressions for the radiative corrections of Yukawa type, as well as
numerically, with a full calculation of one-loop order. These
arguments allowed us to spot several issues, first in the calculation
of the two-loop corrections to the Higgs masses of
$\mathcal{O}(Y_q^4)$, then in the implementation of Higgs decays when
mixing and vertex corrections are included at different orders or via
a mixing matrix defined at the radiative level. The corresponding
inconsistencies become large, admittedly because of the choice of a
heavy-doublet spectrum, but still point at shortcomings in the general
implementation of these observables in all regimes. Observables
measuring $SU(2)_{\text{L}}$-breaking effects, such as mass-splittings
among $SU(2)_{\text{L}}$-partners or heavy-Higgs decays into
electroweak gauge bosons or lighter Higgs states, are particularly
sensitive to the introduction of spurious
$SU(2)_{\text{L}}$-symmetry-violating terms of higher order, so that a
proper control on the symmetries appears as imperative for a
meaningful study of such channels.

As we took the restoration of symmetries as our guiding principle, it
proved more convenient to work with electroweakly-charged
states. However, what we learnt with doublet states can be extended
straightforwardly to more exotic Higgs spectra, and even to other
fields that are not necessarily renormalized on-shell (such as
sfermions or electroweakinos;
see \EG\ \citeres{Heinemeyer:2014yya,Heinemeyer:2015pfa}). Furthermore,
several additional issues that are not constrained by symmetries can
appear in connection with a careless use of mixing formalisms at the
radiative order. Double-counting of mixing corrections can thus emerge
from \EG~hybridizing the mixing formalism with that of effective
couplings for integrated-out SUSY sectors (for the latter formalism,
see \EG\ \citere{Noth:2010jy} and references therein): indeed, SUSY
corrections on the external Higgs line are then potentially
double-counted.

As a concluding word, we believe that, while it is of course still
possible to employ a mixing formalism in calculations of radiative
corrections to the Higgs sector, corresponding results should be
critically analyzed in order to verify whether the effects that they
produce are genuine or just the outcome of symmetry-violating
artifacts.

\vspace{-1ex}
\section*{\tocref{Acknowledgments}}

We thank H.~Bahl, M.\,D.~Goodsell and T.~Hahn for useful
discussions. S.~P. acknowledges support by the BMBF Grant
No.\,05H18PACC2.

\appendix
\vspace{-1ex}
\tocsection{Renormalization of the Higgs potential}\label{ap:renTHDM}

\tocsubsection{Parametrization}

The Higgs potential of the~THDM can be expressed in the following way,
\begin{align}\label{eq:THDMPot}
  \begin{split}
  \mathcal{V}_{\mathrm{THDM}} &=
  m_{H_d}^2\,\lvert H_d\rvert^2 + m_{H_u}^2\,\lvert H_u\rvert^2
  + \left[m^2_{12}\,e^{\imath\,\varphi_{12}}\,H_u\cdot H_d + \mathrm{h.\,c.}\right]\\
  &\quad+ \tfrac{1}{2}\,\lambda_1\,\lvert H_d\rvert^4
  + \tfrac{1}{2}\,\lambda_2\,\lvert H_u\rvert^4
  + \lambda_3\,\lvert H_u\rvert^2\,\lvert H_d\rvert^2
  + \lambda_4\,\lvert H_u\cdot H_d\rvert^2\\
  &\quad+ \left[\tfrac{1}{2}\,\lambda_5\,e^{\imath\,\varphi_5}
  \left(H_u\cdot H_d\right)^2
    + \lambda_6\,e^{\imath\,\varphi_6}\lvert H_d\rvert^2\,H_u\cdot H_d
    + \lambda_7\,e^{\imath\,\varphi_7}\lvert H_u\rvert^2\,H_u\cdot H_d
    + \mathrm{h.\,c.}\right],
  \end{split}
\end{align}
with the real, dimensionless couplings~$\lambda_i$ and the
phases~$\varphi_i$. The Higgs potential of the~MSSM emerges from
\refeq{eq:THDMPot} for specific choices of~$\lambda_i$:
\begin{align}\label{eq:lambdaMSSM}
  \lambda_1^{\text{\tiny MSSM}} &= \frac{M_Z^2}{2\,v^2} =
  \lambda_2^{\text{\tiny MSSM}} =
  -\left(\lambda_3^{\text{\tiny MSSM}} + \lambda_4^{\text{\tiny MSSM}}\right)\,,
  & \lambda_4^{\text{\tiny MSSM}} &= -\frac{M_W^2}{v^2}\,,
  & \lambda_{5,6,7}^{\text{\tiny MSSM}} &= 0\,.
\end{align}
To simplify the contact of the~THDM with the~MSSM, we
reparametrize~\mbox{$\lambda_{i} = \lambda_{i}^{\text{\tiny
MSSM}}+\ell_{i}$}. We also define $\ell_k^r\equiv\ell_k\cos\varphi_k$
and $\ell_k^i\equiv\ell_k\sin\varphi_k$ for $k\in\{5,6,7\}$.

We write the Higgs doublets in terms of v.e.v.-s, \CP-even, \CP-odd
and charged components:
\begin{align}
  H_d &= e^{\imath\,\varphi_d}\begin{pmatrix}
    v\,c_{\beta}+\frac{1}{\sqrt{2}}\left(h_d^0+\imath\,a_d^0\right)\\ H_d^-
  \end{pmatrix},\qquad
  H_u = e^{\imath\,\varphi_u}\begin{pmatrix}
    H_u^+\\ v\,s_{\beta}+\frac{1}{\sqrt{2}}\left(h_u^0+\imath\,a_u^0\right)
\end{pmatrix}.
\end{align}
The phases $\varphi_{d,u}$ can be absorbed in a re-definition of those appearing in the parameters of \refeq{eq:THDMPot} and can thus be discarded.

The tadpole equations can be exploited to express three parameters in
terms of the electroweak vacuum expectation values. Commonly, the
dimensionful parameters are substituted as
\begin{subequations}
\begin{align}
  \begin{split}
  m_{H_d}^2 &= \frac{T_d}{\sqrt{2}\,v\,c_{\beta}}
  + m^2_{12}\,\cos\varphi_{12}\,t_{\beta} - \frac{M_Z^2}{2}\,c_{2\beta}\\
  &\quad- v^2\left[
  \left(\ell_1 - 3\,\ell^r_6\,t_{\beta}\right) c^2_{\beta}
  + \left(\ell_3+\ell_4+\ell^r_5 - \ell^r_7\,t_{\beta}\right)
    s^2_{\beta}\right],
  \end{split}\\
  \begin{split}
  m_{H_u}^2 &= \frac{T_u}{\sqrt{2}\,v\,s_{\beta}}
  + m^2_{12}\,\cos\varphi_{12}\,t^{-1}_{\beta} + \frac{M_Z^2}{2}\,c_{2\beta}\\
  &\quad- v^2\left[\left(\ell_2 - 3\,\ell^r_7\,t^{-1}_{\beta}\right)
  s^2_{\beta}
  + \left(\ell_3+\ell_4+\ell^r_5 - \ell^r_6\,t^{-1}_{\beta}\right)
  c^2_{\beta}\right],
  \end{split}\\
  m^2_{12}\,\sin\varphi_{12} &= \frac{T_a}{\sqrt{2}\,v}
  + \ell^i_5\,v^2\,s_{\beta}\,c_{\beta}
  - \ell^i_6\,v^2\,c^2_{\beta}
  - \ell^i_7\,v^2\,s^2_{\beta}
\end{align}
\end{subequations}
with the tadpole parameters~$T_{d,u,a}$, associated to the
fields~$h_d^0$, $h_u^0$
and~\mbox{$a^0=s_{\beta}\,a_d^0+c_{\beta}\,a_u^0$}, vanishing at the
tree level but kept for the derivation of the counterterms.

The tree-level mass matrix for the charged-Higgs sector reads as
follows in the $(H_d^+,H_u^+)$ basis:
\begin{subequations}
\begin{align}
 \mathcal{M}^2_{\mathrm{C}} &=
  \mathscr{M}_{H^{\pm}}^2
  \begin{pmatrix} s_{\beta}^2 & s_{\beta}\,c_{\beta}\\
    s_{\beta}\,c_{\beta}&c_{\beta}^2\end{pmatrix}
  + \frac{1}{\sqrt{2}\,v}
  \begin{pmatrix} T_d/c_{\beta} & -\imath\,T_a\\
    \imath\,T_a & T_u/s_{\beta}\end{pmatrix}\,,\\
  \label{eq:Hpmass}\mathscr{M}_{H^{\pm}}^2 &\equiv
  \frac{m^2_{12}\,\cos\varphi_{12}}{s_{\beta}\,c_{\beta}}
    + \left[M_W^2 - (\ell_4+\ell^r_5)\,v^2\right]
    +\ell^r_6\,v^2\,t^{-1}_{\beta}+\ell^r_7\,v^2\,t_{\beta}\,.
\end{align}
\end{subequations}
\refeq{eq:Hpmass} can be exploited to
substitute~$\mathscr{M}^2_{H^{\pm}}$ in replacement of the
parameter~$m^2_{12}\,\cos\varphi_{12}$. The diagonalization of this
matrix by a rotation of angle~$\beta_c$ yields the eigenstates and
eigenvalues
\begin{subequations}
\begin{align}
  \begin{pmatrix} G^{\pm}\\ H^{\pm}\end{pmatrix} &\equiv
  \begin{pmatrix} c_{\beta_c} & -s_{\beta_c}\\
    s_{\beta_c} & c_{\beta_c}\end{pmatrix}
  \begin{pmatrix} H_d^{\pm}\\ H_u^{\pm}\end{pmatrix},\\
  m^2_{H^{\pm}} &=
  \mathscr{M}^2_{H^{\pm}}\, c_{\beta-\beta_c}^2
  + \frac{1}{\sqrt{2}\,v} \left[T_u\,\frac{c^2_{\beta_c}}{s_{\beta}}
    + T_d\,\frac{s^2_{\beta_c}}{c_{\beta}}\right],\\
  m^2_{G^{\pm}} &= \frac{1}{\sqrt{2}\,v} \left[T_u\,\frac{s^2_{\beta_c}}{s_{\beta}}
  + T_d\,\frac{c^2_{\beta_c}}{c_{\beta}}\right].
\end{align}
\end{subequations}
After application of the minimization conditions~$T_{d,u,a}=0$, one
finds~\mbox{$m^2_{H^{\pm}}=\mathscr{M}^2_{H^{\pm}}$}, \mbox{$m^2_{G^{\pm}}=0$}
and \mbox{$\beta_c=\beta$} at the tree level.

\needspace{3ex}
The mass matrix for the neutral Higgs sector can be decomposed into
blocks. For the \CP-even sector in the basis~$(h_d^0,h_u^0)$ one has
\begin{subequations}
\begin{align}
  \left.\mathcal{M}^2_{\mathrm{E}}\right|_{11} &=
  m^2_{12}\,\cos\varphi_{12}\,t_{\beta}
  + \left(M_Z^2 + 2\,\ell_1\,v^2\right) c_{\beta}^2
  - 3\,\ell^r_6\,v^2\,s_{\beta}\,c_{\beta}
  + \ell^r_7\,v^2\,\frac{s_{\beta}^3}{c_{\beta}}
  + \frac{T_d}{\sqrt{2}\,v\,c_{\beta}}\,,\\
  \left.\mathcal{M}^2_{\mathrm{E}}\right|_{12} &=
  -m^2_{12}\,\cos\varphi_{12}
  + \left[2\left(\ell_3 + \ell_4 + \ell^r_5\right) v^2
    - M_Z^2\right] s_{\beta}\,c_{\beta}
  - 3 \left(\ell^r_6\,c^2_{\beta}
    + \ell^r_7\,s^2_{\beta}\right) v^2\,,\\
  \left.\mathcal{M}^2_{\mathrm{E}}\right|_{22} &=
  m^2_{12}\,\cos\varphi_{12}\,t^{-1}_{\beta}
  + \left(M_Z^2 + 2\,\ell_2\,v^2\right) s_{\beta}^2
  + \ell^r_6\,v^2\,\frac{c_{\beta}^3}{s_{\beta}}
  - 3\,\ell^r_7\,v^2\,s_{\beta}\,c_{\beta}
  + \frac{T_u}{\sqrt{2}\,v\,s_{\beta}}\,.
\end{align}
\end{subequations}
For the \CP-odd sector in the basis~$(a_d^0,a_u^0)$ the block matrix
reads
\begin{subequations}
\begin{align}
  \mathcal{M}^2_{\mathrm{O}} &=
  \mathscr{M}_A^2\,
  \begin{pmatrix} s_{\beta}^2 & s_{\beta}\,c_{\beta}\\
    s_{\beta}\,c_{\beta} & c_{\beta}^2\end{pmatrix}
    + \frac{1}{\sqrt{2}\,v}\begin{pmatrix} T_d/c_{\beta} & 0\\
      0 & T_u/s_{\beta}
  \end{pmatrix},\\
  \mathscr{M}_A^2 &\equiv \frac{m^2_{12}\,\cos\varphi_{12}}{s_{\beta}\,c_{\beta}}
  - 2\,\ell^r_5\,v^2
  + \ell^r_6\,v^2\,t^{-1}_{\beta}
  + \ell^r_7\,v^2\,t_{\beta}\,.
\end{align}
\end{subequations}
For the off-diagonal block matrix one finds
\begin{align}
  \mathcal{M}^2_{\mathrm{EO}} &=-\ell^i_5\,v^2\begin{pmatrix}
  s_{\beta}^2 & s_{\beta}\,c_{\beta}\\
  s_{\beta}\,c_{\beta} & c_{\beta}^2
  \end{pmatrix}+2\,v^2\begin{pmatrix}
  \ell^i_6\,s_{\beta}c_{\beta} & \ell^i_6\,c_{\beta}^2\\
  \ell^i_7\,s_{\beta}^2 & \ell^i_7\,s_{\beta}c_{\beta}
  \end{pmatrix}+
  \frac{1}{\sqrt{2}\,v}\begin{pmatrix} 0 & T_a\\
    T_a & 0
  \end{pmatrix}.
\end{align}

After application of the minimization conditions, the Goldstone boson
is obtained as the linear
combination~\mbox{$G^0=c_{\beta}\,a_d^0-s_{\beta}\,a_u^0$}. The other
(three) tree-level mass-eigenstates may be
written~\mbox{$h_i^0=(\mathbf{U}_n)_{id}\,h_d^0 +
  (\mathbf{U}_n)_{iu}\,h_u^0+(\mathbf{U}_n)_{ia}\,(s_{\beta}\,a_d^0+c_{\beta}\,a_u^0)$},
where~$\mathbf{U}_n$ is a unitary matrix. In the absence of
\CP-violation, the pseudoscalar sector can be diagonalized separately
with a rotation of angle~$\beta_0$, which is found to coincide
with~$\beta$ after application of the minimization conditions. The
pseudoscalar state~\mbox{$A^0\equiv
s_{\beta_0}\,a_d^0+c_{\beta_0}\,a_u^0$} takes on the mass
\begin{align}
m^2_{A^0}=\mathscr{M}^2_A\,\,c^2_{\beta-\beta_0}
+ \frac{1}{\sqrt{2}\,v}\left[T_u\,\frac{c^2_{\beta_0}}{s_{\beta}}
+ T_d\,\frac{s^2_{\beta_0}}{c_{\beta}}\right].
\end{align}

\needspace{7ex}
The Higgs couplings appearing in loop amplitudes can be summarized as
follows:
\begin{itemize}
\item the neutral symmetric triple-Higgs couplings
\begin{subequations}
\begin{align}
  \Gamma_{h_ih_jh_k} &= \sum\limits_{a,b,c=1}^4\left(\mathbf{U}_n\right)_{ia}
  \left(\mathbf{U}_n\right)_{jb}\left(\mathbf{U}_n\right)_{kc}
  \gamma_{abc}\,\sqrt{2}\,v\,,\\
  \gamma_{\phi_dxy} &= \left(\begin{smallmatrix*}[r]
  3\left(c_\beta\,\lambda_1-s_\beta\,\lambda_6^r\right),&
  s_\beta\left(\lambda_3+\lambda_4+\lambda_5^r\right) - 3\,c_\beta\,\lambda_6^r\,,&
  -s_\beta\,\lambda_6^i\,,&
  s_\beta\,\lambda_5^i - 3\,c_\beta\,\lambda_6^i\\
  \hfill\cdot\hfill&
  c_\beta\left(\lambda_3+\lambda_4+\lambda_5^r\right) - 3\,s_\beta\,\lambda_7^r\,,&
  s_\beta\,\lambda_5^i - c_\beta\,\lambda_6^i\,,&
  c_\beta\,\lambda_5^i - s_\beta\,\lambda_7^i\\
  \hfill\cdot\hfill&\hfill\cdot\hfill&
  c_\beta\,\lambda_1-s_\beta\,\lambda_6^r\,,&
  c_\beta\,\lambda_6^r-s_\beta\,\lambda_5^r\\
  \hfill\cdot\hfill&\hfill\cdot\hfill&\hfill\cdot\hfill&
  c_\beta\left(\lambda_3+\lambda_4-\lambda_5^r\right) - s_\beta\,\lambda_7^r
  \end{smallmatrix*}\right)_{\begin{smallmatrix*}[l]xy\\x,y\in\{\phi_d,\phi_u,\chi_d,\chi_u\}\end{smallmatrix*}}\hspace{-4em},\\
  \gamma_{\phi_uxy} &= \left(\begin{smallmatrix*}[r]
  3\left(s_\beta\,\lambda_2-c_\beta\,\lambda_7^r\right),&
  c_\beta\,\lambda_5^i - 3\,s_\beta\,\lambda_7^i\,,&
  -\,c_\beta\,\lambda_7^i\\
  \hfill\cdot\hfill&
  s_\beta\left(\lambda_3+\lambda_4-\lambda_5^r\right) - c_\beta\,\lambda_6^r\,,&
  s_\beta\,\lambda_7^r - c_\beta\,\lambda_5^r\\
  \hfill\cdot\hfill&\hfill\cdot\hfill&
  s_\beta\,\lambda_2 - c_\beta\,\lambda_7^r
  \end{smallmatrix*}\right)_{\begin{smallmatrix*}[l]xy\\x,y\in\{\phi_u,\chi_d,\chi_u\}\end{smallmatrix*}}\hspace{-3em},\\
  \gamma_{\chi_dxy} &= \left(\begin{smallmatrix*}[r]
  -3\,s_\beta\,\lambda_6^i\,,&
  -s_\beta\,\lambda_5^i-c_\beta\,\lambda_6^i\\
  \hfill\cdot\hfill& -c_\beta\,\lambda_5^i-s_\beta\,\lambda_7^i
  \end{smallmatrix*}\right)_{\begin{smallmatrix*}[l]xy\\x,y\in\{\chi_d,\chi_u\}\end{smallmatrix*}},\\
  \gamma_{\chi_u\chi_u\chi_u} &= -c_\beta\,\lambda_7^i\,.
\end{align}
\end{subequations}
\item the neutral symmetric quadruple-Higgs couplings
\begin{subequations}
\begin{align}
  \Gamma_{h_ih_jh_kh_l} &= \sum\limits_{a,b,c,d=1}^4\left(\mathbf{U}_n\right)_{ia}
  \left(\mathbf{U}_n\right)_{jb}\left(\mathbf{U}_n\right)_{kc}
  \left(\mathbf{U}_n\right)_{ld} \gamma_{abcd}\,,\span\span\\
  \gamma_{\phi_d\phi_dxy} &= \left(\begin{smallmatrix*}[r]
  3\,\lambda_1\,,& -3\,\lambda_6^r,& 0\,, & -3\,\lambda_6^i\\
  \hfill\cdot\hfill&
  \lambda_3+\lambda_4+\lambda_5^r\,, & -\lambda_6^i\,, & \lambda_5^i\\
  \hfill\cdot\hfill&\hfill\cdot\hfill& \lambda_1\,,& \lambda_6^r\\
  \hfill\cdot\hfill&\hfill\cdot\hfill&\hfill\cdot\hfill&
  \lambda_3+\lambda_4-\lambda_5^r
  \end{smallmatrix*}\right)_{\begin{smallmatrix*}[l]xy\\x,y\in\{\phi_d,\phi_u,\chi_d,\chi_u\}\end{smallmatrix*}}\hspace{-4em},&
  \gamma_{\phi_d\chi_dxy} &= \left(\begin{smallmatrix*}[r]
  0\,,& -\lambda_6^i\\
  \hfill\cdot\hfill& -\lambda_5^i
  \end{smallmatrix*}\right)_{\begin{smallmatrix*}[l]xy\\x,y\in\{\chi_d,\chi_u\}\end{smallmatrix*}}\hspace{-2em},\\
  \gamma_{\phi_d\phi_uxy} &= \left(\begin{smallmatrix*}[r]
  -3\,\lambda_7^r,& \lambda_5^i\,,& -\lambda_7^i\\
  \hfill\cdot\hfill& -\lambda_6^r\,,& -\lambda_5^r\\
  \hfill\cdot\hfill&\hfill\cdot\hfill& -\lambda_7^r
  \end{smallmatrix*}\right)_{\begin{smallmatrix*}[l]xy\\x,y\in\{\phi_u,\chi_d,\chi_u\}\end{smallmatrix*}}\hspace{-3em},&
  \gamma_{\phi_d\chi_u\chi_u\chi_u} &= -3\,\lambda_7^i\,,\\
  \gamma_{\phi_u\phi_uxy} &= \left(\begin{smallmatrix*}[r]
  3\,\lambda_2,& -3\,\lambda_7^i\,,& 0\\
  \hfill\cdot\hfill& \lambda_3 + \lambda_4 - \lambda_5^r\,,& \lambda_7^r\\
  \hfill\cdot\hfill&\hfill\cdot\hfill& \lambda_2
  \end{smallmatrix*}\right)_{\begin{smallmatrix*}[l]xy\\x,y\in\{\phi_u,\chi_d,\chi_u\}\end{smallmatrix*}}\hspace{-3em},&
  \gamma_{\phi_u\chi_dxy} &= \left(\begin{smallmatrix*}[r]
  -3\,\lambda_6^i\,, & i\,\lambda_5^i\\
  \hfill\cdot\hfill& -\lambda_7^i
  \end{smallmatrix*}\right)_{\begin{smallmatrix*}[l]xy\\x,y\in\{\chi_d,\chi_u\}\end{smallmatrix*}}\hspace{-2em},\\
  \gamma_{\phi_u\chi_u\chi_u\chi_u} &= 0\,,\\
  \gamma_{\chi_d\chi_dxy} &= \left(\begin{smallmatrix*}[r]
  3\,\lambda_1\,, & 3\,\lambda_6^r\\
  \hfill\cdot\hfill& \lambda_3 + \lambda_4 + \lambda_5^r
  \end{smallmatrix*}\right)_{\begin{smallmatrix*}[l]xy\\x,y\in\{\chi_d,\chi_u\}\end{smallmatrix*}}\hspace{-2em},&
  \gamma_{\chi_u\chi_u\chi_u\chi_u} &= 3\,\lambda_2\,.
\end{align}
\end{subequations}
\item the Hermitian couplings of one neutral and two charged fields
\begin{subequations}
\begin{align}
  \Gamma_{h_ih^-_jh^+_k} &= \sum\limits_{a=1}^4\sum\limits_{b,c=1}^2
  \left(\mathbf{U}_n\right)_{ia}\left(\mathbf{U}_\pm\right)_{jb}
  \left(\mathbf{U}_\pm\right)_{kc}\,
  \gamma^\pm_{abc}\,\sqrt{2}\,v\,,\\
  \gamma^\pm_{\phi_dxy} &= \left(\begin{smallmatrix*}[r]
  c_\beta\,\lambda_1 - s_\beta\,\lambda_6^r\,,&
  -\frac{1}{2}\,s_\beta\left(\lambda_4 + \lambda_5\right)
  + c_\beta\,\lambda_6\\
  \hfill\cdot\hfill& c_\beta\,\lambda_3 - s_\beta\,\lambda_7^r
  \end{smallmatrix*}\right)_{xy},&
  \gamma^\pm_{\chi_dxy} &= \left(\begin{smallmatrix*}[r]
  -s_\beta\,\lambda_6^i\,,&
  \frac{1}{2}\,s_\beta\left(\lambda_4 - \lambda_5\right)\\
  \hfill\cdot\hfill& -s_\beta\,\lambda_7^i
  \end{smallmatrix*}\right)_{xy},\\
  \gamma^\pm_{\phi_uxy} &= \left(\begin{smallmatrix*}[r]
  s_\beta\,\lambda_3 - c_\beta\,\lambda_6^r\,,&
  -\frac{1}{2}\,c_\beta\left(\lambda_4 + \lambda_5\right)
  + s_\beta\,\lambda_7\\
  \hfill\cdot\hfill& s_\beta\,\lambda_2 - c_\beta\,\lambda_7^r
  \end{smallmatrix*}\right)_{xy},&
  \gamma^\pm_{\chi_uxy} &= \left(\begin{smallmatrix*}[r]
  -c_\beta\,\lambda_6^i\,,&
  \frac{1}{2}\,c_\beta\left(\lambda_4 - \lambda_5\right)\\
  \hfill\cdot\hfill& -c_\beta\,\lambda_7^i
  \end{smallmatrix*}\right)_{xy}
\end{align}
\end{subequations}
with $(x,y)\in\{\phi^-_d,\phi^-_u\}\otimes\{\phi^+_d,\phi^+_u\}$.
\needspace{3ex}
\item the couplings of two neutral and two charged fields
(symmetric in the neutral and Hermitian in the charged fields)
\begin{subequations}
\begin{align}
  \Gamma_{h_ih_jh^-_kh^+_l} &= \sum\limits_{a,b=1}^4\sum\limits_{c,d=1}^2
  \left(\mathbf{U}_n\right)_{ia}\left(\mathbf{U}_n\right)_{jb}
  \left(\mathbf{U}_\pm\right)_{kc}\left(\mathbf{U}_\pm\right)_{ld}\,
  \gamma^\pm_{abcd}\,,\span\span\\
  \gamma^\pm_{\phi_d\phi_dxy} &= \gamma^\pm_{\chi_d\chi_dxy} =
  \left(\begin{smallmatrix*}[r]
  \lambda_1\,,& \lambda_6\\
  \hfill\cdot\hfill& \lambda_3
  \end{smallmatrix*}\right)_{xy},&
  \gamma^\pm_{\phi_d\phi_uxy} &= -\gamma^\pm_{\chi_d\chi_uxy} =
  \left(\begin{smallmatrix*}[r]
  -\lambda_6^r\,,& -\frac{1}{2}\left(\lambda_4+\lambda_5\right)\\
  \hfill\cdot\hfill& -\lambda_7^r
  \end{smallmatrix*}\right)_{xy},\\
  \gamma^\pm_{\phi_u\phi_uxy} &= \gamma^\pm_{\chi_u\chi_uxy} =
  \left(\begin{smallmatrix*}[r]
  \lambda_3\,,& \lambda_7\\
  \hfill\cdot\hfill& \lambda_2
  \end{smallmatrix*}\right)_{xy},&
  \gamma^\pm_{\phi_d\chi_uxy} &= \gamma^\pm_{\phi_u\chi_dxy} =
  \left(\begin{smallmatrix*}[r]
  \lambda_6^i\,,& i\,\frac{1}{2}\left(\lambda_4-\lambda_5\right)\\
  \hfill\cdot\hfill& \lambda_7^i
  \end{smallmatrix*}\right)_{xy},\\
  \gamma^\pm_{\phi_d\chi_dxy} &= \gamma^\pm_{\phi_u\chi_uxy} = \mathbf{0}\,,\quad
  \text{with } (x,y)\in\{\phi^-_d,\phi^-_u\}\otimes\{\phi^+_d,\phi^+_u\}\,.
  \span\span
\end{align}
\end{subequations}
\item the charged quadruple-Higgs couplings (symmetric in same-charged
and Hermitian in opposite-charged fields)
\begin{subequations}
\begin{align}
  \Gamma_{h_i^-h_j^+h^-_kh^+_l} &= \sum\limits_{a,b,c,d=1}^2
  \left(\mathbf{U}_\pm\right)_{ia}\,\left(\mathbf{U}_\pm\right)_{jb}\,
  \left(\mathbf{U}_\pm\right)_{kc}\,\left(\mathbf{U}_\pm\right)_{ld}\,
  \gamma^{\pm\pm}_{abcd}\,,\\
  \gamma^{\pm\pm}_{\phi^-_d\phi^+_dxy} &=
  \left(\begin{smallmatrix*}[r]
  2\,\lambda_1\,,& 2\,\lambda_6\\
  \hfill\cdot\hfill& \lambda_3 + \lambda_4
  \end{smallmatrix*}\right)_{xy},\qquad
  \gamma^{\pm\pm}_{\phi^-_u\phi^+_uxy} =
  \left(\begin{smallmatrix*}[r]
  \lambda_3 + \lambda_4\,,& 2\,\lambda_7\\
  \hfill\cdot\hfill& 2\,\lambda_2
  \end{smallmatrix*}\right)_{xy},\\
  \gamma^{\pm\pm}_{\phi^-_d\phi^+_u\phi^-_d\phi^+_u} &=
  \left(\gamma^{\pm\pm}_{\phi^-_u\phi^+_d\phi^-_u\phi^+_d}\right)^* = 2\,\lambda_5\,,\quad
  \text{with } (x,y)\in\{\phi^-_d,\phi^-_u\}\otimes\{\phi^+_d,\phi^+_u\}\,.
\end{align}
\end{subequations}
\end{itemize}

\tocsubsection{Renormalization conditions}

Below, we state the renormalization conditions that we impose on the
free parameters at the one-loop order:
\begin{itemize}
\item tadpoles: the one-point functions are requested to vanish at all
  orders. Thus, expanding the
  parameters~\mbox{$T_{d,u,a}=T^{(0)}_{d,u,a}+\delta T_{d,u,a}$} and
  using the tree-level minimization conditions~$T^{(0)}_{d,u,a}=0$,
  one identifies~$\delta T_{d,u,a}=-\mathcal{T}_{d,u,a}$
  where~$\mathcal{T}_{d,u,a}$ denotes the one-loop tadpole diagrams.
\item charged Higgs mass: it is renormalized on-shell, 
\begin{subequations}
\begin{align}
  \delta m^2_{H^{\pm}} &\equiv
  \frac{\delta (m^2_{12}\,\cos\varphi_{12})}{s_{\beta}\,c_{\beta}}
  + \delta M^2_{W}  - \left[m^2_{12}\,\cos\varphi_{12}\,\frac{c_{2\beta}}{s^2_{\beta}}+\ell^r_6\,v^2\,t^{-2}_{\beta}-\ell^r_7\,v^2\right]\delta t_{\beta}
 \notag\\
  &\quad+ \frac{1}{\sqrt{2}\,v} \left[\frac{c^2_{\beta}}{s_{\beta}}\,\delta T_u
  + \frac{s^2_{\beta}}{c_{\beta}}\,\delta T_d\right]
  - \delta(\ell_4\,v^2) - \delta(\ell^r_5\,v^2)
  + \delta(\ell^r_6\,v^2)\,t^{-1}_{\beta} + \delta(\ell^r_7\,v^2)\,t_{\beta}\\
    &= \Sigma_{H^+H^-}{\left(m_{H^{\pm}}^2\right)}\,,
\end{align}
\end{subequations}
thus, also~$\delta (m^2_{12}\,\cos\varphi_{12})$ is fixed.
\item weak gauge-boson masses: the $W$- and $Z$-boson masses are
  also renormalized on-shell, so that the associated counterterms
  cancel out with the transverse $W$ and $Z$~self-energies evaluated
  at~\mbox{$p^2=M_V^2$}, \IE\ \mbox{$\delta
  M^2_{VV}=\Sigma^{\mathrm{T}}_{VV}(M_V^2)$} for~\mbox{$V=W,\,Z$}.
\item the angle~$\beta$: $t_{\beta}$ is renormalized in the
  $\overline{\text{DR}}$-scheme,
\begin{align}
  \delta t_{\beta} &= \frac{t_{\beta}}{2}\left[
    \frac{d\Sigma_{h_dh_d}}{dp^2}{\left(p^2\right)}
    - \frac{d\Sigma_{h_uh_u}}{dp^2}{\left(p^2\right)}\right]_{\text{UV}}\,.
\end{align}
The associated contributions are purely of Yukawa type,
see \citere{Sperling:2013eva} for details.
\item the THDM shifts~$\delta(\ell_i\,v^2)$, for~$i=1,\dots,7$, are as
  yet undetermined. In the~MSSM, they are set equal to~$0$, since they
  are not needed to achieve renormalizability (and could spoil the
  SUSY relations for non-shifted momenta).
\end{itemize}

\tocsubsection{Counterterms in the neutral Higgs sector}

The renormalized self-energies (ignoring field renormalization for
now) for the neutral Higgs bosons can be expressed in the following
way (in the gauge eigenbasis):
\begin{subequations}\label{eq:masscounterterms}\allowdisplaybreaks
\begin{align}
\mathbf{\hat{\Sigma}}_{\mathrm{E}}{\left(p^2\right)} &=
\mathbf{\Sigma}_{\mathrm{E}}{\left(p^2\right)}
- \delta \mathcal{M}_{\mathrm{E}}^2\,,\quad
\mathbf{\hat{\Sigma}}_{A\mathrm{E}}{\left(p^2\right)} =
\mathbf{\Sigma}_{A\mathrm{E}}{\left(p^2\right)}
- \delta \mathcal{M}_{A\mathrm{E}}^2\,,\quad
\hat{\Sigma}_{AA}{\left(p^2\right)} =
\Sigma_{AA}{\left(p^2\right)}
- \delta m^2_{A}\\
\delta m^2_{A} &= \mathrm{C}_A + \delta m^2_{H^{\pm}} - \delta M^2_{W}\,,\qquad
\mathrm{C}_A\equiv \delta{\left(\ell_4\, v^2\right)}-\delta{\left(\ell^r_5\, v^2\right)}.\\
\begin{split}
\delta \mathcal{M}_{\mathrm{E}}^2 &\equiv \mathbf{C}_{\mathrm{E}}
+ \left(\delta m_{H^\pm}^2 - \delta M_W^2\right)
\begin{pmatrix} s_\beta^2 & -s_\beta\,c_\beta\\
-s_\beta\,c_\beta & c_\beta^2\end{pmatrix}
+ \delta M_Z^2
\begin{pmatrix} c_\beta^2 & -s_\beta\,c_\beta\\
-s_\beta\,c_\beta & s_\beta^2\end{pmatrix}\\
&\quad+ \frac{\delta T_{d}}{\sqrt{2}\,v\,c_\beta}
\begin{pmatrix} 1 - s_\beta^4 & s_\beta^3\,c_\beta\\
s_\beta^3\,c_\beta & -s_\beta^2\,c_\beta^2\end{pmatrix}
+ \frac{\delta T_{u}}{\sqrt{2}\,v\,s_\beta}
\begin{pmatrix} -s_\beta^2\,c_\beta^2 & s_\beta\,c_\beta^3\\
s_\beta\,c_\beta^3 & 1 - c_\beta^4\end{pmatrix}\\
&\quad- \delta t_\beta\,c^2_\beta
\begin{aligned}[t] \,\bigg\{
  & \!\!\begin{pmatrix}
  -s_{2\beta}\left[m_{H^\pm}^2 + \left(\lambda_4+\ell^r_5 - 2\,\lambda_1\right) v^2\right] &
  c_{2\beta} \left[m_{H^\pm}^2 - \left(2\,\lambda_3 + \lambda_4+ \ell^r_5\right) v^2\right]\\[.2ex]
  c_{2\beta} \left[m_{H^\pm}^2 - \left(2\,\lambda_3 + \lambda_4+ \ell^r_5\right) v^2\right] &
  s_{2\beta} \left[m_{H^\pm}^2 + \left(\lambda_4+ \ell^r_5 - 2\,\lambda_2\right) v^2\right]
  \end{pmatrix}\hspace{-1em}\\
  &+\begin{pmatrix}
  4\,\ell^r_6\,v^2\,c_{2\beta} &
  -2(\ell^r_6-\ell^r_7)v^2\,s_{2\beta}\\
  -2(\ell^r_6-\ell^r_7)v^2\,s_{2\beta} &
  4\,\ell^r_7\,v^2\,c_{2\beta}
  \end{pmatrix}\!\!\bigg\}\,,
  \end{aligned}
\end{split}\\
\begin{split}
  \mathbf{C}_{\mathrm{E}} &\equiv
  \begin{pmatrix}
    2\,\delta(\ell_1\,v^2)\, c_{\beta}^2
    + \delta((\ell_4 + \ell^r_5)\,v^2)\, s_{\beta}^2
    & \left[2\,\delta(\ell_3\,v^2)
        + \delta((\ell_4 + \ell^r_5)\,v^2)\right] s_{\beta}\,c_{\beta}\\
    \left[2\,\delta(\ell_3\,v^2)
      + \delta((\ell_4 + \ell^r_5)\,v^2)\right] s_{\beta}\,c_{\beta}
    & 2\,\delta(\ell_2\,v^2)\, s_{\beta}^2
    + \delta((\ell_4 + \ell^r_5)\,v^2)\, c_{\beta}^2 \end{pmatrix}\\
  &\quad-2 \begin{pmatrix}
  2\,\delta(\ell^r_6\,v^2)\, s_\beta\, c_\beta
  & \delta(\ell^r_6\,v^2)\, c^2_{\beta}
  + \delta(\ell^r_7\,v^2)\, s^2_{\beta}\\
  \delta(\ell^r_6\,v^2)\, c^2_{\beta}
  + \delta(\ell^r_7\,v^2)\, s^2_{\beta}
  & 2\,\delta(\ell^r_7\,v^2)\, s_\beta\, c_\beta
\end{pmatrix},
\end{split}\\
\delta \mathcal{M}_{A\mathrm{E}}^2 &\equiv \mathbf{C}_{A\mathrm{E}}
+ \frac{\delta T_a}{\sqrt{2}\,v}\begin{pmatrix}c_{\beta}\\s_{\beta}\end{pmatrix}
-\delta t_{\beta}\,v^2\,c_{\beta}^2
\begin{pmatrix}\ell^i_5\,c_{\beta}+2\,\ell^i_6\,s_{\beta}\\
-\ell^i_5\,s_{\beta}-2\,\ell^i_7\,c_{\beta}\end{pmatrix},\quad
\mathbf{C}_{A\mathrm{E}}\equiv
\begin{pmatrix}2\,\delta(\ell_6^i\,v^2)c_{\beta}-\delta(\ell_5^i\,v^2)s_{\beta}\\
2\,\delta(\ell_7^i\,v^2)s_{\beta}-\delta(\ell_5^i\,v^2)c_{\beta}\end{pmatrix}.
\end{align}
\end{subequations}
The renormalized diagonal self-energies of the neutral sector in the
mass basis are obtained after rotation by~$\mathbf{U}_n$:
\begin{align}
\hat{\Sigma}_{h_ih_i}(m^2_{h_i}) &\equiv
\left[\mathbf{U}_n\cdot\begin{pmatrix}
\mathbf{\hat{\Sigma}}_{\mathrm{E}}{\left(m^2_{h_i}\right)}&\mathbf{\hat{\Sigma}}_{A\mathrm{E}}{\left(m^2_{h_i}\right)}\\
\mathbf{\hat{\Sigma}}_{A\mathrm{E}}^{T}{\left(m^2_{h_i}\right)}&\mathbf{\hat{\Sigma}}_{AA}{\left(m^2_{h_i}\right)}
\end{pmatrix}\cdot\mathbf{U}^T_n\right]_{ii}.
\end{align}

From the explicit calculation of the self-energies and tadpoles, it is
possible to extract the $\xi$-dependence of these expressions:
\begin{subequations}\label{eq:xidepsigma}\allowdisplaybreaks
\begin{align}
\left[\hat{\Sigma}_{h_ih_i}(m^2_{h_i})\right]_{\xi} &=
(\mathbf{U}_n)_{ip}\,(\mathbf{U}_n)_{iq}\left\{
\frac{\mathbf{H}_{\xi}}{16\pi^2}\left[
A_0(\xi M_W^2)+\frac{1}{2}\,A_0(\xi M_Z^2)\right]
-\left[\mathbf{C}\right]_{\xi}\right\}_{pq}\,,\\
\mathbf{H}_{\xi} &\equiv \begin{pmatrix}
  2\,\ell_1\,c_{\beta}^2 + (\ell_4+\ell^r_5)\,s_{\beta}^2
  & (2\,\ell_3+\ell_4+\ell^r_5)\,s_{\beta}\,c_{\beta}& -\ell^i_5\,s_{\beta}\\
  (2\,\ell_3+\ell_4+\ell^r_5)\,s_{\beta}\,c_{\beta}
  & 2\,\ell_2\,s_{\beta}^2 + (\ell_4+\ell^r_5)\,c_{\beta}^2& -\ell^i_5\,c_{\beta}\\
  -\ell^i_5\,s_{\beta}& -\ell^i_5\,c_{\beta} &\ell_4-\ell^r_5
\end{pmatrix}\nonumber\\
&\quad-2 \begin{pmatrix}
  2\,\ell^r_6\, s_\beta\,c_\beta
  & \ell^r_6\,c^2_{\beta}+\ell^r_7\,s^2_{\beta}& -\ell^i_6\,c_{\beta}\\
  \ell^r_6\,c^2_{\beta}+\ell^r_7\,s^2_{\beta}
  & 2\,\ell^r_7\, s_\beta\,c_\beta& -\ell^i_7\,s_{\beta}\\ -\ell^i_6\,c_{\beta} & -\ell^i_7\,s_{\beta} & 0
\end{pmatrix}\,,\quad
\mathbf{C}\equiv\begin{pmatrix*}[l]
\mathbf{C}_{\mathrm{E}}&\mathbf{C}_{A\mathrm{E}}\\
\mathbf{C}_{A\mathrm{E}}^T&\mathrm{C}_{A}\end{pmatrix*}.
\end{align}
\end{subequations}
The conditions that $\hat{\Sigma}_{h_ih_i}(m_{h_i}^2)$ and
$\hat{\Sigma}_{h_ih_j}\big(\frac{1}{2}(m_{h_i}^2+m_{h_j}^2)\big)$ are
UV-finite determine the UV-divergences of the $\delta(\ell_i\,v^2)$
counterterms as well.

\tocsection{Mapping of the MSSM onto a THDM+SUSY\label{ap:mapping}}

\tocsubsection{Definition of the potential}

We want to map the MSSM onto a THDM with SUSY field content, so that
the loop-corrected Higgs masses of the MSSM are tree-level masses in
the THDM. Here, the purpose is to define Higgs self-energies that are
$\xi$-independent and only differ by a shift of higher (two-loop)
order from those in the~MSSM. The parameters~$m^2_{H^{\pm}}$,
$t_{\beta}$ and~$\mathbf{U}_n$ are kept identical between the two
models. Only the neutral Higgs masses in the~THDM differ by a shift of
one-loop order from their counterparts in the~MSSM. All the quantities
defined in appendix\,\ref{ap:renTHDM} take formally similar forms in
the~MSSM and in the~THDM. However, they differ in that the Higgs
sectors are different in both models
(unless~\mbox{$\ell_{1,\cdots,7}\equiv0$}). Below, we will indicate
this distinction by a bar~$\bar{\ }$ placed above the parameters of
the~THDM, \EG\ the Higgs masses~$\bar{m}^2_{h_i}$, or by an explicit
superscript.

The first difficulty is that the~THDM is under-constrained: the system
of neutral Higgs masses of the~MSSM fixes six degrees of freedom while
there are seven~$\lambda_i$-s and three phases---the tadpoles and
charged Higgs mass are already used to identify the quadratic THDM
parameters. We can thus extract six constraints on the~$\lambda_i$-s:
\begin{subequations}\label{eq:lambdavsmass}\allowdisplaybreaks
\begin{align}
\lambda_1+\lambda_5\,\cos\varphi_5\,t_{\beta}^2
- 2\,\lambda_6\,\cos\varphi_6\,t_{\beta} &= \frac{\bar{m}^2_{h_i^0}}{2\,v^2}\left[
  \frac{(\mathbf{U}_n)_{id}^2}{c^2_{\beta}} - (\mathbf{U}_n)_{ia}^2\,t^2_{\beta}
\right],\\
\lambda_2 +\lambda_5\,\cos\varphi_5\,t_{\beta}^{-2}
- 2\,\lambda_7\,\cos\varphi_7\,t^{-1}_{\beta} &=
\frac{\bar{m}^2_{h_i^0}}{2\,v^2}\left[
\frac{(\mathbf{U}_n)_{iu}^2}{s^2_{\beta}}
- (\mathbf{U}_n)_{ia}^2\,t^{-2}_{\beta}\right],\\
\lambda_3 + \lambda_4 - \lambda_6\,\cos\varphi_6\,t^{-1}_{\beta}
- \lambda_7\,\cos\varphi_7\,t_{\beta} &= \frac{\bar{m}^2_{h_i^0}}{2\,v^2}\left[
\frac{(\mathbf{U}_n)_{id}\,(\mathbf{U}_n)_{iu}}{s_{\beta}\,c_{\beta}}
+ (\mathbf{U}_n)_{ia}^2\right],\\
\lambda_4 - \lambda_5\,\cos\varphi_5 &= \frac{1}{v^2}\left[
\bar{m}^2_{h_i^0}(\mathbf{U}_n)_{ia}^2 - m^2_{H^{\pm}}\right],\\
\lambda_5\,\sin\varphi_5
- \left(\lambda_6\,\sin\varphi_6 + \lambda_7\,\sin\varphi_7\right) s_{2\beta} &=
-\frac{\bar{m}^2_{h_i^0}}{v^2}(\mathbf{U}_n)_{ia}\left[
(\mathbf{U}_n)_{id}\,s_{\beta} + (\mathbf{U}_n)_{iu}\,c_{\beta}\right],\\
\lambda_6\,\sin\varphi_6\,c^2_{\beta} - \lambda_7\,\sin\varphi_7\,s^2_{\beta} &=
\frac{\bar{m}^2_{h_i^0}}{2\,v^2}(\mathbf{U}_n)_{ia}\left[
(\mathbf{U}_n)_{id}\,c_{\beta} - (\mathbf{U}_n)_{iu}\,s_{\beta}\right].
\end{align}
\end{subequations}
A possible simplification consists in
imposing~$\lambda_{5,6,7}\stackrel{!}{=}0$ (as in the~MSSM). Such a
choice is only as good as the radiative corrections of
type~$\lambda_{5,6,7}$ are small in the~MSSM. For instance, large
values of the $\mu$~parameter can generate sizable contributions to
\EG~$\lambda_5$, translating in a mass-splitting between the neutral
\CP-even and \CP-odd heavy-doublet states that cannot be efficiently
mapped onto a THDM unless~\mbox{$\lambda_5\neq0$}. The exact choice of
the Higgs potential in the~THDM affects the Higgs-to-Higgs corrections
and could have a sizable impact if \EG~one of the~$\lambda_i$-s
becomes non-perturbative. Thus the exact definition of the tree-level
parameters remains an open issue in general, but the constraints of
\refeqs{eq:lambdavsmass} are fundamental in ensuring that the
masses~$\bar{m}_{h_i^0}$ (identified with the loop-corrected masses in
the~MSSM) are tree-level masses in the~THDM.

\tocsubsection{Definition of the THDM counterterms}

Now our problem rests with the determination of the
counterterms~$\bar{\delta}(\ell_i\,v^2)$ in the~THDM, or at least of
those linear combinations that enter the Higgs mass~matrix. The
extension should be smooth in the sense that
each~$\bar{\delta}(\ell_i\,v^2)$ remains an object of two-loop order
from the perspective of the counting in the~MSSM. The renormalization
scheme should also ensure gauge~invariance at the level of the
renormalized Higgs self-energies evaluated at the corresponding
tree-level Higgs mass in the~THDM, so that the definition of the Higgs
masses is gauge~invariant.

It is obvious that all the counterterms~$\delta C$ that are already
fixed (masses, tadpoles and~$t_{\beta}$) differ by terms of two-loop
order between the two models: in the~MSSM, the renormalization
constants are defined in terms of one-loop one- or two-point
functions~$\mathcal{Q}^{\text{\tiny MSSM}}_{i}$ that are evaluated
with parameters~$\lambda_i^{\text{\tiny MSSM}}$ and~$m_{h_i}^2$. In
the~THDM, the~$\mathcal{Q}^{\text{\tiny THDM}}_{i}$ are computed
from~$\lambda_i$ and~$\bar{m}_{h_i}^2$ that differ from
the~MSSM~parameters by terms of one-loop order. Formally:
\begin{subequations}
\def\stackcond{\begin{array}{@{}>{\scriptstyle}r@{}>{\scriptstyle}l@{}}\lambda_i&\to\lambda_i^{\text{\tiny MSSM}}\\[-1ex]\bar{m}^2_i&\to m_i^2\end{array}}
\setlength{\abovedisplayskip}{1.6ex}
\setlength{\belowdisplayskip}{1.6ex}
\begin{align}
  \bar{\delta}C - \delta C &= \sum_j
  \mathcal{Q}^{\text{\tiny THDM}}_j - \mathcal{Q}^{\text{\tiny MSSM}}_j\,,\\
  \mathcal{Q}^{\text{\tiny THDM}}_j &=
  \Bigg(\mathcal{Q}^{\text{\tiny THDM}}_j\bigg|_{\stackcond}\Bigg)
  + \sum_{x\in\{\lambda_k,\,\bar{m}^2_{h_l}\}}
  \Bigg(\frac{\partial\mathcal{Q}^{\text{\tiny THDM}}_j}{\partial x}\bigg|_{\stackcond}\Bigg)
  \left(x - x^{\text{\tiny MSSM}}\right)\notag\\
  &\quad+ \sum_{x,x^\prime\in\{\lambda_k,\,\bar{m}^2_{h_l}\}}
  \mathcal{O}{\left(\frac{\partial^2\mathcal{Q}^{\text{\tiny THDM}}_j}{\partial x\,dx^\prime}
  \left(x - x^{\text{\tiny MSSM}}\right)
  \left(x^\prime - x^{\prime\text{ \tiny MSSM}}\right)\right)} + \ldots\notag\\[-.9ex]
  &= \mathcal{Q}^{\text{\tiny MSSM}}_j
    + \sum_{x\in\{\lambda_k,\,\bar{m}^2_{h_l}\}}
  \Bigg(\frac{\partial \mathcal{Q}^{\text{\tiny THDM}}_j}{\partial x}\bigg|_{\stackcond}\Bigg)
  \left(x - x^{\text{\tiny MSSM}}\right)
  + \ldots\,,\\
  \omit\span\Rightarrow \bar{\delta}C - \delta C =
  \mathcal{O}{\left(\frac{\partial \mathcal{Q}^{\text{\tiny THDM}}_j}{\partial x}
    \left(x - x^{\text{\tiny MSSM}}\right)\right)} = \mathcal{O}(\text{2L})\,.
\end{align}
\end{subequations}
Similarly, the Higgs self-energies~$\Sigma_{h_ih_j}^{\text{\tiny
    THDM}}(\bar{m}^2_{h_i})$ and~$\Sigma_{h_ih_j}^{\text{\tiny
    MSSM}}(m^2_{h_i})$ differ by a shift of two-loop order.
 
If we wish to preserve the $\xi$-independence of the Higgs
self-energies evaluated in the~THDM at on-shell values of the external
momentum, then \refeqs{eq:xidepsigma} explicitly constrain the
$\xi$-dependence of the~$\bar{\delta}(\ell_i\,v^2)$-s. Similarly, the
UV-divergences can be worked out explicitly from the condition that
the renormalized self-energies (without field renormalization)
evaluated at on-shell external momentum are UV-finite in both the~MSSM
and the~THDM. The differences between the self-energies in the two
models originate from different sources: shifted external momentum;
shifted Higgs couplings; shifted Higgs masses in internal lines. It is
convenient to split these contributions between bosonic diagrams where
the modified Higgs sector directly intervenes, and fermionic
contributions where only the shifted external momentum matters:
\begin{itemize}[leftmargin=*]
\item UV-divergences from bosonic contributions
($\Delta_{\text{UV}}\equiv\left(2-\frac{D}{2}\right)^{-1}$ with
dimension $D$):
\begin{subequations}
\begin{align}
&\!\!\left[\hat{\Sigma}^{\text{\tiny THDM}}_{h_ih_i}{\left(\bar{m}^2_{h_i}\right)}
- \hat{\Sigma}^{\text{\tiny MSSM}}_{h_ih_i}{\left(m^2_{h_i}\right)}
\right]^{\text{bos.}}_{\text{UV-div}} = (\mathbf{U}_n)_{ip}\,(\mathbf{U}_n)_{iq}\left\{
-\frac{\Delta_{\text{UV}}}{16\pi^2}\,\mathbf{H}^{\text{bos.}}_{\text{UV-div}}
- \left[\mathbf{C}\right]^{\text{bos.}}_{\text{UV-div}}\right\}_{pq}\,,\\
&\mathbf{H}_{\text{UV-div}}^{\text{bos.}} = \mathbf{H}_{\xi}\left[
\frac{\xi M_Z^2}{2} + \xi M_W^2\right] - M_W^2\,\mathbf{H}_{\ell W}
- M_Z^2\,\mathbf{H}_{\ell Z} - v^2\left[
\mathbf{H}_{\ell\ell} + 2\,\mathbf{H}_{\ell\ell_6} + 2\,\mathbf{H}_{\ell\ell_7}
\right].
\end{align}
\end{subequations}
\needspace{3ex}
The matrices $\mathbf{H}_{\xi}$ and $\mathbf{C}$ were defined
in \refeqs{eq:xidepsigma} while the (symmetric) others are determined
by the following entries in the basis $(h_d^0,h_u^0,a^0)$:
\begin{subequations}\allowdisplaybreaks
\begin{align}
(\mathbf{H}_{\ell W})_{dd} &=
(\ell_1+\ell_2+4\,\ell_3+3\ell_4+5\,\ell_5^r)\,s^2_{\beta}
+ (6\,\ell_1-4\,\ell_3)\,c^2_{\beta} - 2\,(4\,\ell_6^r-\ell_7^r)\,s_{2\beta}\,,\\*
(\mathbf{H}_{\ell W})_{uu} &=
(\ell_1+\ell_2+4\,\ell_3+3\,\ell_4+5\,\ell_5^r)\,c^2_{\beta}
+ (6\,\ell_2-4\,\ell_3)\,s^2_{\beta} + 2\,(\ell_6^r-4\,\ell_7^r)\,s_{2\beta}\,,\\*
(\mathbf{H}_{\ell W})_{ud} &=
-(3\,\ell_1+3\,\ell_2-2\,\ell_3-7\,\ell_4-5\,\ell_5^r)\,s_{\beta}\,c_{\beta}
+ 2\,\ell_6^r\,(1-5\,c^2_{\beta}) + 2\,\ell_7^r\,(1-5\,s^2_{\beta})\,,\\*
(\mathbf{H}_{\ell W})_{aa} &= \ell_1+\ell_2+4\,\ell_3+3\,\ell_4-5\,\ell_5^r\,,\\*
(\mathbf{H}_{\ell W})_{da} &=
-5\,\ell_5^i\,s_{\beta}+2\,(4\,\ell_6^i-\ell_7^i)\,c_{\beta}\,,\\*
(\mathbf{H}_{\ell W})_{ua} &=
-5\,\ell_5^i\,c_{\beta}-2\,(\ell_6^i-4\,\ell_7^i)\,s_{\beta}\,,\\[.5ex]
\hline\notag\\[-3ex]
\nextParentEquation
(\mathbf{H}_{\ell Z})_{dd} &= (-9\,\ell_1+4\,\ell_3+2\,\ell_4)\,c^2_{\beta}
+\tfrac{5}{2}\,(\ell_4+\ell_5^r)\,s^2_{\beta}-3\,\ell_7^r\,s_{2\beta}\,,\\*
(\mathbf{H}_{\ell Z})_{uu} &= (-9\,\ell_2+4\,\ell_3+2\,\ell_4)\,s^2_{\beta}
+\tfrac{5}{2}\,(\ell_4+\ell_5^r)\,c^2_{\beta}-3\,\ell_6^r\,s_{2\beta}\,,\\*
(\mathbf{H}_{\ell Z})_{ud} &=
\big(3\,\ell_1+3\,\ell_2+\ell_3+\tfrac{1}{2}\,\ell_4+\tfrac{5}{2}\,\ell_5^r\big)
\,s_{\beta}\,c_{\beta}-3\,\ell_6^r\,s^2_{\beta}-3\,\ell_7^r\,c^2_{\beta}\,,\\*
(\mathbf{H}_{\ell Z})_{aa} &= \tfrac{5}{2}\,(\ell_4-\ell_5^r)\,,\\*
(\mathbf{H}_{\ell Z})_{da} &=
-\tfrac{5}{2}\,\ell_5^i\,s_{\beta}+3\,\ell_7^i\,c_{\beta}\,,\\*
(\mathbf{H}_{\ell Z})_{ua} &=
-\tfrac{5}{2}\,\ell_5^i\,c_{\beta}+3\,\ell_6^i\,s_{\beta}\,,\\[.5ex]
\hline\notag\\[-3ex]
\nextParentEquation
(\mathbf{H}_{\ell\ell})_{dd} &=
- 2\,\ell_4\,(2\,\ell_3+\ell_4) - 4\,(3\,\ell_1^2+\ell_3^2)\,c^2_{\beta}
- \ell_4\,(\ell_1+\ell_2)\,s^2_{\beta}\notag\\*
&\quad -\ell_5^r\,(\ell_1+\ell_2+4\,\ell_3+6\,\ell_4)\,s^2_{\beta}
- 2\,\lvert\ell_5\rvert^2\,(1+s^2_{\beta})\,,\\*
(\mathbf{H}_{\ell\ell})_{uu} &=
-2\,\ell_4\,(2\,\ell_3+\ell_4) - 4\,(3\,\ell_2^2+\ell_3^2)\,s^2_{\beta}
- \ell_4\,(\ell_1+\ell_2)\,c^2_{\beta}\notag\\*
&\quad -\ell_5^r\,(\ell_1+\ell_2+4\,\ell_3+6\,\ell_4)\,c^2_{\beta}
- 2\,\lvert\ell_5\rvert^2\,(1+c^2_{\beta})\,,\\*
(\mathbf{H}_{\ell\ell})_{ud} &=-\big[3\,(\ell_1+\ell_2)\,(2\,\ell_3+\ell_4)
+4\,(\ell_3^2+\ell_3\,\ell_4+\ell_4^2)\big]\,s_{\beta}\,c_{\beta}\notag\\*
&\quad -\ell_5^r\,(\ell_1+\ell_2+4\,\ell_3+6\,\ell_4)\,s_{\beta}\,c_{\beta}
- 3\,\lvert\ell_5\rvert^2\,s_{2\beta}\,,\\*
(\mathbf{H}_{\ell\ell})_{aa} &=
-\ell_4\,(\ell_1+\ell_2+4\,\ell_3+2\,\ell_4)
+ \ell_5^r\,(\ell_1+\ell_2+4\,\ell_3+6\,\ell_4)-4\,\lvert\ell_5\rvert^2\,,\\*
(\mathbf{H}_{\ell\ell})_{da} &=
\ell_5^i\,(\ell_1+\ell_2+4\,\ell_3+6\,\ell_4)\,s_{\beta}\,,\\*
(\mathbf{H}_{\ell\ell})_{ua} &=
\ell_5^i\,(\ell_1+\ell_2+4\,\ell_3+6\,\ell_4)\,c_{\beta}\,,\\[.5ex]
\hline\notag\\[-3ex]
\nextParentEquation
(\mathbf{H}_{\ell\ell_6})_{dd} &=
\ell^r_6\,(6\,\ell_1+3\,\ell_3+4\,\ell_4+5\,\ell^r_5)\,s_{2\beta}
- \ell^{r\,2}_6\,(5+7\,c^2_{\beta})\notag\\*
&\quad- 2\,\ell^r_6\,\ell^r_7\,s^2_{\beta}
+ 5\,\ell^i_5\,\ell^i_6\,s_{2\beta} - 12\,\ell^{i\,2}_6\,c^2_{\beta}\,,\\*
(\mathbf{H}_{\ell\ell_6})_{uu} &=
\ell^r_6\,(3\,\ell_3+2\,\ell_4+\ell^r_5)\,s_{2\beta} - 5\,\ell^{r\,2}_6\,c^2_{\beta}
+ \ell^i_5\,\ell^i_6\,s_{2\beta}\,,\\*
(\mathbf{H}_{\ell\ell_6})_{ud} &=
\ell^r_6\,\big[6\,\ell_1\,c^2_{\beta}+3\ell_3+2\,\ell_4\,(1+c^2_{\beta})
-(7\,\ell^r_6+5\,\ell^r_7)\,s_{\beta}\,c_{\beta}\big]\notag\\*
&\quad+(\ell^r_5\,\ell^r_6+\ell^i_5\,\ell^i_6)\,(1+4\,c^2_{\beta})
-\ell^i_6\,(\ell^i_6+2\,\ell^i_7)\,s_{2\beta}\,,\\*
(\mathbf{H}_{\ell\ell_6})_{aa} &= -\ell^i_6\,(5\,\ell^i_6+\ell^i_7)\,,\\*
(\mathbf{H}_{\ell\ell_6})_{da} &=
5\,(\ell^r_5\,\ell^i_6-\ell^i_5\,\ell^r_6)\,c_{\beta}
+ \ell^r_6\,(5\,\ell^i_6+\ell^i_7)\,s_{\beta}
- \ell^i_6\,(6\,\ell_1+3\,\ell_3+4\,\ell_4)\,c_{\beta}\,,\\*
(\mathbf{H}_{\ell\ell_6})_{ua} &= (\ell^r_5\,\ell^i_6-\ell^i_5\,\ell^r_6)\,s_{\beta}
+ \ell^r_6\,(5\,\ell^i_6+\ell_i^7)\,c_{\beta}
- \ell^i_6\,(3\,\ell_3+2\,\ell_4)\,s_{\beta}\,,\\[.5ex]
\hline\notag\\[-3ex]
\nextParentEquation
(\mathbf{H}_{\ell\ell_7})_{dd} &=
\ell^r_7\,(3\,\ell_3+2\,\ell_4+\ell^r_5)\,s_{2\beta}
- 5\,\ell^{r\,2}_7\,s^2_{\beta} + \ell^i_5\,\ell^i_7\,s_{2\beta}\,,\\*
(\mathbf{H}_{\ell\ell_7})_{uu} &=
\ell^r_7\,(6\,\ell_2+3\,\ell_3+4\,\ell_4+5\,\ell^r_5)\,s_{2\beta}
- \ell^{r\,2}_7\,(5+7\,s^2_{\beta})\notag\\*
&\quad-2\,\ell^r_6\,\ell^r_7\,c^2_{\beta}
+ 5\,\ell^i_5\,\ell^i_7\,s_{2\beta}-12\,\ell^{i\,2}_7\,s^2_{\beta}\,,\\*
(\mathbf{H}_{\ell\ell_7})_{ud} &=
\ell^r_7\,\big[6\,\ell_2\,s^2_{\beta}+3\,\ell_3+2\,\ell_4\,(1+s^2_{\beta})
-(5\,\ell^r_6+7\,\ell^r_7)\,s_{\beta}\,c_{\beta}\big]\notag\\*
&\quad+(\ell^r_5\,\ell^r_7+\ell^i_5\,\ell^i_7)\,(1+4\,s^2_{\beta})
- \ell^i_7\,(2\,\ell^i_6+\ell^i_7)\,s_{2\beta}\,,\\*
(\mathbf{H}_{\ell\ell_7})_{aa} &= -\ell^i_7\,(\ell^i_6+5\,\ell^i_7)\,,\\*
(\mathbf{H}_{\ell\ell_7})_{da} &=(\ell^r_5\,\ell^i_7-\ell^i_5\,\ell^r_7)\,c_{\beta}
+\ell^r_7\,(\ell^i_6+5\,\ell^i_7)\,s_{\beta}
-\ell^i_7\,(3\,\ell_3+2\,\ell_4)\,c_{\beta}\,,\\*
(\mathbf{H}_{\ell\ell_7})_{ua} &=5\,(\ell^r_5\,\ell^i_7-\ell^i_5\,\ell^r_7)\,s_{\beta}
+\ell^r_7\,(\ell^i_6+5\,\ell_i^7)\,c_{\beta}
-\ell^i_7\,(6\,\ell_2+3\,\ell_3+4\,\ell_4)\,s_{\beta}\,.
\end{align}
\end{subequations} 
\item UV-divergences from fermionic contributions (only terms of the third
generation are displayed):
\begin{align}
&\!\!\left[\hat{\Sigma}^{\text{\tiny THDM}}_{h_ih_i}{\left(\bar{m}^2_{h_i}\right)}
- \hat{\Sigma}^{\text{\tiny MSSM}}_{h_ih_i}{\left(m^2_{h_i}\right)}
\right]^{\text{ferm.}}_{\text{UV-div}} =\notag\\
&{-}(\mathbf{U}_n)_{ip}\,(\mathbf{U}_n)_{iq}\,\Bigg\{
\!\!\left[\mathbf{C}\right]^{\text{ferm.}}_{\text{UV-div}}
+\frac{\Delta_{\text{UV}}}{16\pi^2}\,\Bigg[\begin{aligned}[t]
&\frac{3\,m_t^2}{s^2_{\beta}}\,\mathbf{H}^{t}_{\text{UV-div}}
+\frac{3\,m_b^2+m^2_{\tau}}{c^2_{\beta}}\,\mathbf{H}^{b}_{\text{UV-div}}
+\left(M_Z^2+2\,M_W^2\right)\mathbf{H}^{t+b}_{\text{UV-div}}\\
&+\frac{s_{2\beta}}{4}\left(\frac{3\,m_t^2}{s^2_{\beta}}
-\frac{3\,m_b^2+m^2_{\tau}}{c^2_{\beta}}\right) \mathbf{H}^{\beta}_{\text{UV-div}}
\Bigg]\!\Bigg\}_{pq}\,,\taghere\end{aligned}
\end{align}
where the symmetric matrices are defined in terms of the $\ell_i$-s as
\begin{subequations}
\begin{align}
(\mathbf{H}^{t}_{\text{UV-div}})_{dd} &=0\,, &
(\mathbf{H}^{t}_{\text{UV-div}})_{aa} &= (\ell_4-\ell^r_5)\,c_{\beta}^2\,,\\
(\mathbf{H}^{t}_{\text{UV-div}})_{uu} &=
2\,\ell_2\,s^2_{\beta}+(\ell_4+\ell^r_5)\,c^2_{\beta}-2\,\ell^r_7\,s_{2\beta}\,, &
(\mathbf{H}^{t}_{\text{UV-div}})_{da} &= \big[{-}\tfrac{1}{2}\,\ell^i_5\,s_{\beta}
+\ell^i_6\,c_{\beta}\big]\,c_{\beta}^2\,,\\
(\mathbf{H}^{t}_{\text{UV-div}})_{ud} &=
\tfrac{1}{2}\,(2\,\ell_3+\ell_4+\ell^r_5)\,s_{\beta}\,c_{\beta}
-\ell^r_6\,c^2_{\beta}-\ell^r_7\,s^2_{\beta}\,, &
(\mathbf{H}^{t}_{\text{UV-div}})_{ua} &= \big[{-}\tfrac{1}{2}\,\ell^i_5\,c_{\beta}
+\ell^i_7\,s_{\beta}\big]\,(1+c_{\beta}^2)\,,\\[.5ex]
\hline\notag\\[-3ex]
\nextParentEquation
(\mathbf{H}^{b}_{\text{UV-div}})_{dd} &=
2\,\ell_1\,c^2_{\beta}+(\ell_4+\ell^r_5)\,s^2_{\beta}-2\,\ell^r_6\,s_{2\beta}\,, &
(\mathbf{H}^{b}_{\text{UV-div}})_{aa} &= (\ell_4-\ell^r_5)\,s_{\beta}^2\,,\\
(\mathbf{H}^{b}_{\text{UV-div}})_{uu} &= 0\,, &
(\mathbf{H}^{b}_{\text{UV-div}})_{da} &=
\big[{-}\tfrac{1}{2}\,\ell^i_5\,s_{\beta}+\ell^i_6\,c_{\beta}\big]\,(1+s_{\beta}^2)\,,\\
(\mathbf{H}^{b}_{\text{UV-div}})_{ud} &=
\tfrac{1}{2}\,(2\,\ell_3+\ell_4+\ell^r_5)\,s_{\beta}\,c_{\beta}
-\ell^r_6\,c^2_{\beta}-\ell^r_7\,s^2_{\beta}\,, &
(\mathbf{H}^{b}_{\text{UV-div}})_{ua} &= \big[{-}\tfrac{1}{2}\,\ell^i_5\,c_{\beta}
+\ell^i_7\,s_{\beta}\big]\,s_{\beta}^2\,,\\[.5ex]
\hline\notag\\[-3ex]
\nextParentEquation
\mathbf{H}^{t+b}_{\text{UV-div}} &\equiv
\mathbf{H}^{t}_{\text{UV-div}}+\mathbf{H}^{b}_{\text{UV-div}}\,,\tag{\theparentequation}\\[.5ex]
\hline\notag\\[-3ex]
\nextParentEquation
(\mathbf{H}^{\beta}_{\text{UV-div}})_{dd} &=
s_{2\beta}\,(\ell_4+\ell^r_5-2\,\ell_1)-4\,c_{2\beta}\,\ell^r_6\,, &
(\mathbf{H}^{\beta}_{\text{UV-div}})_{aa} &= 0\,,\\
(\mathbf{H}^{\beta}_{\text{UV-div}})_{uu} &=
s_{2\beta}\,(2\,\ell_2-\ell_4-\ell^r_5)-4\,c_{2\beta}\,\ell^r_7\,, &
(\mathbf{H}^{\beta}_{\text{UV-div}})_{da} &=
-\ell^i_5\,c_{\beta}-2\,\ell^i_6\,s_{\beta}\,,\\
(\mathbf{H}^{\beta}_{\text{UV-div}})_{ud} &=
c_{2\beta}\,(2\,\ell_3+\ell_4+\ell^r_5)+2\,s_{2\beta}\,(\ell^r_6-\ell^r_7)\,, &
(\mathbf{H}^{\beta}_{\text{UV-div}})_{ua} &=
\ell^i_5\,s_{\beta}+2\,\ell^i_7\,c_{\beta}\,.
\end{align}
\end{subequations}
\end{itemize}

\needspace{3ex}
The linear combinations of the $\bar{\delta}\ell_i$ counterterms
appearing in the Higgs self-energies thus have a determined
$\xi$-dependence and a known UV~divergence (after requiring that the
renormalized self-energies are UV~finite when setting the external
momentum to the tree-level mass). However, their finite part is not
fixed yet, since we have not identified an actual renormalization
condition. The latter seems quite arbitrary. Yet, there are two
`natural' directions.
\begin{enumerate}
\item The first one consists in requiring
  that~\mbox{$\hat{\Sigma}_{h_ih_i}^{\text{\tiny
  THDM}}(\bar{m}^2_{h_i})=\hat{\Sigma}_{h_ih_i}^{\text{\tiny
  MSSM}}(m^2_{h_i})$}, so that the counterterms~$\bar{\delta}\ell_j$
  compensate all the logarithms that are introduced by the
  shifts~\mbox{$\lambda_k^{\text{\tiny MSSM}}\to\lambda_k$}
  and~\mbox{$m^2_{h_l}\to\bar{m}^2_{h_l}$}.
\item The second choice would add new logarithms of the same type as
  those found in~$\delta M_{W,Z}^2$. This appears as a natural
  generalization of the~$\delta M_{W,Z}^2$ counterterms of the~MSSM
  that are losing legitimacy in the~THDM~framework.
\end{enumerate}
In the end, the arbitrariness in the choice of scheme mirrors the
uncertainty of the mapping that replaces the uncertainty associated
with gauge~invariance in the strict~MSSM. We can encode this
uncertainty by defining a `minimal' subtraction where only the
UV~divergence and the $\xi$-dependent terms are included in the
$\bar{\delta}\ell_i$-s. Then, the renormalization~scale included
together with~$\Delta_{\text{UV}}$ serves as a measure of the
arbitrariness introduced with the mapping procedure.

\tocsection{Matching the MSSM with an on-shell THDM+SUSY\label{ap:OSTHDMmatching}}

In the previous section, we have discussed how it was possible to
extend the Higgs self-energies of the~MSSM by a shift of two-loop
order embedded within a~THDM with SUSY matter content. The
corresponding procedure was meant to restore gauge~invariance in the
determination of the loop-corrected Higgs masses. Now, we assume that
these masses have been determined---either in the MSSM via the
truncation method or in a THDM context via the method of
appendix\,\ref{ap:mapping}---and are gauge-invariant quantities, and
we wish to consider particle scattering and decays involving the Higgs
bosons of the~MSSM in a fully on-shell context. This is not possible
in the~MSSM \textit{stricto sensu} since the quartic scalar couplings
are associated to gauge couplings. Yet, since this structure of the
couplings is not preserved by the radiative corrections, we can
instead choose to work in an effective field theory~(EFT) with
identical field content and symmetries, but with Higgs fields that are
renormalized on-shell. This~EFT is obviously a~THDM+SUSY
with~$\lambda_i$-parameters satisfying the conditions of
\refeqs{eq:lambdavsmass}. However, contrarily to the case discussed in
the previous section, the~$\bar{\delta}\ell_i$-s of this~EFT are no
longer shifts of two-loop order, since they absorb the full radiative
corrections to the Higgs masses (on-shell condition). In addition,
this~EFT is not suitable for the determination of the Higgs masses in
the~MSSM since, in the on-shell~THDM+SUSY, these masses are free
input. Yet, the matching conditions between the~MSSM and the
on-shell~THDM+SUSY are akin to the usual scheme-matching conditions
between two renormalization schemes applied to the same model.

\tocsubsection{On-shell renormalization conditions}

Let us first examine the renormalization conditions in
the~THDM+SUSY. We employ the same renormalization scheme as in
the~MSSM for most parameters (fermion, gauge-boson, SUSY masses, gauge
couplings, etc.) with the exception of the Higgs sector. There, the
renormalized self-energies (now including field renormalization) read
\begin{align}
  \hat{\Sigma}^{\text{\tiny THDM}}_{h_ih_j}{\left(p^2\right)}&=
  \Sigma^{\text{\tiny THDM}}_{h_ih_j}{\left(p^2\right)}
  - \bar{\delta}m^2_{h_ih_j}
  + \frac{\bar{\delta}Z_{h_ih_j}}{2} \left(p^2-\bar{m}_{h_i}^2\right)
  + \frac{\bar{\delta}Z_{h_jh_i}}{2} \left(p^2-\bar{m}_{h_j}^2\right).
\end{align}
The mass counterterms~$\bar{\delta}m^2_{h_ih_j}$ can be read from
\refeqs{eq:masscounterterms}. The on-shell conditions on the Higgs
masses and fields provide the following constraints:
\begin{subequations}
\begin{align}
  &\hat{\Sigma}^{\text{\tiny THDM}}_{h_ih_i}(\bar{m}_{h_i}^2) \stackrel{!}{=}0\,,
  & \bar{\delta}Z_{h_ih_i} &= -\frac{d\Sigma^{\text{\tiny THDM}}_{h_ih_i}{\left(\bar{m}_{h_i}^2\right)}}{dp^2}\,,\\
  && \bar{\delta}Z_{h_ih_j} &= -\frac{2}{\bar{m}^2_{h_j}-\bar{m}^2_{h_i}}
  \left[\Sigma^{\text{\tiny THDM}}_{h_jh_i}{\left(\bar{m}^2_{h_j}\right)} - \bar{\delta}m^2_{h_jh_i}\right],\quad
  i\neq j\,.
\end{align}
\end{subequations}
At this level, only the off-diagonal mass counterterms are not fixed
by a renormalization condition. For reasons that will become clear at
the level of the matching conditions, we continue to impose
\begin{align}\label{eq:mixingmasscond}
  \hat{\Sigma}^{\text{\tiny THDM}}_{h_ih_j}{\left(
  \frac{1}{2}\left(\bar{m}_{h_i}^2+\bar{m}_{h_j}^2\right)\right)} &=
  \hat{\Sigma}^{\text{\tiny MSSM}}_{h_ih_j}{\left(
  \frac{1}{2}\left(m_{h_i}^2+m_{h_j}^2\right)\right)}
  + \mathcal{O}(\text{2L})\,,\quad i\neq j\,,
\end{align}
\IE~the off-diagonal counterterm coincides with that of the~MSSM up to
a shift of two-loop order.

In the case where the SUSY~states involve a large mixing, the
tree-level states~$H_k$ of the on-shell model should include the
rotation lifting the degeneracy in the SUSY~model:~\mbox{$H_k =
  S_{ki}\,h_i$}. The procedure remains unchanged otherwise, except
that the SUSY self-energy and counterterms of
\refeq{eq:mixingmasscond} must also be rotated:
\begin{align}
  \hat{\Sigma}^{\text{\tiny THDM}}_{H_kH_l}{\left(
  \frac{1}{2}\left(\bar{m}_{H_k}^2+\bar{m}_{H_l}^2\right)\right)} &=
  S_{ki}\,S_{lj}\,\hat{\Sigma}^{\text{\tiny MSSM}}_{h_ih_j}{\left(
  \frac{1}{2}\left(m_{h_i}^2+m_{h_j}^2\right)\right)}+\mathcal{O}(\text{2L})\,.
\end{align}

\tocsubsection{Matching conditions}

Now we turn to the matching conditions. On the side of the MSSM, the
parameters of the Higgs sector are $M_W^2$, $M_Z^2$, $m_{H^{\pm}}^2$,
$t_{\beta}$ and $G_F$ (encoding the electroweak v.e.v.~$v$). For the
THDM, we have $m_{H^{\pm}}^2$, $\lambda_{1,\cdots,7}$,
$\varphi_{5,6,7}$, $\bar{t}_{\beta}$, $G_F$. The mixing angles (in
the \CP-even sector and/or \CP-violating) are an output of the
potential and thus \textit{a priori} different between the two
models. The matching conditions at the tree level are trivial and
provide $\lambda_i^{\text{\tiny THDM},\,(0)}=\lambda_i^{\text{\tiny
MSSM}}$, with $\lambda_i^{\text{\tiny MSSM}}$ given
in \refeqs{eq:lambdaMSSM}.

At the one-loop order, the requirement that the physical Higgs masses
coincide in both models provide the six matching conditions
of \refeqs{eq:lambdavsmass}. Furthermore, we may consider the
transition \AtoB{H^{\pm}}{t_L\,\bar{b}_R} as a condition determining
$\bar{t}_{\beta}$:
\begin{align}
  \Amp{THDM}{}{\AtoB{H^{\pm}}{t_L\,\bar{b}_R}} &=
  -\imath\,\frac{m_b}{v}\,\bar{t}_{\beta}
  + \Amp{THDM}{1L}{\AtoB{H^{\pm}}{t_L\,\bar{b}_R}}
\end{align}
and similarly for the MSSM with $\bar{t}_{\beta}\to t_{\beta}$. Then,
we observe that $\Amp{THDM}{1L}{}$ and $\Amp{MSSM}{1L}{}$ are
identical up to a shift of two-loop order: indeed, the only difference
between the two amplitudes comes from the modified Higgs potential,
but in a quantity of one-loop order, we can employ
$\lambda_i^{\text{\tiny THDM},\,(0)}$ without spoiling the
expansion. Therefore, we can choose $\bar{t}_{\beta}=t_{\beta}$ as a
matching condition of one-loop order. The same analysis
in \EG~\AtoB{h_i^0}{b\,\bar{b}} shows that we may also identify the
mixing angles provided the off-diagonal mass-counterterms are defined
as in \refeq{eq:mixingmasscond}---otherwise, the tree-level
contributions would disagree by an effect of one-loop order, forcing a
different choice of mixing angles.

At this point, we still have four unconstrained degrees of freedom in
the THDM Higgs sector. Fixing them would require considering
Higgs-to-Higgs transitions, \EG~\AtoB{h^0_2}{h^0_1\,h^0_1}
or \AtoB{H^+\,h_i^0}{H^+\,\gamma}. However, the same argument as for
the mixing angles shows that this choice is arbitrary. Indeed, since
\mbox{$\lambda_i^{\text{\tiny THDM}}-\lambda_i^{\text{\tiny MSSM}}$} is
formally of one-loop order, the distribution of corresponding finite
effects between tree level and counterterms only amounts to a formal
shift of higher order. For instance, if one chooses
\mbox{$\lambda^{\text{\tiny THDM}}_{5,6,7}\equiv0$}, it is always
possible to define $\bar{\delta}\lambda_{5,6,7}$ so that the
Higgs-to-Higgs transitions chosen as matching conditions are
satisfied. Therefore, we are left with the same arbitrariness as in
the previous section with respect to the choice of potential in the
EFT. On the other hand, the renormalization conditions are
well-defined in this scheme-conversion approach, so that the Higgs
transitions that are studied in this framework are no longer
explicitly subject to the renormalization-scale dependence (this
dependence or the gauge one is still implicitly present within the
Higgs masses that are used as input for the matching).

As a final remark, we stress that the transition amplitudes evaluated
in this on-shell THDM+SUSY framework are automatically gauge-invariant
and still a variation of two-loop order with respect to the
corresponding MSSM transition amplitudes.

\tocsection{Singlet--doublet mass-mixing in the NMSSM\label{ap:SHmix}}

We consider a mixing scenario in the \CP-conserving NMSSM using the
following
input: \mbox{$\lambda=0.7$}, \mbox{$\kappa=0.1$}, \mbox{$t_{\beta}=2$},
\mbox{$M_{H^{\pm}}=1$\,TeV}, \mbox{$\mu^{\text{eff}}=410$\,GeV},
\mbox{$\lvert A_{\kappa}\rvert\in[1,361]$\,GeV},
\mbox{$m_{\tilde{F}_{1,2}}=2$\,TeV}, \mbox{$m_{\tilde{F}_{3}}=1.5$\,TeV},
\mbox{$A_f=0$\,TeV}, \mbox{$\mu_{\text{dim}}=m_t$}. Then the two
lightest \CP-even Higgs states, including the SM-like and the singlet
components, take comparable masses and receive a relevant mixing at
the radiative order.

\begin{figure}[p!]
\centering
\includegraphics[width=\textwidth]{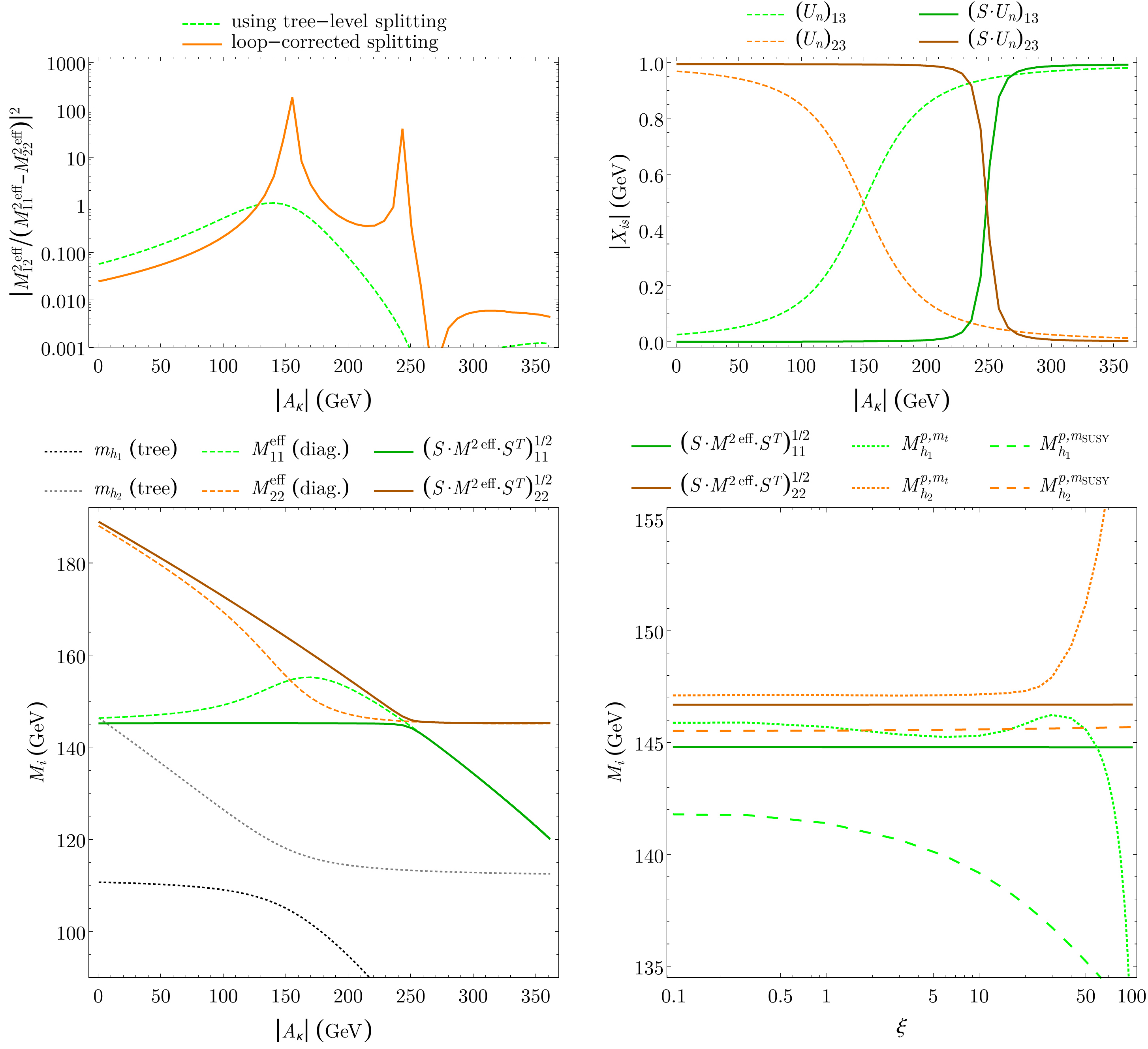}
\caption{The singlet--doublet mixing in the NMSSM is depicted
for \mbox{$\lambda=0.7$}, \mbox{$\kappa=0.1$}, \mbox{$t_{\beta}=2$},
\mbox{$M_{H^{\pm}}=1$\,TeV}, \mbox{$\mu^{\text{eff}}=410$\,GeV} and
varying $\lvert A_{\kappa}\rvert$.
\newline {\em Upper left}: the magnitude of the mixing-effect at
radiative order with respect to the tree-level (dashed green) and
loop-corrected mass-splitting (solid orange) is shown.
\newline {\em Upper right}: the magnitude (squared) of the singlet
component in the Higgs fields before (dashed) and after (solid)
radiative mixing is displayed.
\newline {\em Lower left}: the mass values obtained from the effective
mass-matrix, neglecting the radiative mixing (dashed) or including it
(solid) are shown. The corresponding results are insensitive to the
gauge-fixing parameter. The tree-level masses are represented as
dotted lines.
\newline {\em Lower right}: the gauge-parameter dependence in the mass
values at~\mbox{$\lvert A_{\kappa}\rvert\simeq243$\,GeV} are depicted,
obtained from the effective mass matrix (brown and dark-green solid
horizontal lines), and from an iterative pole-search procedure (orange
and green dotted or dashed curves with
field-renormalization scale at $m_t$ or
$m_{\text{SUSY}}$ respectively).\label{fig:SHmix}}
\end{figure}

Our renormalization scheme in the doublet sector is still determined
by `physical' conditions, connecting the Higgs sector to the masses of
the electroweak gauge bosons. On the other hand, the singlet and
singlet--doublet-mixing parameters are renormalized
$\overline{\text{DR}}$. In these circumstances, it is not completely
trivial whether gauge invariance in the Higgs masses can be discussed
without relating the $\overline{\text{DR}}$ parameters to physical
quantities. As it turns out, this subtlety does not matter for
parameters associated with a singlet state since the latter does not
couple to the electroweak gauge sector. Technically, looking
at \refeq{eq:selfxidep}, $\overline{\text{DR}}$-renormalization
conditions could only spoil the cancellation of $\tilde{C}_A$ at the
level of the renormalized self-energies. However, this cancellation is
fully ensured by the counterterms in the doublet sector, so that,
while our scheme is not `physical' in the strict sense, it still
displays the same properties in view of the gauge dependence.

The variation of $\lvert A_{\kappa}\rvert$ modulates the diagonal mass
input for the singlet component, hence the mass-splitting with the
SM-like component. The radiative mixing also varies with the
spectrum. In \fig{fig:SHmix}, we first study the magnitude of the
mixing effect between the two light \CP-even states (upper left-hand
corner). To this end, we consider the effective mass-matrix
of \refeq{eq:effmass}. Its diagonal entries are gauge invariant as
explained in section\,\ref{sec:gaugedep_mass}. The off-diagonal
self-energy contains a gauge dependence of three-loop order, which we
will not attempt to neutralize here (its effect is numerically
negligible). Then we plot $\lvert\text{mixing}/\text{diagonal
splitting}\rvert^2$: we observe that this quantity is `large' for
\mbox{$100\,\text{GeV}\lesssim\lvert A_{\kappa}\rvert\lesssim260$\,GeV},
hence that the mixing generated at radiative order is important in
this range. It can be safely neglected in the mass determination
outside---even though the mass-splitting is of electroweak order in
the full scenario. In the upper right-hand corner of \fig{fig:SHmix},
we show the magnitude of the singlet composition of the Higgs fields
at the tree level (dashed curves) and after diagonalization of
$\mathcal{M}^{2\,\text{eff}}$ (using a real orthogonal matrix $S$;
solid curves): we see that the maximal singlet--doublet mixing is
displaced from \mbox{$\lvert A_{\kappa}\rvert\sim150$\,GeV} to
$\simord250$\,GeV by the radiative effects.

In the lower left-hand quadrant of \fig{fig:SHmix}, we plot various
(almost) $\xi$-independent definitions of the Higgs masses. The dotted
lines correspond to the tree-level values: the mass of the
mostly-singlet state decreases with increasing $\lvert
A_{\kappa}\rvert$ and crosses the mass of the SM-like state at
\mbox{$\lvert A_{\kappa}\rvert\sim150$\,GeV}, leading to substantial
mixing. The (square root of the) diagonal entries of the effective
mass matrix---exactly gauge independent---are shown with dashed
curves: radiative corrections displace the mass values. Remarkable
points are \mbox{$\lvert A_{\kappa}\rvert\sim150$\,GeV}, where the
tree-level mixing is `un-mixed' (leading to an inversion of the
hierarchy between $\mathcal{M}^{2\,\text{eff}}_{11}$ and
$\mathcal{M}^{2\,\text{eff}}_{22}$), and \mbox{$\lvert
A_{\kappa}\rvert\sim250$\,GeV}, where the actual crossing of singlet
and doublet masses takes place at the one-loop order. As argued
before, this definition of the masses at one-loop order is sufficient
for \mbox{$\lvert A_{\kappa}\rvert\lesssim100$\,GeV} and \mbox{$\lvert
A_{\kappa}\rvert\gtrsim260$\,GeV}, but not in the intermediate regime
where the mixing at radiative order competes with the diagonal
mass-splitting. The masses in the mixing formalism are shown with
solid curves and `meet' at \mbox{$\lvert
A_{\kappa}\rvert\sim250$\,GeV}. As explained above, a small gauge
dependence remains present due to the off-diagonal self-energy. The
impact of the latter on the mass-determination is of order
$\simord2$--$10$\,MeV when varying the gauge-fixing parameter
between~$0.1$ and~$100$, hence completely negligible in view of
higher-order corrections. The `spikes' of the orange curve in the
upper plot can now be understood: they correspond to the points of
maximal tree-level mixing---where a large radiative mixing is needed
to counteract the `fake' tree-level mixing---and maximal mixing at the
radiative order (with a very narrow mass-splitting).

The plot on the lower right-hand corner of \fig{fig:SHmix} displays
the dependence of the mass determination on the gauge-fixing
parameter~$\xi$ for the point \mbox{$\lvert
A_{\kappa}\rvert\simeq243$\,GeV} with near-maximal mixing. The
horizontal brown and dark-green solid lines lines are obtained
with \refeq{eq:effmass} and show negligible variation. The orange and
green curves employ an iterative diagonalization procedure with
variable external momentum---set to one eigenvalue of the mass matrix
at each iteration in an attempt to solve \refeq{eq:massimplicit}: this
procedure is dependent on~$\xi$ and the difference at low~$\xi$ with
respect to the eigenvalues of \refeq{eq:effmass} cannot be viewed as a
genuine improvement in accuracy. The dependence on the field
renormalization, concerning only the iterative diagonalization
procedure, is also shown through a scale variation (dotted vs.\ dashed
curves).

\begin{figure}[tb!]
\centering
\includegraphics[width=\textwidth]{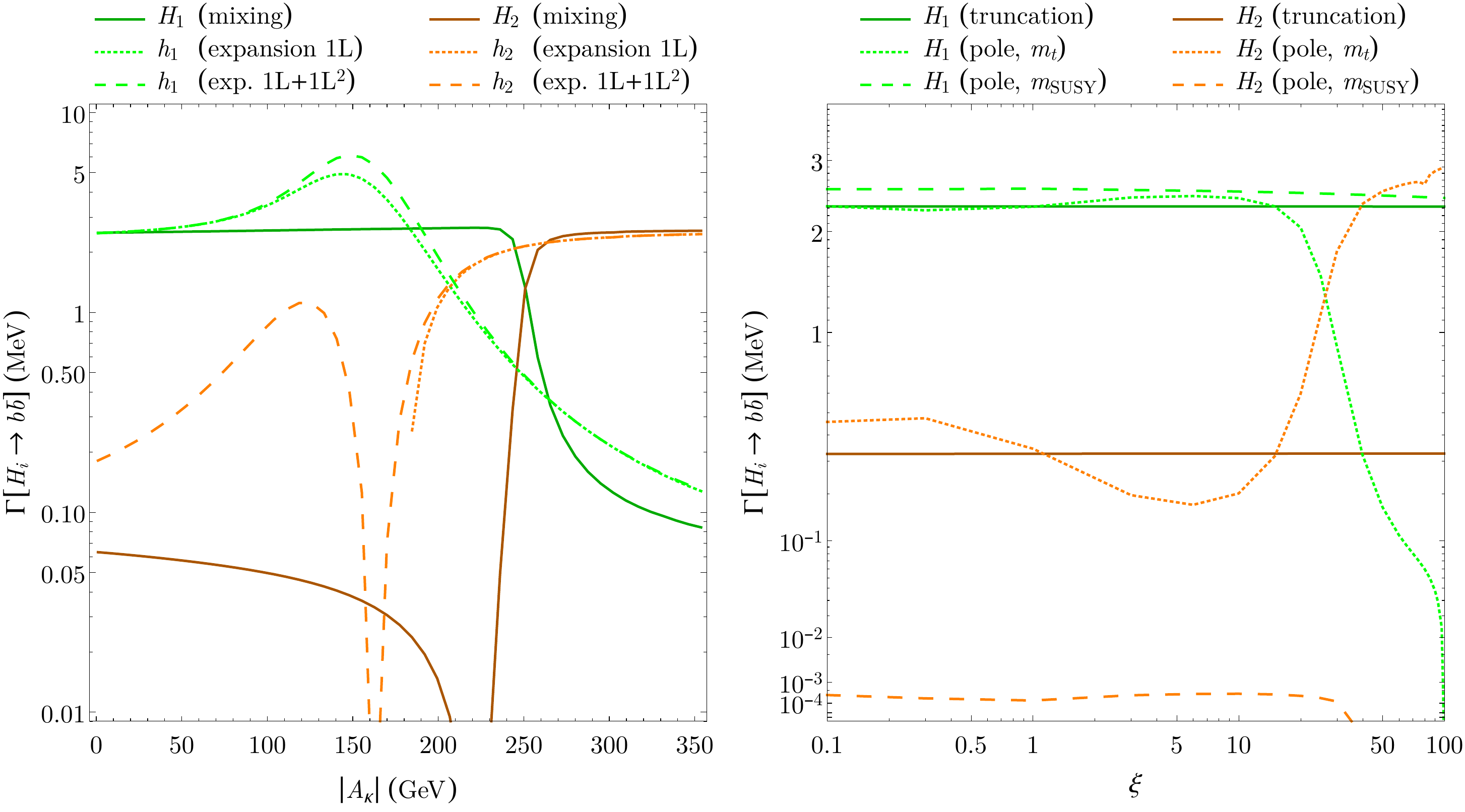}
\caption{The decay widths for the $b\bar{b}$ channel in the scenario of
\fig{fig:SHmix} are displayed.
\newline {\em Left}: the predictions of \refeq{eq:decaymix} for the
decays of the singlet--doublet admixtures $H_{1,2}$ (solid brown and
dark-green) are compared to the perturbative expansion at 1L (without
resumming the mixing, dotted green and orange) and the same
perturbative expansion including a 1L$^2$ term (dashed green and
orange).
\newline {\em Right}: the gauge- and field-renormalization dependence of
the decay widths at~\mbox{$\lvert A_{\kappa}\rvert\simeq243$\,GeV},
obtained with \refeq{eq:decaymix} (solid dark-green and brown) and
with the mixing formalism associated to an iterative pole search
(green and orange, dotted or dashed depending on the
field renormalization) are shown.\label{fig:degSD_dec}}
\end{figure}

Finally, we study the Higgs decay channel into the $b\bar{b}$ final
state. On the left-hand side of \fig{fig:degSD_dec}, we show the decay
widths for the full range of~$A_{\kappa}$. The dotted orange and green
curves correspond to the naive perturbative expansion truncated at 1L
order and ignoring the resummation of mixing effects. For the heavier
state (dotted orange), this width is negative for \mbox{$\lvert
A_{\kappa}\rvert\lesssim180$}\,GeV, highlighting the fact that the
decay is dominated by radiative effects, so that a 1L$^2$ term is
needed. For both states, the mixing contribution dominates in the
vicinity of \mbox{$\lvert A_{\kappa}\rvert\sim160$}\,GeV, showing the
necessity of resumming mixing effects in this regime. The `bump' in
the dotted green curve is related to the simultaneous variation of the
diagonal mass (see the dashed green curve in the lower left-hand plot
of \fig{fig:SHmix}). The dashed curves correspond to the same `naive'
expansion but including a (gauge- and field-counterterm independent)
1L$^2$ term: while a positive width is restored for $h_2$, the
description is still unreliable for \mbox{$\lvert
A_{\kappa}\rvert\in[100,260]$}\,GeV since the mixing dominates in this
parameter range. On the other hand, for \mbox{$\lvert
A_{\kappa}\rvert\lesssim100$}\,GeV and \mbox{$\lvert
A_{\kappa}\rvert\gtrsim260$}\,GeV (with negligible mixing), this
description is \textit{a priori} legitimate. Nevertheless, as 1L
contributions still dominate the width of $h_2$ in the
regime \mbox{$\lvert A_{\kappa}\rvert\lesssim100$}\,GeV, the
corresponding calculation cannot be seen as really predictive (the
width is compatible with $0$), and \textit{a priori} needs the
inclusion of full 2L effects for reliability. The solid brown and
dark-green lines are derived with the mixing formalism
of \refeq{eq:decaymix}. In this description, the `bump' disappears in
the intermediate regime (similarly to what happens to the masses: see
solid lines in the lower left-hand plot of \fig{fig:SHmix}) while, in
the limits \mbox{$\lvert A_{\kappa}\rvert\to0$} and \mbox{$\lvert
A_{\kappa}\rvert\sim350$}\,GeV, the widths converge towards the
predictions of the `naive' expansion---quite clearly in the case of
the doublet-dominated state (width values $\simord2.5$\,MeV) but more
slowly in the case of the singlet-dominated state (sub-MeV
widths). The cancellation at \mbox{$\lvert
A_{\kappa}\rvert\sim220$}\,GeV for the width of the heavy state
corresponds to a destructive interference between tree-level and 1L
decay amplitudes. Given
that \mbox{$m^2_{h_2}-m^2_{h_1}=\mathcal{O}(\text{1L})$} in the whole
considered range of~$A_{\kappa}$, the mixing formalism is also valid
(though not formally needed) for \mbox{$\lvert
A_{\kappa}\rvert\lesssim100$}\,GeV and \mbox{$\lvert
A_{\kappa}\rvert\gtrsim260$}\,GeV. We observe that it seems to capture
the physics of the singlet-dominated state in these regimes and offers
an alternative to the explicit inclusion of a 1L$^2$ term (dashed
lines). Yet it is not completely clear whether the associated
predictions are quantitatively meaningful for any of these
descriptions as both rely on the inclusion of partial 2L effects:
`genuine' 2L corrections may in fact compete with the 1L$^2$
contributions and affect the global properties of the singlet state.
	
On the right-hand side of \fig{fig:degSD_dec}, we compare the decay
widths obtained with \refeq{eq:decaymix} and the pole-search
formalism---\IE~the definition of a mixing matrix $\mathbf{Z}$
associated with the iterative resolution of \refeq{eq:massimplicit},
as described in \EG~\citeres{Williams:2011bu,Domingo:2018uim} and
summarized at the very end of section\,\ref{sec:1Lfermion}---for the
point \mbox{$\lvert A_{\kappa}\rvert\simeq243$}\,GeV with near-maximal
mixing. Again, we observe a significant dependence of this latter
formalism on the gauge-fixing parameter and the field
renormalization. In particular, it is clear that the prediction for
the singlet-dominated state ($H_2$) is non-quantitative, with
variations of~$\mathcal{O}(100\%)$ with the scale employed for the
field renormalization. For the doublet-dominated state ($H_1$), the
widths at low~$\xi$ are more predictive, though including variations
of~$\mathcal{O}(10\%)$. While not fully $\xi$-independent, the
procedure employing \refeq{eq:decaymix} is by construction much more
stable under these variations. Associated results are compatible with
those of the pole search. Yet again, in such a scenario with strong
mixing, 2L corrections may affect the diagonal mass-splitting, hence
the strength of the mixing, so that it is unlikely that the properties
derived at the 1L order for the singlet-dominated state are actually
predictive.

\begingroup
\let\secfnt\undefined
\newfont{\secfnt}{ptmb8t at 10pt}
\setstretch{.5}
\bibliographystyle{h-physrev}
\bibliography{literature}
\endgroup

\end{document}